\documentclass[10pt,twocolumn,twoside]{IEEEtran}

\usepackage{graphicx}
\usepackage{amsfonts,amsthm,color}
\usepackage{mathrsfs,amsmath,arydshln,latexsym}
\usepackage{amssymb}
\usepackage{cite}
\usepackage{url}
\usepackage[bookmarks,colorlinks]{hyperref}
\usepackage[linesnumbered,ruled,vlined]{algorithm2e}
\usepackage{multirow}
\usepackage{stfloats}
\usepackage{enumitem}
\usepackage{pifont}
\usepackage{subcaption}
\usepackage{nomencl}
\usepackage{colortbl}
\usepackage{array, makecell}
\usepackage{diagbox}
\usepackage{balance}

\usepackage{mwe} 
\usepackage[nopar]{kantlipsum} 

\usepackage{bm}

\usepackage[utf8]{inputenc}
\usepackage[T1]{fontenc}
\usepackage{tikz}

\usetikzlibrary{arrows, calc, decorations.markings, positioning}

\makeatletter
\newenvironment{timeline}[6]{%

    \newcommand{\startyear}{#1}
    \newcommand{\tlendyear}{#2}

    \newcommand{\yearcolumnwidth}{#3}
    \newcommand{\rulecolumnwidth}{#4}
    \newcommand{\entrycolumnwidth}{#5}
    \newcommand{\timelineheight}{#6}

    \newcommand{\templength}{}

    \newcommand{\entrycounter}{0}

    \long\def\ifnodedefined##1##2##3{%
        \@ifundefined{pgf@sh@ns@##1}{##3}{##2}%
    }

    \newcommand{\ifnodeundefined}[2]{%
        \ifnodedefined{##1}{}{##2}
    }

    \newcommand{\drawtimeline}{%
        \draw[timelinerule] (\yearcolumnwidth+5pt, 0pt) -- (\yearcolumnwidth+5pt, -\timelineheight);
        \draw (\yearcolumnwidth+0pt, -10pt) -- (\yearcolumnwidth+10pt, -10pt);
        \draw (\yearcolumnwidth+0pt, -\timelineheight+15pt) -- (\yearcolumnwidth+10pt, -\timelineheight+15pt);

        \pgfmathsetlengthmacro{\templength}{neg(add(multiply(subtract(\startyear, \startyear), divide(subtract(\timelineheight, 25), subtract(\tlendyear, \startyear))), 10))}
        \node[year] (year-\startyear) at (\yearcolumnwidth, \templength) {\startyear};

        \pgfmathsetlengthmacro{\templength}{neg(add(multiply(subtract(\tlendyear, \startyear), divide(subtract(\timelineheight, 25), subtract(\tlendyear, \startyear))), 10))}
        \node[year] (year-\tlendyear) at (\yearcolumnwidth, \templength) {\tlendyear};
    }

    \newcommand{\entry}[2]{%

        \pgfmathtruncatemacro{\lastentrycount}{\entrycounter}
        \pgfmathtruncatemacro{\entrycounter}{\entrycounter + 1}

        \ifdim \lastentrycount pt > 0 pt%
            \node[entry] (entry-\entrycounter) [below of=entry-\lastentrycount] {##2};
        \else%
            \pgfmathsetlengthmacro{\templength}{neg(add(multiply(subtract(\startyear, \startyear), divide(subtract(\timelineheight, 25), subtract(\tlendyear, \startyear))), 10))}
            \node[entry] (entry-\entrycounter) at (\yearcolumnwidth+\rulecolumnwidth+10pt, \templength) {##2};
        \fi

        \ifnodeundefined{year-##1}{%
            \pgfmathsetlengthmacro{\templength}{neg(add(multiply(subtract(##1, \startyear), divide(subtract(\timelineheight, 25), subtract(\tlendyear, \startyear))), 10))}
            \draw (\yearcolumnwidth+2.5pt, \templength) -- (\yearcolumnwidth+7.5pt, \templength);
            \node[year] (year-##1) at (\yearcolumnwidth, \templength) {##1};
        }

        \draw ($(year-##1.east)+(2.5pt, 0pt)$) -- ($(year-##1.east)+(7.5pt, 0pt)$) -- ($(entry-\entrycounter.west)-(5pt,0)$) -- (entry-\entrycounter.west);
    }

    \newcommand{\plainentry}[2]{

        \pgfmathtruncatemacro{\lastentrycount}{\entrycounter}
        \pgfmathtruncatemacro{\entrycounter}{\entrycounter + 1}

        \ifdim \lastentrycount pt > 0 pt%
            \node[entry] (entry-\entrycounter) [below of=entry-\lastentrycount] {##2};
        \else%
            \pgfmathsetlengthmacro{\templength}{neg(add(multiply(subtract(\startyear, \startyear), divide(subtract(\timelineheight, 25), subtract(\tlendyear, \startyear))), 10))}
            \node[entry] (entry-\entrycounter) at (\yearcolumnwidth+\rulecolumnwidth+10pt, \templength) {##2};
        \fi

        \ifnodeundefined{invisible-year-##1}{%
            \pgfmathsetlengthmacro{\templength}{neg(add(multiply(subtract(##1, \startyear), divide(subtract(\timelineheight, 25), subtract(\tlendyear, \startyear))), 10))}
            \draw (\yearcolumnwidth+2.5pt, \templength) -- (\yearcolumnwidth+7.5pt, \templength);
            \node[year] (invisible-year-##1) at (\yearcolumnwidth, \templength) {};
        }

        \draw ($(invisible-year-##1.east)+(2.5pt, 0pt)$) -- ($(invisible-year-##1.east)+(7.5pt, 0pt)$) -- ($(entry-\entrycounter.west)-(5pt,0)$) -- (entry-\entrycounter.west);
    }

    \begin{tikzpicture}
        \tikzstyle{entry} = [%
            align=left,%
            text width=\entrycolumnwidth,%
            node distance=8.6mm,%
            anchor=west]
        \tikzstyle{year} = [anchor=east]
        \tikzstyle{timelinerule} = [%
            draw,%
            decoration={markings, mark=at position 1 with {\arrow[scale=1.5]{latex'}}},%
            postaction={decorate},%
            shorten >=0.4pt]

        \drawtimeline
}
{
    \end{tikzpicture}
    \let\startyear\@undefined
    \let\tlendyear\@undefined
    \let\yearcolumnwidth\@undefined
    \let\rulecolumnwidth\@undefined
    \let\entrycolumnwidth\@undefined
    \let\timelineheight\@undefined
    \let\entrycounter\@undefined
    \let\ifnodedefined\@undefined
    \let\ifnodeundefined\@undefined
    \let\drawtimeline\@undefined
    \let\entry\@undefined
}
\makeatother

\makenomenclature

\allowdisplaybreaks[4]

\newcommand{\tabincell}[2]{\begin{tabular}{@{}#1@{}}#2\end{tabular}}

\begin{document}
%

\title{Holographic MIMO Communications:\\ Theoretical Foundations, Enabling Technologies, and Future Directions}
%
%
%

\author{Tierui Gong,~\IEEEmembership{Member,~IEEE,} 
        Panagiotis Gavriilidis,~\IEEEmembership{Graduate Student Member,~IEEE,}
	Ran Ji, \\
	Chongwen Huang,~\IEEEmembership{Member,~IEEE,}
	George C. Alexandropoulos,~\IEEEmembership{Senior Member,~IEEE,}
	Li Wei, 
	Zhaoyang Zhang,~\IEEEmembership{Senior Member,~IEEE,}
	Mérouane Debbah,~\IEEEmembership{Fellow,~IEEE,} 
	H. Vincent Poor,~\IEEEmembership{Life Fellow,~IEEE,} \\
	and Chau Yuen,~\IEEEmembership{Fellow,~IEEE} 
\thanks{
T. Gong, L. Wei, and C. Yuen are with the School of Electrical and Electronics Engineering, Nanyang Technological University, Singapore 639798 (e-mail: trgTerry1113@gmail.com, weili\_xd@163.com, chau.yuen@ntu.edu.sg).
}
\thanks{P. Gavriilidis and G. C. Alexandropoulos are with the Department of Informatics and Telecommunications, National and Kapodistrian University of Athens, 15784 Athens, Greece (e-mails: \{pangavr, alexandg\}@di.uoa.gr).}
\thanks{R. Ji, C. Huang, and Z. Zhang are with the College of Information Science and Electronic Engineering, Zhejiang University, Hangzhou 310027, China, and with the Zhejiang Provincial Key Laboratory of Information Processing, Communication and Networking (IPCAN), Hangzhou 310027, China. C. Huang is also with the State Key Laboratory of Integrated Service Networks, Xidian University, Xi’an 710071, China (e-mails: \{ranji, chongwenhuang, ning\_ming\}@zju.edu.cn).
}
\thanks{M. Debbah is with Khalifa University of Science and Technology, Abu Dhabi 127788, United Arab Emirates, and also with the CentraleSupelec, University ParisSaclay, 91192 Gif-sur-Yvette, France (e-mail: merouane.debbah@ku.ac.ae).}
\thanks{H. V. Poor is with the Department of Electrical and Computer Engineering, Princeton University, Princeton, NJ 08544 USA (e-mail: poor@princeton.edu).}
}

%



\maketitle

\begin{abstract}
    Future wireless systems are envisioned to create an endogenously holography-capable, intelligent, and programmable radio propagation environment, that will offer unprecedented capabilities for high spectral and energy efficiency, low latency, and massive connectivity. A potential and promising technology for supporting the expected extreme requirements of the sixth-generation (6G) communication systems is the concept of the holographic multiple-input multiple-output (HMIMO), which will actualize holographic radios with reasonable power consumption and fabrication cost. The HMIMO is facilitated by ultra-thin, extremely large, and nearly continuous surfaces that incorporate reconfigurable and sub-wavelength-spaced antennas and/or metamaterials. Such surfaces comprising dense electromagnetic (EM) excited elements are capable of recording and manipulating impinging fields with utmost flexibility and precision, as well as with reduced cost and power consumption, thereby shaping arbitrary-intended EM waves with high energy efficiency. The powerful EM processing capability of HMIMO opens up the possibility of wireless communications of holographic imaging level, paving the way for signal processing techniques realized in the EM-domain, possibly in conjunction with their digital-domain counterparts. However, in spite of the significant potential, the studies on HMIMO communications are still at an initial stage, its fundamental limits remain to be unveiled, and a certain number of critical technical challenges need to be addressed. In this survey, we present a comprehensive overview of the latest advances in the HMIMO communications paradigm, with a special focus on their physical aspects, their theoretical foundations, as well as the enabling technologies for HMIMO systems. We also compare the HMIMO with existing multi-antenna technologies, especially the massive MIMO, present various promising synergies of HMIMO with current and future candidate technologies, and provide an extensive list of research challenges and open directions for future HMIMO-empowered wireless applications.
\end{abstract}

\begin{IEEEkeywords}
	Holographic multiple-input multiple-output (HMIMO), holography, near-/far-field communications, channel modeling, performance analysis, electromagnetic information theory, channel estimation, beamforming/beam focusing, reconfigurable intelligent surfaces (RIS).
\end{IEEEkeywords}

%
\IEEEpeerreviewmaketitle

\nomenclature{5G}{Fifth Generation}
\nomenclature{eMBB}{enhanced Mobile Broadband}
\nomenclature{URLLC}{Ultra-Reliable Low Latency Communications}
\nomenclature{mMTC}{massive Machine Type Communications}
\nomenclature{6G}{Sixth Generation}
\nomenclature{mmWave}{millimeter Wave}
\nomenclature{mMIMO}{massive Multiple-Input Multiple-Output}
\nomenclature{UDN}{Ultra-Dense Network}
\nomenclature{RF}{Radio Frequency}
\nomenclature{EM}{ElectroMagnetic}
\nomenclature{$i$D}{$i$-Dimensional with $i = 1, 2, 3, 4$}
\nomenclature{HMIMO}{Holographic MIMO}
\nomenclature{TCA}{Tightly Coupled Antenna}
\nomenclature{LWA}{Leaky-Wave Antenna}
\nomenclature{RIS}{Reconfigurable Intelligent Surfaces}
\nomenclature{BS}{Base Station}
\nomenclature{UE}{User Equipment}
\nomenclature{UAV}{Unmanned Aerial Vehicle}
\nomenclature{RAN}{Radio Access Network}
\nomenclature{CGH}{Computer-Generated Hologarphy}
\nomenclature{OAM}{Orbital Angular Momentum}
\nomenclature{OMP}{Orthogonal Matching Pursuit}
\nomenclature{ISAC}{Integrated Sensing And Communications}
\nomenclature{SLM}{Spatial Light Modulator}
\nomenclature{EOM}{Electro-Optic Modulator}
\nomenclature{UTC-PD}{Uni-Traveling-Carrier PhotoDetector}
\nomenclature{TE}{Transverse Electric}
\nomenclature{TM}{Transverse Magnetic}
\nomenclature{TEM}{Transverse Electric and Magnetic}
\nomenclature{PVF}{Planar Vivaldi Feed}
\nomenclature{AVF}{Antipodal Vivaldi Feed}
\nomenclature{YUF}{Yagi-Uda-Feed}
\nomenclature{SIW}{Substate-Integrated Waveguide}
\nomenclature{TGV}{Through-Glass-Via}
\nomenclature{PCB}{Printed Circuit Board}
\nomenclature{SPP}{Surface Plasmon Polariton}
\nomenclature{CELC}{Complementary Electric Inductive Capacitive}
\nomenclature{DC}{Direct Current}
\nomenclature{LC}{Liquid Crystal}
\nomenclature{PIN}{Positive-Intrinsic-Negative}
\nomenclature{QAM}{Quadrature Amplitude Modulation}
\nomenclature{OFDM}{Orthogonal Frequency-Division Multiplexing}
\nomenclature{VR}{Virtual Reality}
\nomenclature{WPT}{Wireless Power Transfer}
\nomenclature{WEH}{Wireless Energy Harvesting}
\nomenclature{SWIPT}{Simultaneous Wireless and Information Power Transfer}
\nomenclature{WPCN}{Wireless Powered Communication Network}
\nomenclature{RL}{Reinforcement Learning}
\nomenclature{RNN}{Recurrent Neural Network}
\nomenclature{CRLB}{Cram{\'e}r-Rao Lower Bound}
\nomenclature{LEO}{Low-Earth-Orbit}
\nomenclature{CSI}{Channel State Information}

\nomenclature{DoF}{Degrees of Freedom}
\nomenclature{ULA}{Uniform Linear Array}
\nomenclature{UPA}{Uniform Planar Array}
\nomenclature{LoS}{Line-of-Sight}
\nomenclature{NLoS}{Non-Line-of-Sight}
\nomenclature{SNR}{Signal-to-Noise Ratio}
\nomenclature{MRC}{Maximum Ratio Combining}
\nomenclature{MRT}{Maximum Ratio Transmission}
\nomenclature{ZF}{Zero-Forcing}
\nomenclature{MMSE}{Minimum Mean Squared Error}
\nomenclature{LIS}{Large Intelligent Surfaces}
\nomenclature{DMA}{Dynamic Metasurface Antenna}
\nomenclature{XL-MIMO}{Extremely Large-scale MIMO}

\nomenclature{CNN}{Convolutional Neural Network}
\nomenclature{CS}{Compressed Sensing}
\nomenclature{OMP}{Orthogonal Matching Pursuit}
\nomenclature{AO}{Alternating Optimization}
\nomenclature{ADC}{Analog-to-Digital Converter}
\nomenclature{MF}{Matched Filter}
\nomenclature{HDMA}{Holographic-pattern Division Multiple Access}
\nomenclature{WDM}{Wavenumber-Division Multiplexing}
\nomenclature{CPS}{Continuous planar surface}

\printnomenclature[0.8in]
\hfill 

\section{Introduction}
%
%
%
%

The fifth-generation (5G) of wireless networks is becoming a reality and being deployed worldwide \cite{Andrews2014What, Wang2014Cellular, Agiwal2016Next}. It enables various functionalities that are shared among three pillar paradigms: enhanced mobile broadband (eMBB), ultra-reliable and low-latency communications (URLLC), and massive machine type communications (mMTC), each of which is oriented to satisfy different aspects of wireless operation requirements. Specifically, eMBB is intended for supporting high data traffic services, such as video streaming applications and mobile augmented reality, with expected peak data rates of $10$ and $20$ gigabit-per-second, as well as average data rates of $50$ and $100$ megabit-per-second in the downlink and uplink, respectively. The URLLC use case involves mission-critical applications, such as autonomous driving and remote robotic operation, with reliability $99.999\%$ and low latency of the order of $1$ millisecond. Alternatively, mMTC enables massive wireless connectivity applications, e.g., internet of things and internet of vehicles, with a requirement of one million low-cost devices per square kilometer and overall low power consumption. 
The dramatic increases in data rate needs, wireless device plurality, and thus, number of connections, and the envisioned emerging applications with extreme requirements are pushing current wireless networks to a new frontier, motivating the emergence of the sixth-generation (6G) of wireless networks \cite{Zhang20196G, Letaief2019Roadmap, Bi2019Ten, Chen2020Vision, Dang2020What, Saad2020Vision, Tariq2020Speculative, You2020Towards, Tataria20216G,timereversal6g}. It is estimated and foreseen in the 6G era that: \textit{i}) the increase in data traffic will exceed $5000$ exabytes in 2030; \textit{ii}) the services will be expanded to various environments such as space, air, ground and sea, to fulfill globally ubiquitous connections for realizing the internet of everything paradigm; \textit{iii}) emerging applications, such as holographic communications, tactile and haptic internet, fully autonomous driving, as well as high precision manufacturing and automation, will become prevalent and dominant. Under these perspectives, 6G is envisioned to offer extremely immersive experiences, full dimension coverage, extremely low latency, ultra-high reliability, as well as synthesized functionalities of communications, localization/sensing, control, and computing with native intelligence and integrated security. Compared to 5G, the envisioned 6G is expected to provide tremendous performance enhancements, offering \cite{Saad2020Vision, Tariq2020Speculative, You2020Towards, Tataria20216G}: \textit{i}) from $100$ up to $1000$ times and $10$ times higher peak data rates and average data rates, respectively, reaching over $1$ terabit-per-second and $1$ gigabit-per-second, respectively;  \textit{ii}) $10$ million devices per square kilometer connection density, which is $10$ times larger than that of 5G; \textit{iii}) over $99.99999\%$ reliability and less than $0.1$ milliseconds air interface latency; \textit{iv}) $5$ times and from $10$ up to $100$ times higher spectral and energy efficiencies, respectively; \textit{v}) up to $10$ Gigahertz (GHz) bandwidth in millimeter-wave (mmWave) frequencies and $100$ GHz in Terahertz (THz) and visible light frequencies; and \textit{vi}) centimeter-level localization/positioning accuracy and the support of high mobility communications of up to $1000$ kilometers per hour.

To satisfy the requirements of 5G and promote significant performance enhancements, major efforts have been made for boosting system capacity, reducing latency, and broadening connectivity, mainly through the following three key technologies: \textit{i}) massive MIMO (mMIMO) that involves the adoption of access points with large numbers of antennas \cite{Marzetta2010Noncooperative}; \textit{ii}) mmWave that bringd large amounts of unoccupied spectrum resources in the respective frequencies \cite{Rappaport2013Millimeter}; and \textit{iii}) ultra-densification  with densely deployed small cells \cite{Boccardi2014Five}. It is noted that the first two technologies have been verified as adequate for efficient combination, since, on one hand, the large antenna arrays are capable of offering substantial power gains that can combat the severe path loss usually appearing in high frequency signal propagation, and on the other hand, mmWave wavelengths allow the integration of large numbers of antenna elements in limited spaces, thus, compact mmWave transceivers are feasible. For networks beyond 5G, the mMIMO technology is envisioned to continuing playing a profound role, but with a scaled increase in the number of antennas; this has recently led to the extremely large-scale MIMO (XL-MIMO) concept~\cite{Yang20196G, Tataria20216G}, also known as ultra-massive MIMO. In addition, 6G is expected to include THz, or even visible light, frequencies, introducing relevant wireless applications targeting communications as well as localization and sensing. To this end, the concept of small cells currently in 5G will shrink to large amounts of tiny cells in 6G, paving the way for the framework of ultra-dense networks (UDNs).

Despite the proven feasibility of the latter technologies in 5G and their resulting enhancements, as well as their promising potential for 6G, severe problems need still to be overcome in different practical applications. A large number of radio-frequency (RF) chains, essential for supporting mMIMO transmissions, brings in large amounts of power consumption and increased hardware cost, demanding large integration areas, especially when operating at high frequencies \cite{Heath2016Overview}; this leads to unsustainable and energy inefficient system models. With the deployment of tiny cells, inter-cell interference kicks in, which imposes various challenges that need to be efficiently resolved. To this end, the coordination of access points has been proposed as an efficient solution~\cite{coordinated_beamforming}, giving lately rise to the cell-free mMIMO paradigm~\cite{Ngo2017CellFree}, which has been theoretically shown to offer interference management flexibility. However, the potential improvements introduced by latter paradigm are still unclear under realistic conditions~\cite{cooperation_limits}, where there exist imperfections in estimating intended/unintended channels as well as uncoordinated interference~\cite{robustness_CB}. Apart from the above problems, mMIMO/XL-MIMO systems follow a unified paradigm that operates under uncontrollable signal propagation environments. A shift towards an intelligent and software reconfigurable 6G is expected, where the end-to-end wireless systems, including the wireless environment, can be software programmable. It is also noteworthy that mMIMO/XL-MIMO systems achieve their critical beamforming functionality following the beam-space model, which models the spatial dimension via beams in specific angular directions; this is, however, a low dimensional approximation \cite{Rajatheva2020White}. To this end, the approximated optimality is achieved via a set of ideal assumptions: \textit{i}) predefined antenna array geometry with perfect calibrations; \textit{ii}) uncoupled antenna elements (i.e., no mutual coupling); and \textit{iii}) far-field propagation (i.e., no near-field scattering), which will be no longer valid when apertures become larger and shaped in arbitrary geometries and/or covered with dense antenna elements.

To fulfill the requirements of 6G wireless networks, while compensating the shortcomings of existing architectures, new technologies are emerging. In addition, to approach the fundamental limits of wireless environments, there is lately increasing interest in their complete characterization in the domain of the propagation of electromagnetic (EM) fields, which is expected to pave the wave for the full manipulation of EM waves. Holography, which is an innovative technology capable of recording and reconstructing the amplitude and phase of EM wavefronts, offers significant potential for the aforementioned wave manipulation, hence, it can enable holographic-type radios that can substantially contribute in meeting the extreme 6G requirements. Metamaterials and metasurfaces for wireless communications, localization, and sensing has been recently a very active area of research and development~\cite{WavePropTCCN}, serving as a candidate technology for the first releases of 5G-Advanced, as well as providing feasible solutions for supporting the realization of holography for EM wave recording and reconstruction. Metamaterials indicate a class of artificial composite materials capable of interacting with incident EM waves in various effective electric and/or magnetic responses, which are not found in nature \cite{Chen2016Review, Sun2019Electromagnetic,alexandropoulos2021reconfigurable}. They comprise collections of possibly sub-wavelength unit elements, usually termed as meta-atoms, which are metallic or dielectric micro-structures in volumetric configurations, whose effective electric and magnetic responses can be represented by permittivity and permeability, respectively. The design structure and employed material define the EM properties of the metamaterial, enabling pluralities of EM-domain functionalities. In principle, metamaterials are capable of realizing arbitrary values of permittivity and permeability, enabling them to manipulate EM waves. However, the main principle followed by metamaterial-based devices, namely accumulating propagation phases inside the devices via increasing their thickness for achieving desired wave manipulations, may in general lead to bulky structures that can increase the devices' fabrication complexity, while limiting their applicability. Metasurfaces are developed as two-dimensional (2D) equivalents of volumetric metamaterials, whose meta-atoms form an ultra-thin planar structure that can be readily fabricated. Without following the propagation phase accumulating principle of metamaterials, metasurfaces utilize the abrupt phase and amplitude discontinuity of EM waves that occurs at the interface with the meta-atoms. As such, spatially varying EM waves with desired amplitude, phase, and/or polarization can be achieved by the proper arrangement of the meta-atoms or the programmability of their responses. Metasurfaces can be further integrated with programming capabilities, leading to the concept of programmable (reconfigurable or dynamic) metasurfaces. The remarkable features offered by employing programmable metasurfaces bring a broad range of applications, such as metalens \cite{Khorasaninejad2016Metalenses}, metasurface-based holograms \cite{Ni2013Metasurface}, and metasurface-empowered cloaking \cite{Ni2015Cloak}.

\subsection{Holographic MIMO Systems}
Incorporating the powerful capabilities of holography and metasurfaces into future wireless communications, particularly revolutionizing mMIMO and XL-MIMO systems, a paradigm shift towards 6G is expected. Holographic MIMO (HMIMO) surfaces are envisioned as an efficient implementation of large antenna systems, which can interestingly go beyond the original scope of mMIMO/XL-MIMO. To further shed light on the HMIMO concept and its potential for 6G, we define \textit{HMIMO communications} as follows: ``\textit{the HMIMO paradigm refers to the physical process of completely restoring three dimensional (3D) target scenes in a realistic way, which is realized via communication ends equipped with holographic-type radios and EM-domain signal processing, while simultaneously implementing 3D dynamic wireless interactions with people, objects, and their surrounding environments}." HMIMO communications can be actualized via HMIMO surfaces, leveraging their wave manipulation capabilities. 
\begin{figure}[t!]
	\centering
	\includegraphics[height=2.4cm, width=8.8cm]{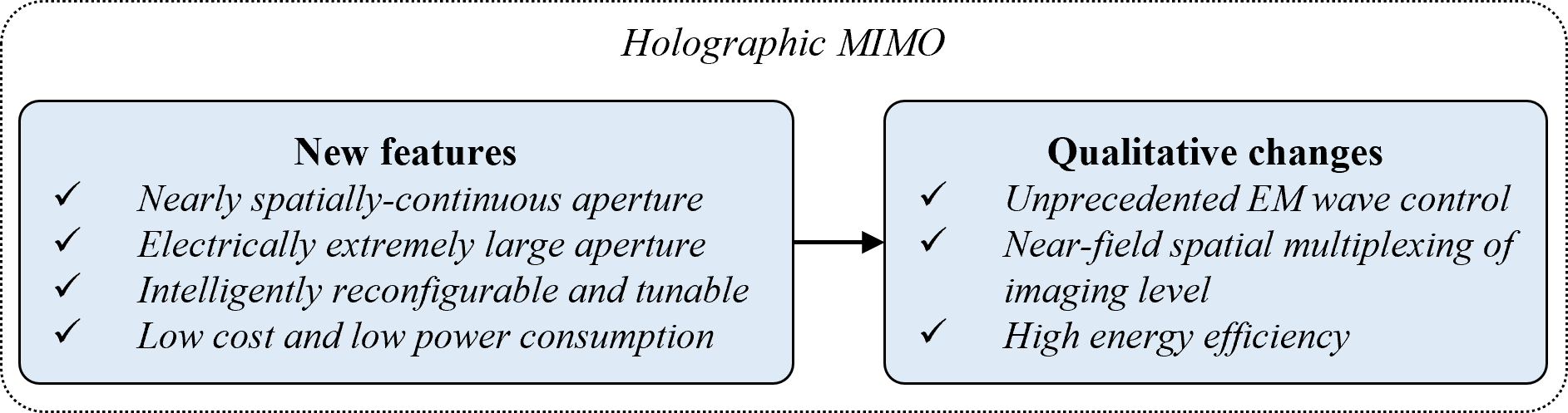}
	\caption{New features and qualitative changes offer by HMIMO systems, as compared with mMIMO systems in 5G.}
	\label{fig:HMIMOSummary}
\end{figure}

Figure~\ref{fig:HMIMOSummary} summarizes the revolutionary items offered by HMIMO, as compared with conventional mMIMO systems in 5G New Radio. Starting with the hardware characteristics, mMIMO transceivers are implemented with antenna arrays including equidistant elements with half-wavelength inter-element spacing. This condition simplifies the transceiver design, excluding mutual coupling among adjacent antenna elements. On the other hand, HMIMO surfaces can be synthesized as almost spatially continuous apertures, since their unit elements can be placed in sub-wavelength distances. This key feature is enabled by their hardware implementation (this is particularly feasible via metamaterials~\cite{Shlezinger2021Dynamic}), making them capable of forming very sharp (also termed as pencil-like) beams with weak sidelobes. More importantly, the nearly continuous apertures of HMIMO systems enable the control and recording of almost continuous phase changes of the wavefront, thus, offering manipulation of EM waves with unprecedented flexibility. The amplitude and phase tuning of HMIMO surfaces is implemented through totally different hardware than mMIMO. The latter's large numbers of costly and power-hungry RF devices are replaced in HMIMO by utilizing the holographic-based leaky-wave antennas (LWAs) \cite{Araghi2021Holographic} or the photonic tightly coupled antenna arrays (TCAs). This simplified and reconfigurable hardware is of reduced size, weight, cost and power consumption, and facilitates signal processing in the analog domain. In addition, it can be scaled in large levels, enabling the fabrication of electrically extremely large HMIMO surfaces, which can be used for combating high path-loss in high frequencies systems (e.g., in the THz frequency band~\cite{Luo_THz_RIS}).

The differences in the hardware structures of mMIMO and HMIMO, as well as their quantitative changes in terms of the placement of antenna elements (from sparse to dense) and aperture size (from small to extremely large), inevitably cause qualitative changes in the way wireless operations are designed with the two technologies. In fact, the hardware features and the resulting working mechanism of HMIMO surfaces necessitate new mathematical models for characterizing their operation and reconfigurability in efficient ways. The models need to conveniently capture the essence of their hardware architectures, while complying with any physical constraints. Interestingly, the new models may inspire new design and optimization approaches for future HMIMO communication systems. Another qualitative change in comparison with mMIMO systems is the denser and denser placement of the HMIMO unit elements to formulate almost spatially continuous apertures. In this case, mutual coupling among adjacent unit elements will kick in. Recall that mutual coupling has been considered as harmful in conventional communication systems, and it was mitigated in mMIMO antenna arrays by placing adjacent antenna elements in distances of a half wavelength. However, the proper exploitation of mutual coupling in HMIMO surfaces can possibly realize super-directivity \cite{Marzetta2019Super}, which can be further pronounced when the number of unit elements becomes larger. In addition, this feature can potentially enhance the received signal-to-noise ratios (SNRs), enlarging the structure's coverage area. All in all, it is important to study the mutual coupling effect and devise convenient mathematical models for coupling-aware wireless designs \cite{alexandg_ESPARs, qian2021mutual, Han2022Coupling}. In addition to mutual coupling, spatially continuous apertures bring forth another significant potential: signal processing can be carried out entirely on the EM-domain instead of the conventional digital domain. To this end, EM-inspired analyses and techniques can be devised to revolutionize existing wireless operation frameworks to (hybrid digital-) EM-domain-based ones, or even hybrid digital- and EM-based ones, paving the way for highly flexible processing, targeting high spatial resolution and low latency wireless communications. For example, communication and channel models can be characterized entirely in the EM-domain \cite{Dardari2020Communicating2, Sanguinetti2021Wavenumber, Pizzo2022Fourier, moustakas_2023}. It is worth mentioning that conventional mMIMO communications built upon Shannon's information theory, ignoring the underlying physical phenomena of radio wave propagation, thereby failing to characterize the ultimate fundamental limits. Blending theories from Shannon and Maxwell, EM information theory is envisioned as the next milestone for guiding wireless analyses and designs~\cite{Chafii2022Twelve}. This field can actually act as an interdisciplinary framework to evaluate the fundamental limits of wireless communications. Last but not least, the potentially large aperture sizes of HMIMO surfaces may lead to wireless operations in the near-field regime (i.e., in the Fresnel region), in contrast to conventional mMIMO systems designed for far-field scenarios, which will enable holographic-type applications. Compared with far-field mMIMO communications that are angle-aware, near-field HMIMO communications are capable of discriminating not only the angle where an object is residing, but also its distance~\cite{FD_HMIMO_2023}. This indicates that conventional angle-aware beamforming in the far field transforms to the distance-angle-aware beam focusing in the near field. This will bring significant performance benefits, such as broadening the degrees of freedom (DoF) of communication systems \cite{Dardari2021Holographic}.

The manipulation of radio waves in the EM domain and the near-field signal propagation potential render HMIMO systems capable of realizing communications of unprecedented spatial DoF and holographic imaging level with ultra-high pixel density, which can be used for extremely large spatial multiplexing \cite{Bjornson2019Massive, Zong20196G, Rajatheva2020White}. This is attributed to the feasibility of realizing HMIMO surfaces of extremely large numbers of metaterial-based antennas, which can reach the asymptotic limit of mMIMO systems, offering wireless links of very large capacity~\cite{Bjornson2018Massive}. It is worth noting that the latter asymptotic results have been primarily reported for HMIMO surfaces used as fully active transceivers~\cite{Hu2018LIS}. However, such surfaces can be also used as almost passive smart reflectors, coinciding with the concept of reconfigurable intelligent surfaces (RISs)~\cite{Huang_GLOBECOM_2019,Huang2019Reconfigurable,Renzo2019Smart} (also known as intelligent reflecting surfaces \cite{Wu2019Intelligent}). In this case, HMIMO surfaces are deployed at positions between the communication ends, enabling the smart wireless environments' paradigm~\cite{RISE-6G-EUCNC,Strinati2021Reconfigurable}, which envisions the transformation of the over-the-air signal propagation (which is conventionally treated as a random process) into a fully programmable process that can be dynamically optimized via conventional and/or artificial intelligence techniques~\cite{Alexandropoulos2022Pervasive}. It is noted for completeness that, apart from the latter versions, several other hardware architecture for HMIMO surfaces have been recently presented, for example, time-orthogonal reflecting/receiving RISs~\cite{hardware2020icassp,receivingRIS}, simultaneous reflecting and sensing RISs~\cite{HRIS_Nature,HRIS_Mag}, simultaneous reflecting and transmitting RISs~\cite{Yuanwei2021}, reflection amplifications RISs~\cite{amplifying_RIS_2022,rao2023active}, and hybrid analog and digital metasurface-based transceivers~\cite{Shlezinger2021Dynamic}.

\begin{figure*}
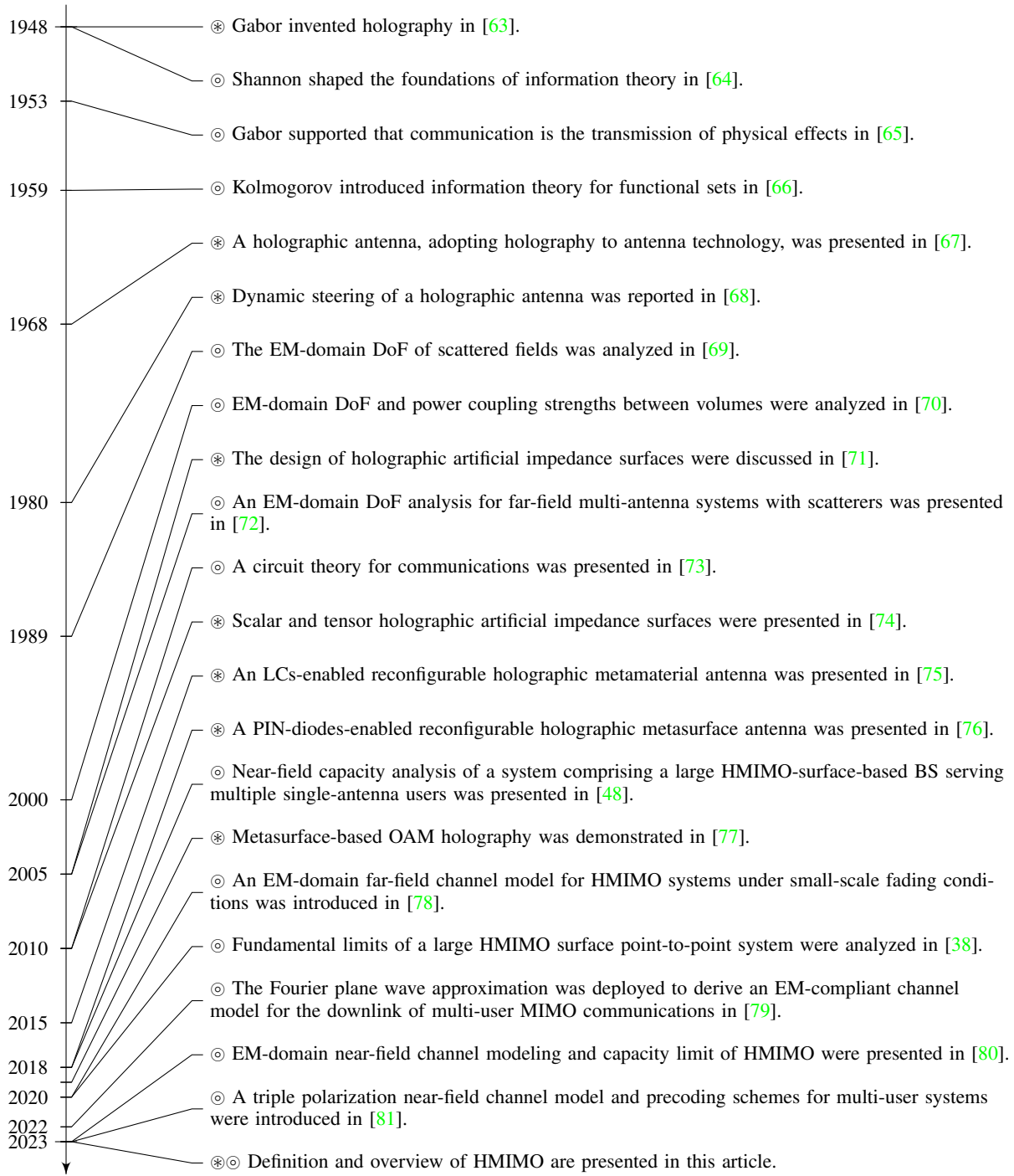

\small
\centering
    \begin{timeline}{1948}{2023}{2cm}{2cm}{12.8cm}{18.6cm}
    \entry{1948}{$\circledast$ Gabor invented holography in \cite{Gabor1948New}.}
    \entry{1948}{$\circledcirc$ Shannon shaped the foundations of information theory in \cite{Shannon1948Mathematical}.}
    \entry{1953}{$\circledcirc$ Gabor supported that communication is the transmission of physical effects in \cite{Gabor1953Communication}.}
    \entry{1959}{$\circledcirc$ Kolmogorov introduced information theory for functional sets in \cite{Kolmogorov1959Varepsilon}.}
    \entry{1968}{$\circledast$ A holographic antenna, adopting holography to antenna technology, was presented in \cite{Checcacci1968Holographic}.}
    \entry{1980}{$\circledast$ Dynamic steering of a holographic antenna was reported in \cite{Farhat1980Holographically}.}
    \entry{1989}{$\circledcirc$ The EM-domain DoF of scattered fields was analyzed in \cite{Bucci1989DOF}.}
    \entry{2000}{$\circledcirc$ EM-domain DoF and power coupling strengths between volumes were analyzed in \cite{Miller2000Communicating}.}
    \entry{2005}{$\circledast$ The design of holographic artificial impedance surfaces were discussed in \cite{Sievenpiper2005Holographic}.}
    \entry{2005}{$\circledcirc$ An EM-domain DoF analysis for far-field multi-antenna systems with scatterers was presented in \cite{Poon2005Degrees}.}
    \entry{2010}{$\circledcirc$ A circuit theory for communications was presented in \cite{Ivrlac2010Circuit}.}
    \entry{2010}{$\circledast$ Scalar and tensor holographic artificial impedance surfaces were presented in \cite{Fong2010Scalar}.} 
    \entry{2015}{$\circledast$ An LCs-enabled reconfigurable holographic metamaterial antenna was presented in \cite{Johnson2015Sidelobe}.}
    \entry{2018}{$\circledast$ A PIN-diodes-enabled reconfigurable holographic metasurface antenna was presented in \cite{Yurduseven2018Dynamically}.}
    \entry{2018}{$\circledcirc$ Near-field capacity analysis of a system comprising a large HMIMO-surface-based BS serving multiple single-antenna users was presented in \cite{Hu2018LIS}.}
    \plainentry{2019}{$\circledast$ Metasurface-based OAM holography was demonstrated in \cite{Ren2019Metasurface}.}
    \entry{2020}{$\circledcirc$ An EM-domain far-field channel model for HMIMO systems under small-scale fading conditions was introduced in \cite{Pizzo2020Spatially}.}
    \entry{2020}{$\circledcirc$ Fundamental limits of a large HMIMO surface point-to-point system were analyzed in \cite{Dardari2020Communicating2}.}
    \entry{2022}{$\circledcirc$ The Fourier plane wave approximation was deployed to derive an EM-compliant channel model for the downlink of multi-user MIMO communications in~\cite{WeiLi2022Multi-user}.}
    \entry{2023}{$\circledcirc$ EM-domain near-field channel modeling and capacity limit of HMIMO were presented in \cite{Gong2023HMIMO}.}
    \entry{2023}{$\circledcirc$ A triple polarization near-field channel model and precoding schemes for multi-user systems were introduced in  \cite{Wei2022Tri}.}   
    \entry{2023}{$\circledast$$\circledcirc$ Definition and overview of HMIMO are presented in this article.}
    \end{timeline}
    \caption{A historical timeline on the evolution of HMIMO, where $\circledast$ and $\circledcirc$ indicate representative studies in antenna technologies and theoretical foundations of HMIMO, respectively.}
    \label{fig:timeline}
\end{figure*}

\subsection{The Evolution of HMIMO}
The HMIMO concept is a seamless blend of advanced antenna technologies, including metamaterials and metasurfaces, with the fundamental theories of communications and EM waves. The former deals with the concept's efficient hardware implementation, whereas the latters are responsible for the theoretical analyses and design guidelines for HMIMO systems. HMIMO surfaces originate from the concept of the holographic antenna which was first suggested in 1968 in \cite{Checcacci1968Holographic}. That antenna was the first application of the holography paradigm which was first proposed in the antenna community by Gabor in \cite{Gabor1948New}. 
The holographic antenna concept was extended in 1980 in \cite{Farhat1980Holographically} by focusing on holographically steered mmWave antennas. Therein, possible implementations and a theory of operation were presented. The latter early holographic-type antennas were mainly based on simple hardware structures, such as strip gratings, and later improved using modulated impedance surfaces. The latter design technology led to the emergence of holographic artificial impedance surfaces in 2005 \cite{Sievenpiper2005Holographic}. The advanced tensor holographic artificial impedance surfaces, capable of performing polarization control, were further presented in \cite{Fong2010Scalar} in 2010. Those types of HMIMO surfaces, exhibiting a nearly continuous aperture, can control the design of the surface's impedance to implement various manipulations of the impinging EM wave. Obviously, the evolution of HMIMO surfaces within those ten years resulted in reconfigurable structures offering powerful control of their EM response, hence, realizing various functionalities, such as for example orbital angular momentum (OAM) multiplexing \cite{Ren2019Metasurface}. The reconfigurability of the surfaces can be realized via different mechanisms, e.g., those based on liquid crystals (LCs) \cite{Johnson2015Sidelobe} and on positive-intrinsic-negative (PIN) diodes \cite{Yurduseven2018Dynamically}. 

The theoretical foundations of HMIMO play a critical role in unveiling the fundamental limits of the technology, while motivating effective 
criteria for system designs. HMIMO can be analyzed via Shannon's information theory established in 1948 \cite{Shannon1948Mathematical}, which views communications from a mathematical perspective. A bit later, in 1953, Gabor elaborated in the physical effects of communications \cite{Gabor1953Communication}. In 1959, Kolmogorov \textit{et al.} in \cite{Kolmogorov1959Varepsilon} established information theory for sets in functional spaces, which offers a different perspective on the information as compared with Shannon's information theory. In 1989, \cite{Bucci1989DOF} studied the DoF of scattered EM fields, revealing that the DoF is equal to the Nyquist number proportional to the spatial bandwidth and the extension of the observation domain. At the beginning of the 21st century, and particularly in \cite{Miller2000Communicating}, Miller evaluated the EM-domain DoF and power coupling strengths between two arbitrary volumes in free-space showcasing that it is proportional to the transmit and receive HMIMO surface areas; the power coupling strengths are proportional to the transmit and receive volumes. Later, Poon \textit{et. al.} presented a signal space approach for analyzing the DoF for far-field multi-antenna systems in scattering environments \cite{Poon2005Degrees}, indicating that it is determined by the effective surface aperture and the angular spread. A circuit-theory-based communications theory was proposed in \cite{Ivrlac2010Circuit}, guaranteeing the consistency between communications and physical principles. In the recent years, inspired by the significant advances of HMIMO, a multitude of studies were presented, such as near-field capacity analysis \cite{Hu2018LIS}, EM-domain far-field channel modeling \cite{Pizzo2020Spatially}, power coupling and DoF analysis \cite{Dardari2020Communicating2}, as well as EM-domain near-field channel modeling and capacity limit analysis for point-to-point HMIMO systems in \cite{Gong2023HMIMO} and for multi-user systems in \cite{WeiLi2022Multi-user,Wei2022Tri}. Representative studies on indicating the evolution timeline in HMIMO research and development are summarized in Fig.~\ref{fig:timeline}.
\begin{figure*}[t!]
	\centering
	\includegraphics[height=9.0cm, width=18.2cm]{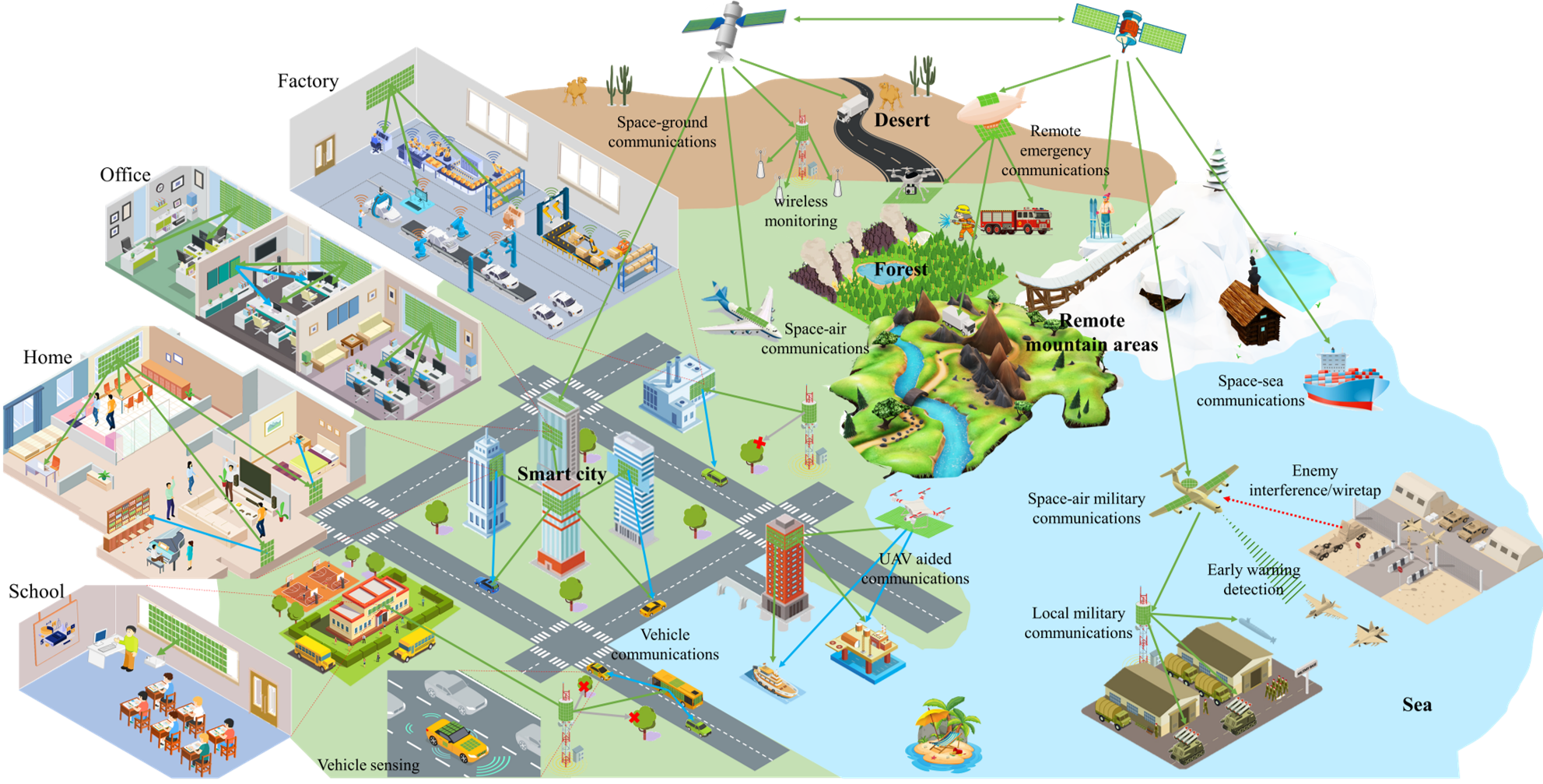}
	\caption{Future holographic-type communications enabled by HMIMO surfaces.}
	\label{fig:vision}
\end{figure*}

\subsection{The Vision for Holographic-Type Applications}
It is foreseen that the HMIMO technology has the potential to drive two key paradigm shifts from conventional 5G networks to future 6G ones. The first is the extremely large spatial multiplexing feature of holographic radios and the second are the HMIMO/RIS-enabled smart wireless environments. Based on paradigm shifts, we envision a broad range of HMIMO applications in future 6G networks, as illustrated in Fig.~\ref{fig:vision}. A generic space-air-ground-sea integrated wireless network is demonstrated that includes several scenarios of HMIMO deployment either as active transceivers or passive reflectors, such as smart cities, remote mountain areas, forests, deserts, and sea. For instance, in outdoor communications of smart cities, HMIMO surfaces can be mounted on building surfaces to serve as base stations (BSs) or passive relays for data transmissions \cite{Shlezinger2019Dynamic,xu_dynamic_2021} and localization \cite{Keykhosravi2022infeasible,RISsmartcity} to user equipments (UEs) located, for example, at offices, homes, schools, and factories, and also for communicating with satellites. They can be also installed on vehicle surfaces~\cite{Ghazalian_RISloc} or carried by unmanned aerial vehicles (UAVs), acting as passive relays for assisting vehicle and ship communications. Those installed on vehicle surfaces can be alternatively utilized for vehicle sensing, positioning, and/or tracking. On the other hand, in indoor communications of smart cities, the HMIMO surfaces can be coated on windows and/or walls for transferring outdoor BS signals to indoor UEs~\cite{3d_io_loc} and/or reflecting existing signals to meet communication requirements of indoor UEs \cite{Alexandropoulos_eucnc2022,Risscenarios}. 

In the envisioned space-air-ground-sea integrated wireless network shown of Fig.~\ref{fig:vision}, HMIMO surfaces can be placed on the solar panels of a satellite, bodies/wings of an airplane/airship, or carried by a flying object, providing a relatively high communication/sensing performance with reduced cost and power consumption. This can be truly beneficial in satisfying communication requirements for emergencies occurring in remote areas, e.g., forest fires, rescues in remote mountain areas, deserts, and sea. Moreover, HMIMO-empowered satellites are capable of assisting wireless monitoring in remote areas, such as desertification monitoring. The benefits of the HMIMO technology can be also significant in promoting physical-layer security~\cite{PLS_Kostas_all,PLS2022_counteracting}. This feature will be particularly critical for realizing secure data exchanges in military applications. For example, an early warning aircraft can utilize HMIMO surfaces to confront enemy interference and wiretaps, offering accurate and timely early warning detection, as well as assisting a timely situational awareness report and military command feedback. Far beyond the above vision, it is expected that the powerful HMIMO technology will be applied to a multitude of cases and scenarios, offering a communications paradigm shift, while supporting newly emerged upper-layer applications.
\begin{table*}[!t]
	\footnotesize
	\renewcommand{\arraystretch}{1.2}
	\caption{\textsc{Literature overview on the HMIMO technology.}}
	\label{tab:surveys}
	\centering
    \resizebox{\linewidth}{!}{
	\begin{tabular}{!{\vrule width0.6pt}c|c|c|l|l!{\vrule width0.6pt}}
	
		\Xhline{0.6pt}
		\rowcolor{yellow} \textbf{Functionality} & \textbf{Ref.} & \textbf{Date} & \qquad \qquad \quad \qquad \qquad \qquad \textbf{Main contents} & \qquad \qquad \textbf{Main contributions} \\
		
		\Xhline{0.6pt}
		\multirow{38}{*}{\tabincell{c}{Communications}} & \cite{Bjornson2019Massive} & 2019 & \tabincell{l}{Common holographic principles: Recording and reconstruction;\\ Approach for realizing approximately continuous apertures;\\ Vision and open problems.} & \tabincell{l}{Envisions and analyzes HMIMO as one of\\ the most promising directions.} \\
		
		\cline{2-5}
		& \cite{Zong20196G} & 2019 & \tabincell{l}{Holographic radio: Uplink/downlink field imaging/synthesizing;\\  Photodiode tightly coupled antenna arrays as {HMIMO surfaces}.} & \tabincell{l}{Introduces an implementation and system\\ architecture of all-photonic RANs for\\ computational holographic radios.} \\
		
		\cline{2-5}
		& \cite{Huang2020Holographic} & 2020 & \tabincell{l}{Taxonomy by power consumption: Active/passive {HMIMO surfaces};\\ Taxonomy by hardware structure: Continuous/discrete {HMIMO surfaces};\\ Fabrication methodologies;\\ Operation modes: Active transceivers/passive reflectors;\\ Functionality, characteristics, communication applications;\\ Design challenges and opportunities.} & \tabincell{l}{Overviews HMIMO surfaces from different\\ physical perspectives, lists their applications\\ and studies their performance for positioning\\ and communications.} \\
		
		\cline{2-5} 
		& \cite{Shlezinger2021Dynamic} & 2021 & \tabincell{l}{{HMIMO surfaces} for wireless communications: Passive/Active;\\ {HMIMO surfaces} for mMIMO communications: Hardware/characteristics;\\ Open research challenges.} & \tabincell{l}{Shows advanced analog signal processing\\ capabilities as well as details operations and\\ equivalent signal models of {HMIMO surfaces}\\ for transmission and reception.} \\
		
		\cline{2-5} 
		& \cite{Dardari2021Holographic} & 2021 & \tabincell{l}{Information-theoretic optimal communications of HMIMO;\\ Communication modes and power scaling laws;\\ Research directions.} & \tabincell{l}{Analyzes information-theoreticallly optimal\\ communications of HMIMO, emphasizing\\ on near-field holographic communications\\ in the large {HMIMO surfaces} regime.} \\
		
		\cline{2-5}
		& \cite{Araghi2021Holographic} & 2021 & \tabincell{l}{LWA structures and EM holographic principle;\\ Design considerations/implementations of LWA-based {HMIMO surfaces};\\ Evolution trends and summary.} & \tabincell{l}{Introduces the basic physical working principles\\ of LWA-based {HMIMO surfaces} and presents\\ various implementations for them.} \\
		
		\cline{2-5} 
		& \cite{Deng2021Reconfigurable} & 2021 & \tabincell{l}{Hardware structure: Feed, waveguide, and radiation element;\\ Holographic principle: Adjust amplitudes of radiation elements;\\ Fabrication methodologies: PIN/varactor diodes and LCs;\\ {HMIMO surfaces} aided communications: System structure/beamforming;\\ Key challenges.} & \tabincell{l}{Details the structure and holographic tuning\\ principle of {HMIMO surfaces}, presents full-wave\\ analysis, and proposes hybrid holographic\\ and digital beamforming.} \\

            \cline{2-5} 
		& \cite{Wang2022Extremely} & 2023 & {\tabincell{l}{Fundamental theories for HMIMO with respect to hardware desing,\\ channel modeling, performance analysis, and signal processing; \\ Key challenges, solutions, and future directions.}} & {\tabincell{l}{Offers a timely overview and classification\\ on different hardware and channel modeling,\\ presents an effective DoF analysis and several\\ signal processing schemes.}} \\


            \cline{2-5} 
		& \cite{Zhang20236G} & 2023 & {\tabincell{l}{Radiating near-field physical features and channel modeling;\\ Near-field beam focusing and relevant applications;\\ Design challenges and research directions.}} & {\tabincell{l}{Introduces the transition from far-field to near-\\field in terms of wave propagation and channel\\ modeling, and emphasizes the benefits and\\ applications of beam focusing.}} \\

         \cline{2-5} 
		& \cite{Zhu2022Electromagnetic} & 2022 & {\tabincell{l}{Fundamentals of information and EM wave theories;\\ Basic modeling and analysis approaches for EM information theory;\\ Applications of EM information theory.}} & {\tabincell{l}{Proposes a preliminary treatment on the\\ EM information theory and presents its\\ modeling principles, analysis, and applications.}} \\

        \cline{2-5} 
		& \cite{An2023Tutorial-1,An2023Tutorial-2,An2023Tutorial-3} & 2023 & {\tabincell{l}{Tutorial on channel modeling and channel estimation for HMIMO;\\ Tutorial on performance analysis and beamforming of HMIMO;\\ Open opportunities and challenges of HMIMO.}} & {\tabincell{l}{Presents a simple tutorial on HMIMO in terms\\ of channel modeling, performance analysis,\\ and signal processing.}} \\

		\Xhline{0.6pt}
		localization & \cite{Elzanaty2023Towards} & 2023 & \tabincell{l}{localization history and applications;\\ Holographic localization enabling technologies;\\ Performance limits and enabling algorithms;\\ Future directions.} & \tabincell{l}{Presents enabling technologies, performance\\ limits and enabling algorithms for holographic\\ localization; numerical investigation of\\ the estimation error lower bound.} \\
		\Xhline{0.6pt} 
	\end{tabular}
    }
\end{table*}

\subsection{Literature Overview}
The appealing benefits brought by HMIMO paradigm have already attracted numerous research interests. Amongst the available overview studies, the vast majority is focusing on the concept of smart radio environments, which is enabled by HMIMO surfaces mainly deployed as almost passive refonfigurable reflectors, i.e., the RIS technology. Following this mode of HMIMO operation, extensive studies have been presented~\cite{Liaskos2018New, Renzo2019Smart, Basar2019Wireless, Alghamdi2020Intelligent, Bjornson2020Reconfigurable, Sena2020What, Renzo2020Smart, ElMossallamy2020Reconfigurable, Gong2020Toward, Kisseleff2020Reconfigurable, Tang2020Wireless, Wu2020Towards, Basar2021Reconfigurable, Basharat2021Reconfigurable, Liu2021Reconfigurable, Pan2021Reconfigurable, RISE-6G-EUCNC, Strinati2021Reconfigurable, Taha2021Enabling, hardware2020icassp, Wang2021Interplay, Wu2021Intelligent, You2021Enabling, Yuan2021Reconfigurable, Bjornson2022Reconfigurable, Alexandropoulos2022Pervasive, OnlineRIS, Renzo2022Communication, faqiri2022physfad, tacit, Zhang2022Toward, Huang2022Reconfigurable, Liang2022Backscatter, Ding2022State, Liu2022Path, Mei2022Intelligent, Noh2022Channel, Pan2022Overview, Swindlehurst2022Channel, Zheng2022Survey, Ye2022Non,Shojaeifard2022,CE_overview_2022,RIScontrol2023,Risscenarios}, ranging from theoretical foundations, sophisticated hardware architectures, and operation schemes/algorithms to enabling technologies for various wireless systems in miscellaneous scenarios. We recommend readers to refer to the latter multitude of RIS overview, survey, and tutorial papers for the latest advances in the technology. On the other hand, the investigation of HMIMO surfaces as active transceivers, i.e., as the technology to implement the RF front ends of uMIMO/XL-MIMO systems, to enable the holographic-type radios is still in its infancy, and their full potential remains to be unveiled. In this survey, we mainly focus on the research area of active HMIMO surfaces, and we next summarize the latest relevant overview papers. 

In \cite{Bjornson2019Massive}, the authors outlined several new research directions beyond mMIMO, in which HMIMO was listed as one of the most promising enabling technologies. The basic principles of holography were introduced and approaches for realizing spatially continuous apertures were discussed. A general HMIMO vision was also provided and open problems with HMIMO communications were discussed. In \cite{Zong20196G}, a promising implementation of all-photonic radio access networks (RANs) was described for realizing computational holographic radios. This system is enabled by {HMIMO surfaces} implemented by photonic TCAs and optical processing. In \cite{Huang2020Holographic}, the authors overviewed HMIMO surfaces with respect and categorized the available designs in terms of hardware structure, power consumption, and fabrication methodologies, as well as operation modes. A certain number of functionalities and characteristics together with a series of communication applications were highlighted and discussed, respectively. Several challenges of HMIMO communications were also presented. 
A detailed analysis of the operations and the equivalent signal path models with dynamic metasurface antennas (DMAs), which serve as a metasurface-based implementation solution for HMIMO transceivers, was carried out in \cite{Shlezinger2019Dynamic,Shlezinger2021Dynamic}, where also their benefits for realizing XL-MIMO systems was showcased. The authors of those papers further analyzed their advantages and capabilities for 6G communications, and provided a list of challenges emerging from their extensive potential applications. Considering a different perspective, the authors in \cite{Dardari2021Holographic} put an emphasis on near-field HMIMO communications, motivated by the potential extremely large size of future HMIMO surfaces. In this regard, the authors of this paper advocated that conventional wireless propagation models are no longer applicable, and new spherical wave propagation models should be built, which will open up new communication opportunities. From a more hardware design point of view, the work in \cite{Araghi2021Holographic} introduced LWA-based {HMIMO surfaces} discussing their basic physical working principles up to various practical implementations. Moreover, in \cite{Deng2021Reconfigurable}, a basic hardware structure of HMIMO surfaces and the holographic principle for constructing holographic patterns were presented. Taking advantage of this holographic principle, the authors suggested a hybrid holographic and digital beamforming scheme for multi-user communication systems. Additionally, in \cite{Wang2022Extremely}, a contemporary overview on HMIMO was presented with an emphasis on hardware design, channel modeling, effective DoF analysis, and signal processing. Particularly focusing on near-field communications realized by large HMIMO surfaces, the authors in \cite{Zhang20236G} showcased near-field physical features, channel modeling, and beam focusing benefits of envisioned HMIMO-based 6G networks as well as discussed their possible applications. The work \cite{Zhu2022Electromagnetic} focused on HMIMO inspired EM information theory and presented a preliminary principles of this interdisciplinary theory, namely modeling, analysis, and applications. The very recent works \cite{An2023Tutorial-1, An2023Tutorial-2, An2023Tutorial-3} presented a simple tutorial on HMIMO in terms of channel modeling, performance analysis, signal processing with emphasis to channel estimation and holographic beamforming, as well as listed future opportunities and challenges. Beyond the objective of communications, the authors in \cite{Elzanaty2023Towards} studied facilitated HMIMO surfaces for realizing wireless holographic localization. In addition, the authors in~\cite{FD_HMIMO_2023} designed an angle and range estimation approach for multiple users for an HMIMO-surface-based receiver, as a part of a full duplex HMIMO integrated sensing and communications (ISAC) system operating in the near-field regime. For the ease of presentaton and comparison, we list the details of the aforedescribed HMIMO papers in Table \ref{tab:surveys}, where we also summarize their main contents and contributions.

\subsection{Contributions}
Even though the existing papers on the HMIMO paradigm overview the research area from different perspectives, spanning from physical aspects of the the concept to information-theoretic foundations as well as hardware implementation technologies and operation mechanisms, such as HMIMO beamforming, each of them focuses on  limited number of aspects providing a limited scope and details. For example, the holographic principle introduced in \cite{Bjornson2019Massive, Araghi2021Holographic, Deng2021Reconfigurable} is rather basic and focuses only on EM-based foundations and the holographic configuration for HMIMO surfaces. In addition, these papers do not detail holographic-type applications, neither include any technological roadmaps. The studies \cite{Bjornson2019Massive, Zong20196G, Huang2020Holographic, Shlezinger2021Dynamic, Deng2021Reconfigurable, Dardari2021Holographic,Wang2022Extremely} are positioned around hardware designs and lack a more panoramic view of the HMIMO communications potential. Furthermore, although the works \cite{Dardari2021Holographic,Deng2021Reconfigurable,Wang2022Extremely,Zhang20236G,Zhu2022Electromagnetic,An2023Tutorial-1,An2023Tutorial-2,An2023Tutorial-3} include theoretical foundations for HMIMO as well as enabling technologies, such as channel modeling, communication modes, EM information theory principles, channel estimation, beamforming and beam focusing, the treatment is limited. Motivated by this status on the open HMIMO literature, we present a extended overview of the HMIMO technology that covers relevant various aspects in detail, aiming to not only provide a useful reference to the interested readers, but also to serve as a holistic survey on the topic inspiring valuable future research and innovation works. The major contributions of this survey are summarized as follows. 
\begin{itemize}
	\item We present various holographic-type applications for future wireless communications, which are categorized in terms of several aspects, including entertainment, education, medical healthcare, and production. Different technology enablers of holography for realizing the envisioned HMIMO communications paradigm are introduced. Several aspects of holography, ranging from the original optical holography to the computer-generated holography (CGH) and EM holography, are presented, and their differences are discussed in depth.
	\item A comprehensive systematic overview of the physical aspects of HMIMO surfaces, which constitute a critical enabler of EM holography, is provided to help the readers understand their underlying working principle. The physical aspects with respect to hardware structures, holographic design methodologies, tuning mechanisms, aperture shapes, and typical functionalities with such surfaces are presented. We also overview the recent practical deployments of HMIMO surfaces, including a list of representative hardware prototypes as well as HMIMO trials used for communications. 
	\item The recent theoretical foundations of HMIMO communications are overviewed. In particular, representative HMIMO channel models under line-of-sight (LoS) and non-LoS (NLoS) conditions are presented together with performance analyses. To this end, characterizations of the DoF and system capacity of HMIMO communications in LoS and NLoS environments are included. We also review the EM wave sampling and EM information theory, and for the latter, we discuss the effective analysis and design tools of EM wave and circuit theories.
	\item We provide a contemporary survey on the latest progress in HMIMO channel estimation, beamforming, and beam focusing, emphasizing, among other aspects, on the near-field regime and hybrid-field schemes. The state-of-the-art is listed in tables with respect to different system models and channel types used, as well as to estimation and optimization approaches.
	\item We present the distinctive features of HMIMO communication systems, as compared with conventional MIMO and mMIMO systems as well as RISs, and detail relevant enabling physical-layer technologies. The comparison with mMIMO is particularly focused on the aspects of hardware components, directivity, coverage, capacity,  and energy efficiency. 
       \item A detailed discussion on the research challenges and future directions of the HMIMO technology  with respect to hardware design and experimentation, fundamental limits, and operation mechanisms is provided.
\end{itemize}

\begin{figure*}[t!]
	\centering
	\includegraphics[height=15.0cm, width=18.2cm]{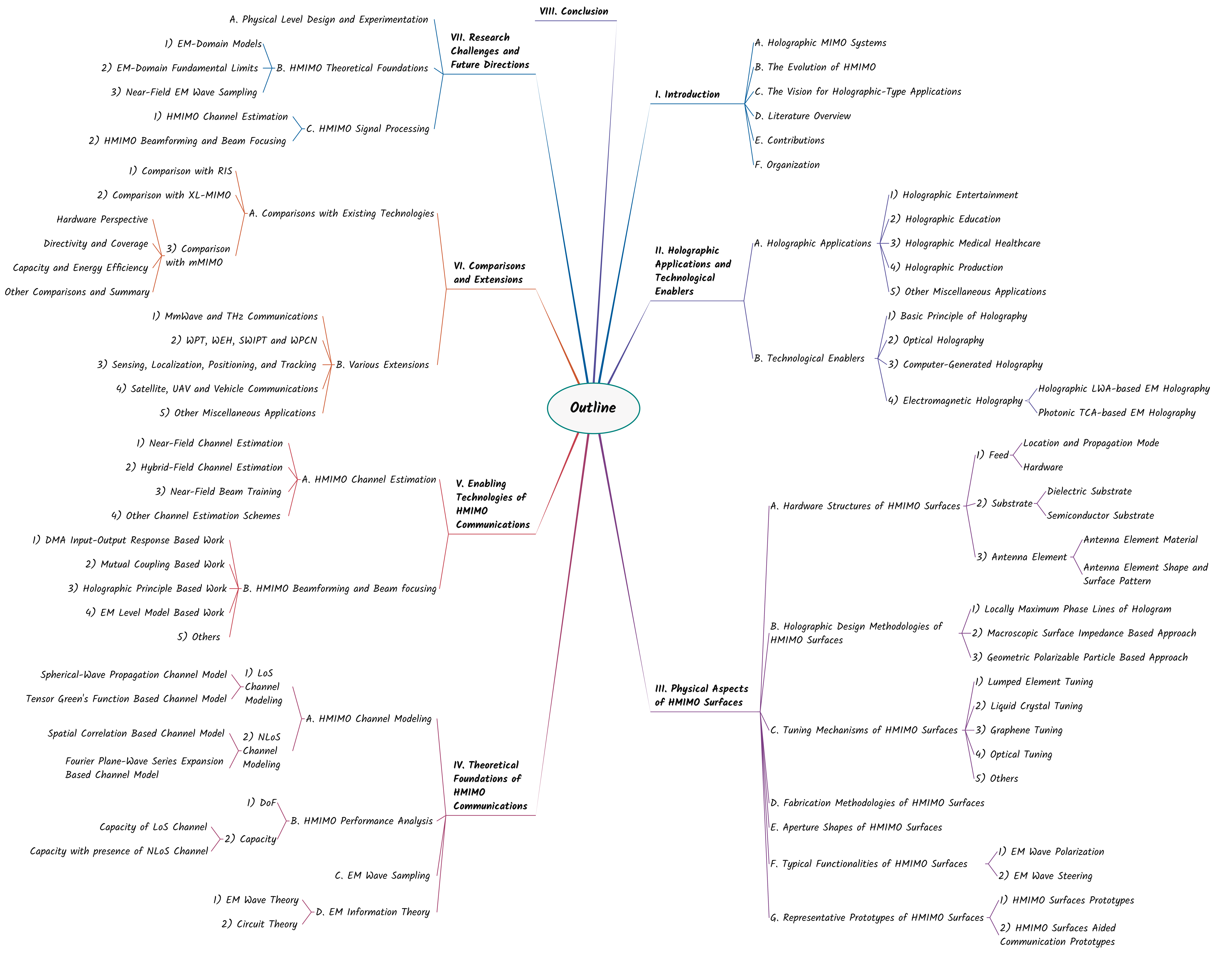}
	\caption{The organization of the current survey.}
	\label{fig:outline}
\end{figure*}

\subsection{Organization}
The rest of the survey is organized as follows: Section~\ref{SectionII} presents holographic-type applications and relevant technology enablers, while Section~\ref{SectionIII} details various physical aspects of HMIMO surfaces. Theoretical foundations of HMIMO communications are provided in Section~\ref{SectionIV}, while relevant enabling technologies are discussed in Section~\ref{SectionV}. Comparisons of HMIMO surfaces for communications with existing MIMO technologies, as well as extensions of HMIMO to various topics, are included in Section~\ref{SectionVI}. Research challenges and future directions of HMIMO surfaces are presented in Section~\ref{SectionVII}, while the conclusion of the survey are included in Section~\ref{SectionVIII}. The main structure of the survey is schematically illustrated in Fig.~\ref{fig:outline}.


\section{Holographic Applications\\ and Technological Enablers}
\label{SectionII}
Holography is an innovative methodology for high-fidelity 3D imaging that exploits both wave amplitude (intensity) and phase information. It was first invented to improve the resolution of electron microscopy by Dennis Gabor in 1948 \cite{Gabor1948New}. Compared with the conventional geometrical-optics-based photography, which records only the wave intensity to obtain a 2D image, holography utilizes coherent light sources to record the complete wave information in order to derive a 3D image. Holography includes two typical stages, namely, the \textit{recording} and \textit{reconstruction} phases, which follow respectively the interference and diffraction principles of waves. Interference occurs when one or more wavefronts are superimposed, whereas diffraction takes place when a wavefront encounters an object. Completely different from the point-to-point mapping of conventional photography, in holography, each object point is recorded by the whole recording surface of the holographic device, forming a point-to-surface mapping rule. Each point on the recording surface of the holography captures the information from all the points of the object to be mapped. The detailed differences between holography and conventional photography are summarized in Table~\ref{tab:holo_photo_comp}.

With the emergence of holography, many applications and technologies were spurred and developed. It is envisioned that the future 6G technologies will provide extremely immersive experience in entertainment, education, medical healthcare, production, and so forth, by means of the holography paradigm~ \cite{Dang2020What,Tariq2020Speculative,Tataria20216G}. With holography, it is feasible to naturally reconstruct a realistic scenery, breaking the barriers between the virtual and actual scenes as well as blending the virtual and real worlds seamlessly. In this section, we first describe potential holographic-type applications for 6G networks, and then, present the main holographic technology enablers to satisfy the requirements for the envisioned holographic-type applications. 
\begin{table*}[!t]
	\footnotesize
	\renewcommand{\arraystretch}{1.2}
	\caption{\textsc{Differences between holography and conventional photography.}}
	\label{tab:holo_photo_comp}
	\centering
	\begin{tabular}{!{\vrule width0.6pt}c|c|c!{\vrule width0.6pt}}
	
		\Xhline{0.6pt}
		\rowcolor{yellow} 
		\textbf{Metrics/Features} & \textbf{Holography} & \textbf{Photography} \\
		
		\Xhline{0.6pt}
		Image dimension & 3D & 2D \\
		
		\hline 
		Information recorded & Wave amplitude (intensity) and phase & Wave amplitude (intensity) \\
		
		\hline 
		Mapping rule & Point-to-surface mapping
		& Point-to-point mapping
		\\
		\hline 
		Light source requirement & Coherent light source & Non-coherent light source \\
		\hline 
		Principle followed & Wave interference/diffraction for recording/reconstruction & Geometrical optics \\
		\Xhline{0.6pt}
	\end{tabular}
\end{table*}

\subsection{Holographic Applications} 
\begin{figure*}[t!]
	\centering
	\includegraphics[height=13.6cm, width=18cm]{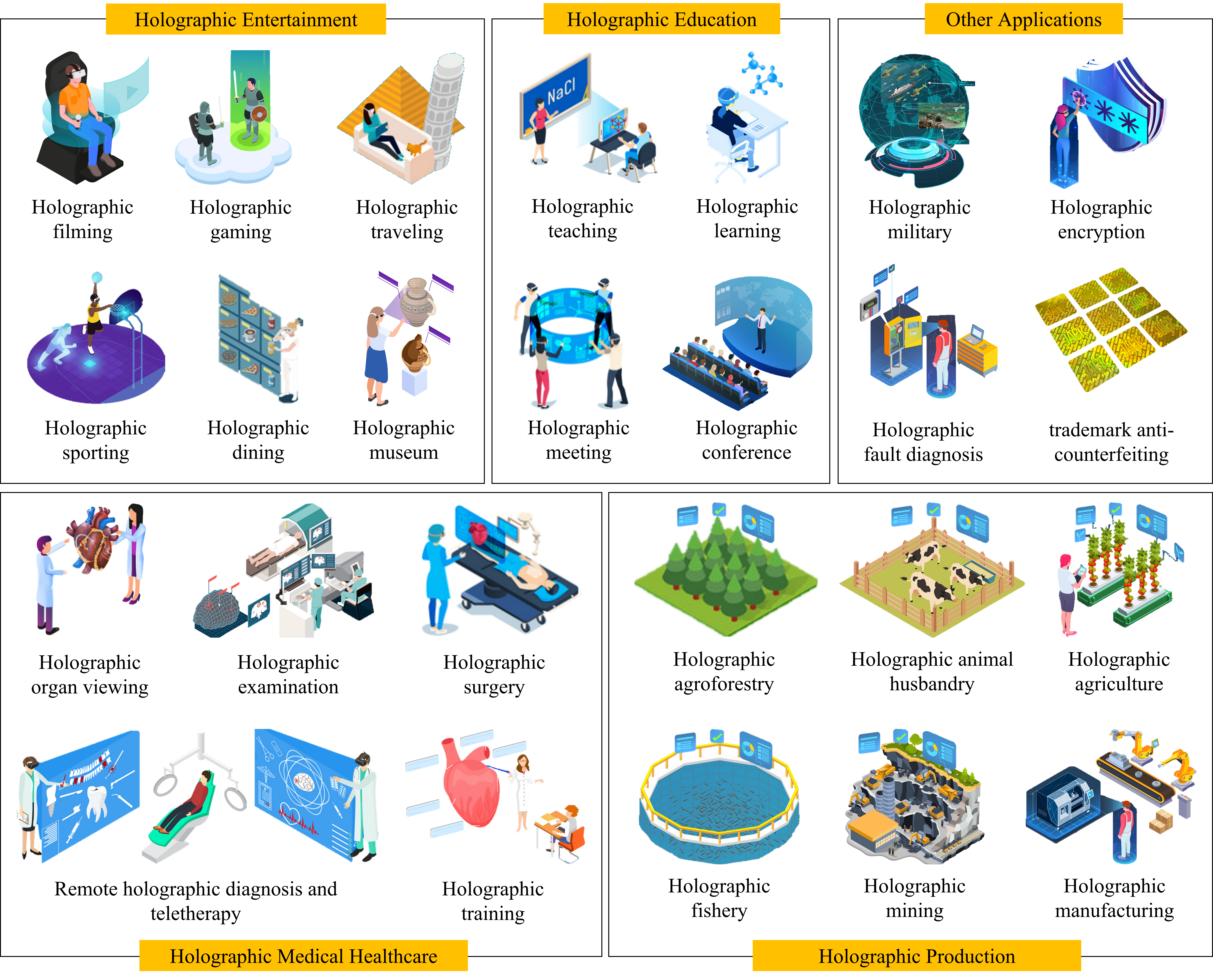}
	\caption{Holographic applications: (a) entertainment, (b) education, (c) medical healthcare, (d) production, and (e) others.}
	\label{fig:holographic_applications}
\end{figure*}

We present several holographic-type applications for 6G wireless networks, encompassing holographic entertainment, holographic education, holographic medical healthcare, and holographic production. In each application, possible perspectives are proposed. Their details are further depicted in Fig.~\ref{fig:holographic_applications} with corresponding illustrations.

\subsubsection{Holographic Entertainment}
Filming, gaming, sports, traveling, dining, and cultural cultivation, to name a few, are expected to be empowered and fully revolutionized by holography. People watching a 3D movie will not be limited to traditional 3D video using binocular parallax, but immersed in a realistic viewing experience provided by holography. For strongly interactive entertainment (e.g., holographic gaming, sporting, and traveling), a fusion of holograms with multi-dimensional senses (such as senses of sight, hearing, and feeling) will further construct both deeply interactive and extremely abundant connections between UEs and environments. A multiplicity of interactively immersive experiences can therefore be perfectly achieved. Holographic restaurants can be built where the senses of smell, taste, and sight will be fully integrated and clients will be able to experience the whole process of ordering, waiting, and eating tailored to their own requirements and preferences. For people who are interested in cooking, the platform will be able to include them in the production process of various dishes seamlessly. People can also entertain themselves with holographic cultural experiences, such as understanding etiquette and customs of a country, watching drama, and listening to concerts through a holographic virtual stage, as well as visiting museums and taking part in auctions. To date, Microsoft Hololens2 \cite{Mircrosoft_HoloLens} and Hypervsn \cite{Hypervsn} are two representative products for supporting holographic entertainment.

\subsubsection{Holographic Education}
Holography can make learning and teaching more efficient, immersive, and consistent; this would actually be a service of paramount importance during the recent pandemic. Students from different locations will be able to attend the same mixed-reality classes. This will provide a realistic learning experience for them, but without the need to relocate. Teachers will also benefit from teaching holographic classrooms. They will be able to demonstrate historic events, and also perform complex experiments that would be impossible in conventional classrooms. Simultaneously teaching multiple classrooms will be enabled, something very important for underdeveloped countries. For people who conduct scientific research, and target at sharing their latest research advances with colleagues via international conferences or workshops, live holographic video conferences will enable fully immersive talks, presentations, and collaborations. Holography will be employed to promote the popularization of science and technology, enriching people's scientific understanding and curiosity.

\subsubsection{Holographic Medical Healthcare}
Holography will provide doctors the ability to visualize the human body with high resolution. This will be used by radiology, planning of surgeries, and precise human tissue reconstruction. Surgeons will be able to carry out sophisticated surgeries in distant hospitals and medical professionals will treat their patients remotely. It will also improve the access to specialized consultants for underdeveloped remote areas, where patients are not able to physically visit a doctor. Medical students can get hands-on training on realistic hologram patients and equipment with the help of mixed-reality systems. A recent thrilling advance, a world-first in medical training, was achieved at Addenbrooke’s Hospital in Cambridge~\cite{HolographicTraining}.

\subsubsection{Holographic Production}
Introducing holography to industries like agroforestry, animal husbandry, and fishery, which often operate over large areas, will allow them to remotely supervise crops, woods, livestock, fish, etc.. This will simplify the full operation and management processes and reduce operating costs. For industries operating in harsh environments, e.g., mining, coal, and nuclear power industry, holography will assist staff with full perception of the environments, as well as it will help during emergencies. Holography will be a promising technology for building digital twins in manufacturing in several aspects~\cite{RIS_DT_2023}, such as improving system designs, testing new products, monitoring and predictive maintenance, as well as lifecycle management.

\subsubsection{Other Miscellaneous Applications}
Holography has a great potential in military applications, and particularly, in the design of defense systems. Assisted with internet of battlefield things, full-domain intelligence will be captured and a full-domain holographic battlefield will be reconstructed, enabling holographic situation awareness and holographic combat commanding \cite{Reding2020Science}. Holography has been also widely applied to encryption and decryption for information security. A $10$-bit OAM-multiplexing hologram for high-security holographic encryption was proposed in \cite{Fang2020OAM}. Image encryption based on interleaved computer-generated holograms was presented in \cite{Kong2018Image}. Optical encryption was realized by reprogrammable metasurface-based holograms in \cite{Qu2020Reprogrammable}. In addition, holography generalizes its applicability to fault diagnosis, such as microwave holographic diagnosis for antennas \cite{Rahmatsamii1988Application}, and holographic techniques for determining antenna radiation characteristics and imaging aperture fields \cite{Schejbal2008Accuracy}. Besides, holography has been widely employed in packaging and trademark anti-counterfeiting \cite{Vather2018Serialized} as well as holographic metrology \cite{Dong2016Digital}.

\subsection{Technological Enablers}
\label{subsec:HTR}
Holography is the method of generating 3D images and has been primarily realized in~\cite{Leith1962Reconstructed} via optical holographic techniques that relied on the emergence of lasers and coherent light sources, i.e. sources emitting light waves of the same frequency with constant phase difference. This seminal work dating back to 1962 was extended in 1967 to the more flexible CGH \cite{Lohmann1967Binary}, which was enabled by computers. The first attempt for realizing holography in the EM domain was presented for X-band imaging in \cite{Dooley1965Xband}. This work motivated the consideration of the combination of holography with antenna technologies to achieve holographic-type wireless communications \cite{Checcacci1970Holographic}. In the following, a brief description of the basic principle of holography is presented. Then, the three main technology enablers for realizing holography are discussed, namely the conventional optical holography, the CGH, and the EM-based holography.

\subsubsection{Basic Principle of Holography} 
As previously mentioned, the realization of holography mainly includes two steps: recording and reconstruction. During the recording process, a recording medium or device, such as holographic plates, charge coupled devices, or complementary metal oxide semiconductor cameras, is employed to track the intensities of the \textit{hologram}, which is obtained by superimposing a known reference wave with the wave emitted by the desired object generating an interference pattern. In the reconstruction process, the recording medium or device is illuminated by a replica of the reference wave, thereby reconstructing the object's wave perfectly. The main idea of this implementation relies on the interference between two coherent waves which captures their phase differences; this fact can be leveraged for the perfect reconstruction of the object's wave. A mathematical description of this two-step procedure is provided in the sequel. 
We first define the wavefront of a desired object, which is the result of scattering from that object when illuminated by a wave source generator, as follows:
\begin{equation}
	\begin{aligned}
		\mathcal{O} \triangleq |\mathcal{O}| e^{i\theta},
		\label{ObjectWave}
	\end{aligned}
\end{equation}
where $|\mathcal{O}|$ represents the wave's amplitude (intensity) and $\theta$ is the wave's phase. Likewise, we define the reference wave illuminating the recording medium or device as:
\begin{equation}
	\begin{aligned}
		\mathcal{R} \triangleq |\mathcal{R}| e^{i\phi},
		\label{RefWave}
	\end{aligned}
\end{equation}
having the amplitude $|\mathcal{R}|$ and the phase $\phi$. Assuming that the recording medium or device is intensity sensitive (alternatively, it can only records the phase information), the intensity of the hologram (i.e., the intensity of the superposition of waves $\mathcal{O}$ and $\mathcal{R}$) is given by:
\begin{equation}
	\begin{aligned}
		\mathcal{I} \triangleq |\mathcal{O} + \mathcal{R}|^{2} = |\mathcal{O}|^{2} + |\mathcal{R}|^{2} + \mathcal{O} \mathcal{R}^{*} + \mathcal{O}^{*} \mathcal{R},
		\label{InterferWave}
	\end{aligned}
\end{equation}
where $*$ indicates the complex conjugate operation. It can be observed from this expression that the last two terms include the phase difference between the reference and object waves, which is actually critical for the object's wave reconstruction. By retaining the intensity of the hologram in the recording medium or device and further illuminating it with a replica of the reference wave, the following wave is resulting:
\begin{equation}
	\begin{aligned}
		\mathcal{I} \mathcal{R} = |\mathcal{O}|^{2} \mathcal{R} + |\mathcal{R}|^{2} \mathcal{R} + \mathcal{O} |\mathcal{R}|^{2} + \mathcal{O}^{*} \mathcal{R}^{2}.
		\label{RecObjectWave}
	\end{aligned}
\end{equation}
The last two terms of this expression indicate that it is possible to completely reconstruct the wave of the desired object, which includes both amplitude and phase information.
\begin{figure*}[t!]
	\centering
	\includegraphics[height=4.5cm, width=16cm]{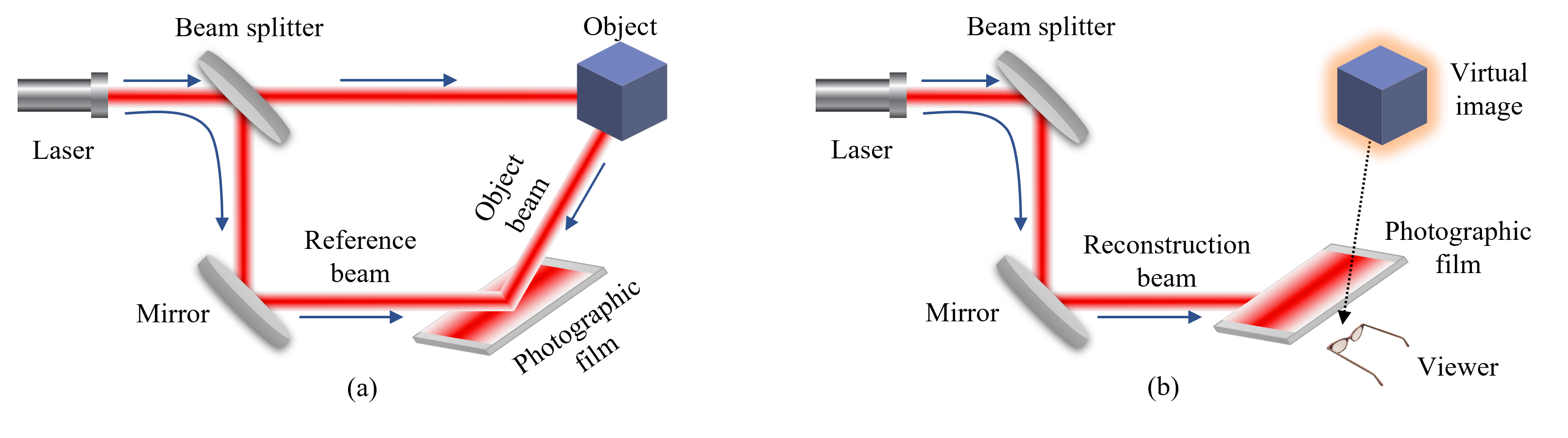}
	\caption{Schematic of the optical holography: the (a) recording and (b) reconstruction processes.}
	\label{fig:optical_holography}
\end{figure*}

\subsubsection{Optical Holography}
In optical holography technology, coherent light sources are generated via a laser. A realization of the recording process with this kind of holography is illustrated in Fig.~\ref{fig:optical_holography}(a). As demonstrated, a laser beam steers with a reference beam a beam splitter whose role is to divide its single light input into two output coherent light beams. The one beam is intended to propagate towards a desired object to enable the beam's scattering towards a recording medium; in this figure, the medium is a photographic film. The other output of the beam splitter is guided via a mirror (i.e., via reflection) to the photographic film to get superposed with the scattered beam from the object. In this way, a hologram will be recorded by the photographic film. 

By using the setup depicted in Fig.~\ref{fig:optical_holography}(b), the second step of the optical holography, i.e., object reconstruction, is performed. During this step, the laser emits again a replica of the reference beam to the beam splitter, which is now configured to direct the entire beam towards the mirror, while ensuring blocking completely its propagation to the object. The mirror reflects the reference beam towards the photographic film to illuminate the hologram recorded in the previous step. The latter process yields the reconstructed object at the film as a virtual 3D image. 

It is noted that there exist various optical holography techniques, such as the (compressive) optical scanning holography and the phase-shifting holography, with each being implemented via dedicated hardware components. The reader may refer to \cite{Tsang2016Review}, and references therein, for further details.
\begin{figure}[t!]
	\centering
	\includegraphics[height=3.1cm, width=8.5cm]{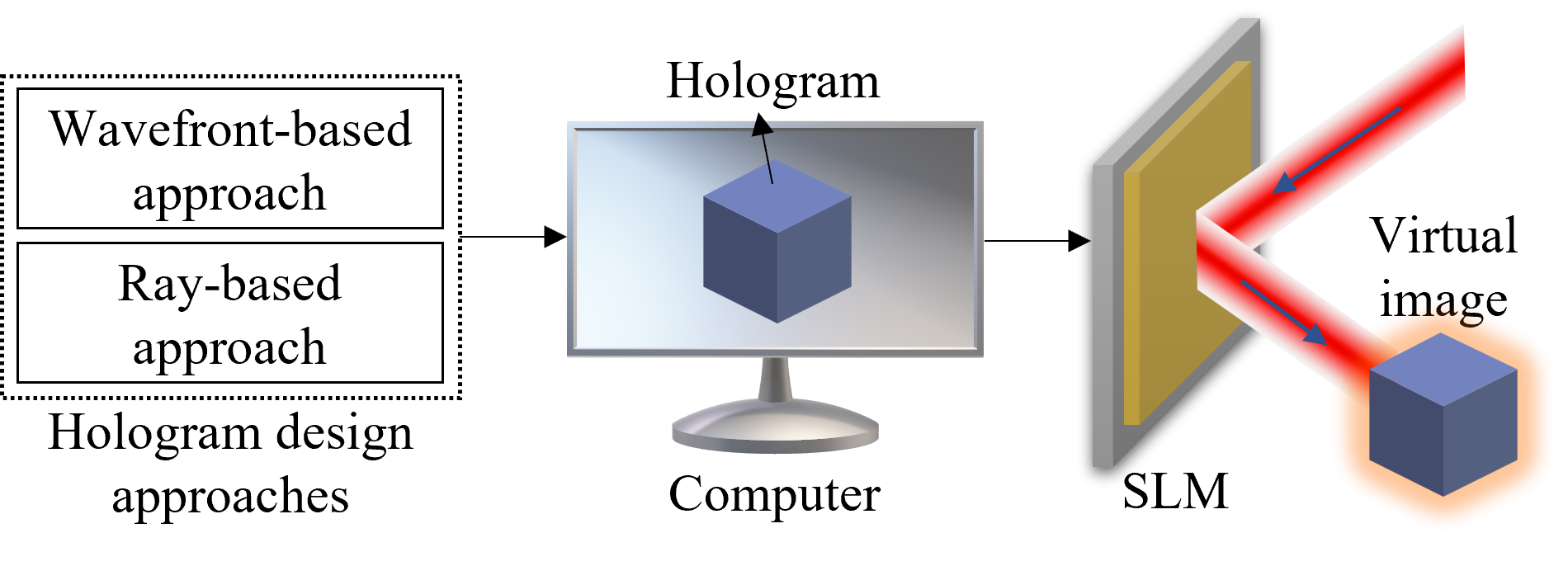}
	\caption{Schematic of the computer-generated holography.}
	\label{fig:computer_generated_holography}
\end{figure}

\subsubsection{Computer-Generated Holography}
Conventional optical holography requires a vast amount of complicated optical components for the recording and reconstruction processes. It also necessitates the existence of a real object for the acquisition of holograms, which can be inefficient, limiting the generalization of the imaging process to other types of objects. With the advent of computers and spatial light modulators (SLMs), CGH was proposed in \cite{CHG_1967} in 1967 for mitigating the problems encountered in optical holography. According to the proposed framework, instead of realizing holography in a hardware-dependent optical manner, both the recording and reconstruction processes are implemented in software in an electronic device. A holographic image can be generated, for example, by digitally computing a holographic interference pattern and printing it onto a mask or film for subsequent illumination by a suitable coherent light source. CGH, not only reduces the requirement for complicated hardware components, but also facilitates imaging of various virtual objects. Computer-generated holograms have the advantage that the objects which one wants to show do not have to possess any physical reality at all, i.e., CGH performs completely synthetic hologram generation. The schematic of CGH is depicted in Fig.~\ref{fig:computer_generated_holography}. The two main groups of CGH methods are mentioned in the figure: \textit{i}) wavefront-based and \textit{ii}) ray-based methods \cite{Sahin2020Computer}. Through simulating the wave diffraction process, wavefront-based methods numerically calculate the 3D wave fields of a given object/scenery, as well as its 2D distribution on the hologram plane. The most widely adopted schemes belonging in this group and utilizing 3D positional information are based on the point cloud model, polygon model, and the layer-based representation of 3D objects/scenes. Differently, ray-based methods generate holograms by capturing incoherent 2D images of given 3D objects/scenes based on the transformation from ray-based representations to wavefront-based holographic information. Two typical categories exist for the group of ray-based methods: the holographic stereogram and the multiple viewpoint projection. With arbitrarily computer-generated holograms illuminated by the reconstruction beam, high-fidelity virtual 3D images can be reconstructed. 

It is finally noted that CHG finds various applications in holographic computer displays that are used for a wide range of applications, ranging  from computer-aided design to gaming, holographic video and TV programs, as well as automotive and communication applications (e.g., cell phone displays). The recent advances of CGH are discussed in detail in the surveys \cite{Yaras2010State} and \cite{Sahin2020Computer}.
\begin{figure*}[t!]
	\centering
	\includegraphics[height=5.6cm, width=16.8cm]{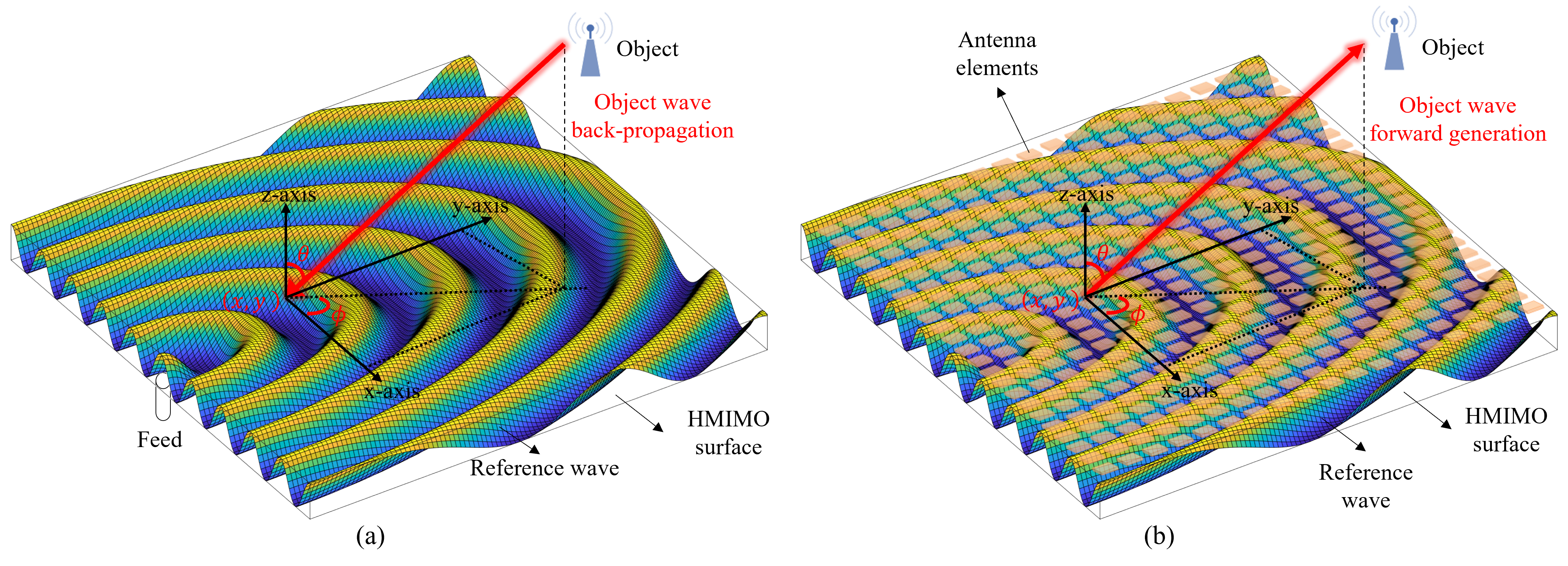}
	\caption{Schematic of the holographic LWA-based EM holography: the (a) recording and (b) reconstruction processes.}
	\label{fig:EM_holography}
\end{figure*} 

\subsubsection{Electromagnetic Holography}
As previously mentioned, holography aims to realize 3D imaging of a given object, which is equivalent to the reconstruction of the wave field illuminated by that object. In principle, this goal can be also achieved via EM waves, instead of the light waves used in the optical holography. To this end, EM holography deploys antenna technologies to realize holography. In the following, we present two approaches for EM holography, namely, holographic LWA-based EM holography~\cite{Araghi2021Holographic} and photonic TCA-based EM holography~\cite{Zong20196G, Prather2017}. 

\emph{\underline{Holographic LWA-based EM Holography}}: According to this holographic approach, an EM wave source is used for reference wave generation (replacing the laser in optical holography), and an EM antenna aperture, which is henceforth referred to as an HMIMO surface, plays the role of the photographic film that is used in optical holography. One or more objects need to be imaged, or in the case of holographic MIMO communications, the objects are the nodes to provide communications links. It is noteworthy that, in EM Holography, the EM wave source can be embedded to the HMIMO surface, instead of being a separate source. This is one of the key differences compared to optical holography structures. A typical example of this EM-source-integrated case is illustrated in Fig.~\ref{fig:EM_holography}, where the recording and reconstruction processes with an LWA are depicted. As shown, the LWA-based HMIMO surface consists of a substrate with one or more wave feeding elements, which are integrated in the substrate, and antenna elements that are printed on the substrate's surface (see Section~\ref{SectionIII} that follows for more hardware details). The substrate serves as a waveguide allowing the reference waves emitted from the feed(s) to propagate along it. It is noted that each feed requires a wave generator, which could be a transmission RF chain. The antenna elements are designed to construct various holograms with explicit textures, or to approach holograms via various tuning mechanisms without possessing any specific texture. With the configured holograms, specific radiations, with the radiating signals generated in response to the the reference waves, can be realized. 

During the recording process, the hologram is realized by the HMIMO surface. A wave emitted by an object from a given direction (either generated by the object or scattered by it) reaches the HMIMO surface that generates a hologram. This is the result of the superposition of the object and reference waves, with the latter being illuminated by the LWA's feed, as depicted in Fig.~\ref{fig:EM_holography}(a). To describe this process in mathematical terms, we assume that the following reference wave is excited by a point source (the feed) within a lossless substrate:
\begin{equation}
	\begin{aligned}
		E_{rw} \triangleq A_{r}e^{-i \beta_r d_r (x, y)},
		\label{EMSurfaceWave}
	\end{aligned}
\end{equation}
where $A_{r}$ denotes the wave's amplitude, which is constant in a lossless substrate, $\beta_r$ represents the wavenumber of the reference wave,  and $d_r (x, y)$ indicates the distance between the $(x,y)$-coordinate position of the HMIMO surface and the point source. Assuming that this source lies in the origin of the coordinate system, it holds that $d_r (x, y) = \sqrt{x^2 + y^2}$. We also use the notation $d_o (x, y)$ for the distance between the object and the position of the HMIMO surface, $\beta_0$ denotes the free-space wavenumber, and $\phi$ and $\theta$ represent the azimuth and elevation angles of the object, respectively, in the coordinate system in Fig.~\ref{fig:EM_holography} having as origin the position point of the HMIMO surface. The phase variation due to the distance $d_o (x, y)$ can be described through the projection of the free-space wavenumber $\beta_0$ on the $x$ and $y$ axes. This is represented as the inner product: $[\beta_0 \sin \theta \cos \phi, \beta_0 \sin \theta \sin \phi] [x, y]^{T} = \beta_0 (x \sin \theta \cos \phi + y \sin \theta \sin \phi)$. Hence, the wave from the object on the HMIMO surface is given by the expression: 
\begin{equation}
	\begin{aligned}
		E_{obj} \triangleq A_{o}e^{i \beta_0 d_o (x, y)} = A_{o}e^{i \beta_0 (x \sin \theta \cos \phi + y \sin \theta \sin \phi)},
		\label{EMObjectWave}
	\end{aligned}
\end{equation}
where $A_{o}$ indicates the wave's amplitude. Similar to the optical holography, at any given $(x,y)$-coordinate, the hologram (i.e., the superposition of the waves $E_{rw}$ and $E_{obj}$) can be expressed as $E_{int} \triangleq E_{obj} E_{rw}^{*}$, yielding the expression:
\begin{equation}
	\begin{aligned}
		E_{int} = A_{o} A_{r} e^{i \left(\beta_0 (x \sin \theta \cos \phi + y \sin \theta \sin \phi) + \beta_r \sqrt{x^2 + y^2} \right)}.
		\label{EMInterferWave}
	\end{aligned}
\end{equation}
The latter process summarizes the way to design the HMIMO surface to capture holograms, i.e., to realize the recording process of EM holography. It is noted that HMIMO surfaces can be implemented with various ways in both fixed or reprogrammable configurations; this will be the subject of Section~\ref{SectionIII} that follows.

In the reconstruction process, with the hologram being implemented/retained by the HMIMO surface via \eqref{EMInterferWave}, a wave toward the $(\phi, \theta)$-direction can be generated once the reference wave is illuminated by the feed and then travels along the HMIMO surface's substrate, as demonstrated in Fig.~\ref{fig:EM_holography}(b). The mapping rule from the reference to the object wave in the HMIMO surface follows the LWA theory \cite{Monticone2015Leaky,Jackson2012Leaky,Jackson2011Fundamental}, which obeys the wave diffraction principle.

\begin{figure}[t!]
	\centering
	\includegraphics[height=3.2cm, width=8.6cm]{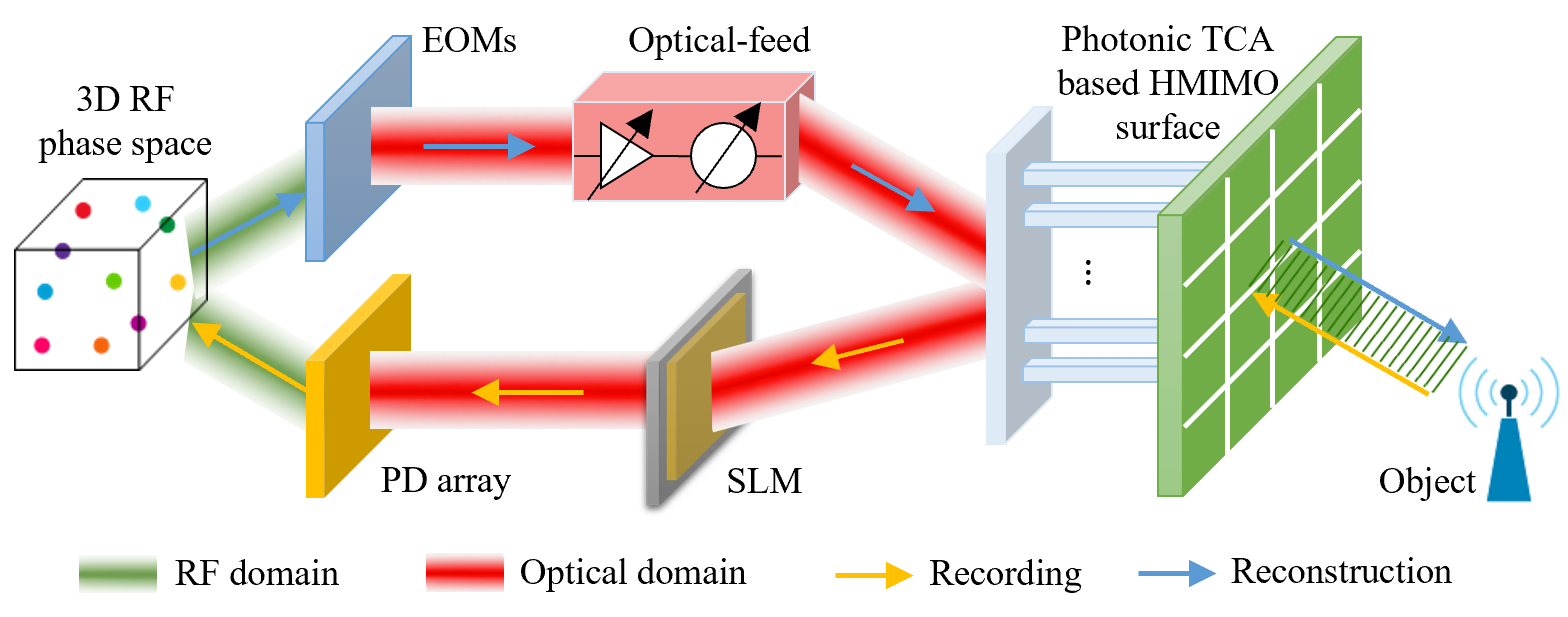}
	\caption{Schematic of the photonic TCA-based EM holography including the recording and reconstruction processes.}
	\label{fig:EM_holography_TCA}
\end{figure}

\emph{\underline{Photonic TCA-based EM Holography}}: This kind of EM holography is implemented via optical-domain signal processing using photonic TCA-based HMIMO surfaces that realize holographic RF-optical mapping, which is capable to realize signal transformations from RF to optical beams, and vice versa, as demonstrated in Fig.~\ref{fig:EM_holography_TCA}. The latter transformation is enabled by electro-optic modulators (EOMs) and uni-traveling-carrier photodetectors (UTC-PDs), which carry out the transformation of the RF received signals to optical beams and the transformations of optical beams to RF signals for transmission, respectively. As shown in the figure, each EOM, that is connected to a distinct antenna element, upconverts the received RF signal to the optical domain that is further propagated in the structure via an optical-fiber bundle. It is noted that each UTC-PD is bonded to adjacent antenna elements following the flip chip technology. UTC-PDs are very efficient in converting optical beams to electrical signals with high power and large bandwidth, which enables them to directly drive the antenna elements with a very large bandwidth ($\ge 40$ GHz). 

The optical-domain signal processing is facilitated by an optical-feed that drives the photonic TCA-based HMIMO surface for signal transmission, and by an SLM that performs massive spatial processing (i.e., the optical Fourier transform) to the optical beams upon signal reception. It is noted that the optical processing takes place in the speed of light, a fact that reduces latency contributing in achieving real-time processing. The optical feed includes two phase-locked lasers whose frequencies are offset with a specific value that serves as the RF carrier to be transmitted. The combined output of the lasers is divided into a certain amount of fibers that connect to UTC-PDs in a point-to-point mapping. This optical-feed design is capable of realizing RF waves that are tunable in both amplitude and phase. In a similar manner, the SLM serves as a `lens' that images the optical beams originating from the RF signals through the EOMs onto a PD array; each PD corresponds to a unique spatial signature of the received RF signal. 

The recording (bottom signal flow) and reconstruction (top signal flow) processes of the photonic TCA-based EM holography are described in Fig.~\ref{fig:EM_holography_TCA}. In the recording process, the object wave is received by the photonic TCA-based HMIMO surface whose output RF signals are transformed to optical beams via upconverted EOMs. The resulting optical beams are then directed to an optical-fiber bundle and propagated to the SLM for spatial processing and imaging. Then, the images are presented onto a PD array with accurate separations to finally construct a 3D constellation of the object in an RF phase space. Such a space provides an accurate feedback for synthesizing spatial RF wave fields for highly focalized transmissions. In the reconstruction process, the electrical signals guided by the 3D RF phase space are first transformed to optical beams via EOMs and then inserted to an optical-feed for the purpose of achieving amplitude and phase control. The outputs of that feed are then routed to the photonic TCA-based HMIMO surface by driving UTC-PDs for directly exciting the surface's antenna elements. This mechanism yields a desired synthesis of the spatial RF wave field to the intended object.

\begin{figure}[t!]
	\centering
	\includegraphics[height=6.8cm, width=8.6cm]{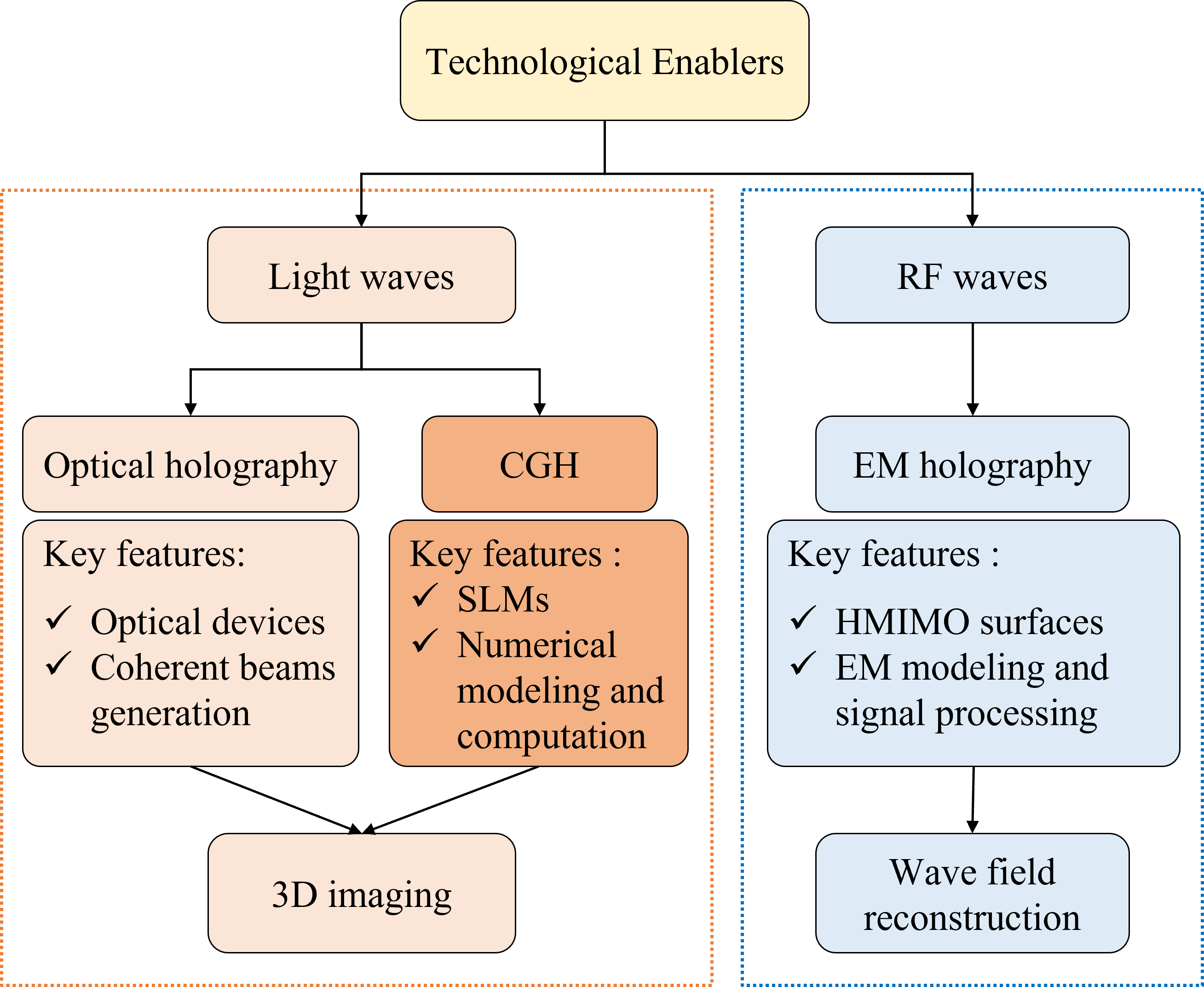}
	\caption{The enabling technologies for holography, their key features, and their intended functionalities.}
	\label{fig:HoloTechRoadmaps}
\end{figure}

The enabling technologies for holography, their key features, and their intended functionalities are summarized in Fig.~\ref{fig:HoloTechRoadmaps}. As shown in the figure, both optical holography and CGH are based on light waves since they manipulate light beams to implement holograms. On the other hand, EM holography mainly relies on RF waves. It is noted that optical holography and CGH are commonly applied for realizing holographic 3D imaging, while EM holography is mainly used for RF wave field reconstruction, which enables holographic communications. Optical holography relies on optical devices and the generation of coherent light beams, CGH deploys SLMs to carry out computer-generated holograms which require efficient numerical modeling and computations, while EM holography necessitates HMIMO surfaces which serve as the physical entity to support the corresponding EM modeling and signal processing. In the remaining of this survey , we mainly focus on EM holography and its promising applications for 6G wireless communications.


\section{Physical Aspects of {HMIMO Surfaces}}
\label{SectionIII}




EM holography can be understood macroscopically as the superposition of two EM waves, the object and reference waves, which superimpose to create the interference wave. Under a certain reference wave, the formed hologram establishes a point-to-point mapping to the object wave, as demonstrated in holographic LWA based EM holography. It is however different in photonic TCA based EM holography that can mostly be considered as persisting the principle of conventional phased array antennas while shifting the feed processing from the RF domain to the optical domain via UTC-PDs. If we only focus on the antenna aperture, we can see that the photonic TCA based {HMIMO surfaces} comply with the conventional phased array antennas, where each antenna element is independently tunable. 
Based on this insight, in this section, we mainly focus on the LWA based {HMIMO surfaces} for inherent differences. 

In order to realize the reconstruction process of EM holography, it is necessary to employ one or more feeds to generate the reference waves, and to carry the hologram on a specifically designed {HMIMO surface}, such that the desired radiation can be achieved when reference waves illuminate the surface. It is worth noting that the feed can be placed in different positions: (P1) integrated into the {HMIMO surface}, and (P2) located on the exterior of the {HMIMO surface}. 
In the (P1) setting, the feed position can be divided into three cases: Surface-fed, bottom-fed, and edge-fed. In addition, there are different hardware structures of the feed that excite various propagation modes of a reference wave, which will be detailed next. Otherwise, in the (P2) setting, the feed can be placed behind the {HMIMO surface} to form a lens-like radiation through signal refraction. It can also be located in front of the {HMIMO surface} to form a desired specular-like radiation by signal reflection. In such a case, the feed can be implemented by a horn antenna, producing the required near/far-field reference wave for illuminating the surface to generate the object wave. 
Lastly, the antenna elements composing an {HMIMO surface}, can be of various shapes and materials, and their responses can be controlled through different mechanisms such as electro-mechanical and lumped element tunability etc. Moreover, holography can be achieved through distinct design methodologies, all the aforementioned details will be explained in the following.


\subsection{Hardware Structures of {HMIMO Surfaces}}

\begin{figure*}[t!]
	\centering
	\includegraphics[height=8.2cm, width=16cm]{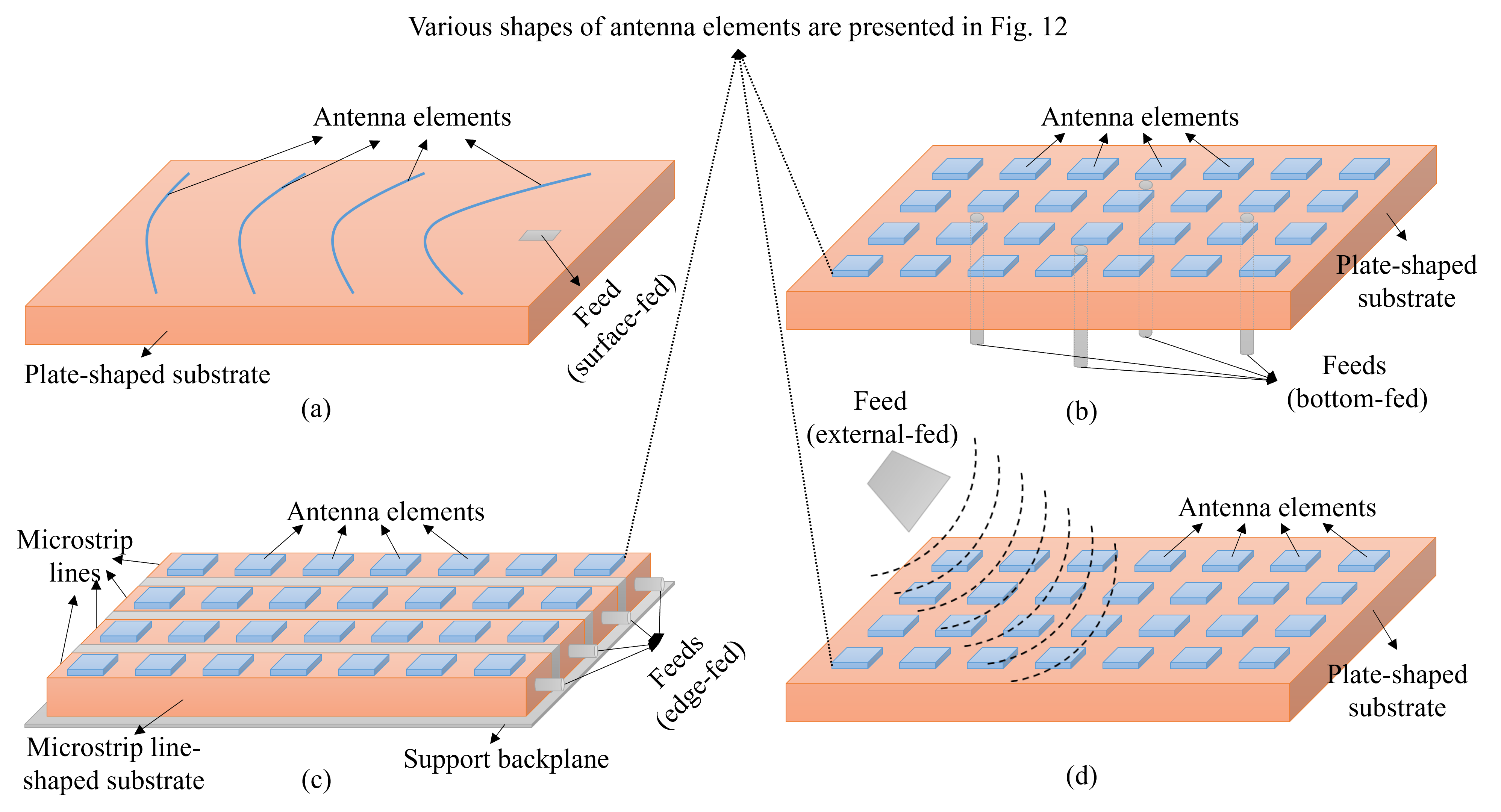}
	\caption{Structure schematics of an {HMIMO surface}: (a) Plate-shaped substrate covered with curve-shaped strip grating antenna elements excited by a surface-fed feed; (b) plate-shaped substrate covered with square-shaped antenna elements excited by bottom-fed feeds; (c) microstrip line-shaped substrate covered with square-shaped antenna elements excited by edge-fed feeds; (d) plate-shaped substrate covered with square-shaped antenna elements excited by external-fed feed.}
	\label{fig:HMIMOS_structure}
\end{figure*}

The building blocks of an {HMIMO surface} mainly include three components that can be macroscopically summarized as feed, substrate, and antenna element. The structure schematics of an {HMIMO surface} with four possible structures are shown in Fig. \ref{fig:HMIMOS_structure}. It is emphasized that this figure only exhibits representative schematics, while the feed, substrate, and antenna element can be in different types that will be detailed as follows.

\subsubsection{Feed}
The feed is utilized to generate a reference wave to propagate along the {HMIMO surface}, thus exciting a desired object wave accordingly. It has different forms in accordance with its locations ((P1) and (P2)) and propagation mode of the reference wave that it supports, as well as depending on the material of substrate sometimes. Its realizations can be implemented by various hardware. We discuss each content as follows. 

\emph{\underline{Location and Propagation Mode}}:  
As seen from Fig. \ref{fig:HMIMOS_structure}, the feed position can be located on the surface displayed by schematic (a), and can be located at the bottom or the edge of substrate illustrated by schematic (b) and (c), respectively. The surface-fed form is used for supporting a transverse electric (TE) propagation mode of reference wave, where the electric field is transverse to the propagation direction while the magnetic field is perpendicular to the propagation direction. Oppositely, the bottom-fed form facilitates a transverse magnetic (TM) propagation mode of reference wave, where the magnetic field is transverse to the propagation direction while the electric field is perpendicular to the propagation direction. Furthermore, the edge-fed form matches the quasi transverse electric and magnetic (TEM) propagation mode of reference wave. Specifically, in the TEM mode, both electric and magnetic fields are perpendicular to each other and also perpendicular to the propagation direction. The microstrip line structure, depicted in Fig. \ref{fig:HMIMOS_structure}(c), leads to the quasi TEM mode that makes the propagation resemble the TEM mode, with the difference that the components of the electric and magnectic field along the propagation direction are much smaller than the transverse but nonzero. The schematics in Fig. \ref{fig:HMIMOS_structure}(a)(b)(c) belong to (P1), in which the reference wave propagates along the aperture surface and can be interpreted as a surface wave, while the feed can be called as surface wave launcher accordingly.
Antithetically, the reference wave can be externally fed as depicted by Fig. \ref{fig:HMIMOS_structure}(d) that belongs to (P2). In this case, the reference wave propagates in free space and is termed as space wave. If the space wave further forms a plane wave, it follows a TEM propagation mode.

\emph{\underline{Hardware}}:  
Apart from the location configuration previously described, the hardware of the feed can be different as well. For the TE propagation mode, dipoles \cite{Petosa2004Microwave}, planar Vivaldi feed (PVF), antipodal Vivaldi feed (AVF), and dipole based Yagi-Uda-feed (YUF) \cite{Rusch2016Holographic, Movahhedi2019Multibeam}, are capable of exciting such a propagation. Furthermore, an array of dipole sources was employed as the hardware structure for exciting the TE mode reference wave in \cite{Cheng2016Real}. The array enables a planar wavefront of the reference wave instead of a circular one generated by a single source. As such, the curve-shaped strip grating pattern driven by one feed as seen from Fig. \ref{fig:HMIMOS_structure}(a) will become a line-shaped strip grating pattern (e.g., one-dimensional (1D) periodic metallic strip pads employed in \cite{Cheng2016Real}). Alternatively, the TE propagation mode can be excited by multiple substrate-integrated waveguide (SIW) horns within a substrate layer \cite{Yurduseven2021Multibeam, Boyarsky2021Electronically}. 
For the TM mode propagation, single or multiple bottom-fed vertical monopoles are widely adopted in designing artificial impedance surfaces \cite{Sievenpiper2005Holographic, Fong2010Scalar, Ramalingam2019Polarization, Ren2022Graphene}, enabling a single or multiple beam radiations. The feed can be alternatively replaced by a bottom-fed coaxial probe, as in \cite{Yurduseven2017Dual, Yurduseven2017Design}. Depending on the employed glass-based substrate, the feed is realized by a through-glass-via (TGV) \cite{Galler2020High}.
Furthermore, the coaxial probe is not only adopted for exciting the TM propagation mode, but also utilized for generating the quasi TEM propagation mode \cite{Yurduseven2018Dynamically}. In such a setup, the reference wave is excited by an edge-fed coaxial probe, and then propagates along the microstrip line which is terminated by a matched load eventually. 
For the generation of external-fed reference wave, the feed can be implemented by a horn antenna \cite{Liu2016Horn, Wu2021Millimeter}. 

\begin{table}[!t]
	\footnotesize
	\renewcommand{\arraystretch}{1.2}
	\caption{\textsc{Configurations of a feed.}}
	\label{tab:SWL_config}
	\centering
	\begin{tabular}{!{\vrule width0.6pt}c|c|c|c!{\vrule width0.6pt}}
		\Xhline{0.6pt}
		
		\rowcolor{yellow} \textbf{\tabincell{c}{Equivalent\\ name}} & \textbf{\tabincell{c}{Propagation\\ mode}} & \textbf{Hardware} & \textbf{Location} \\ 
		\Xhline{0.6pt}
		
		\multirow{5}{*}{\tabincell{c}{Surface wave}} & TE & \tabincell{c}{Dipole (array)\\ PVF\\ AVF\\ YUF\\ SIW} & Surface-fed \\
		\cline{2-4} 
		& TM & \tabincell{c}{Monopole\\ Coaxial probe\\ TGV} & Bottom-fed \\
		\cline{2-4} 
		& Quasi TEM & \tabincell{c}{Coaxial probe} & Edge-fed \\
		\Xhline{0.6pt}
		
		Space wave & \tabincell{c}{TEM\\ (if plane wave)} & \tabincell{c}{Horn antenna} & External-fed \\
		\Xhline{0.6pt}
	\end{tabular}
\end{table}

To provide a complete reference of the feed configurations, we summarize and list the equivalent names of reference wave, and their propagation modes, as well as the feed hardware and corresponding locations in Table \ref{tab:SWL_config}.

\subsubsection{Substrate}
The substrate is basically utilized as a waveguide that enables the reference wave injected by the feed to propagate along it. When integrating periodic partially reflecting surface on the top of a substrate, the reference wave can be leaked out to free space. The main criterion for substrate selection is that it can achieve LWA-based EM holography by exploiting its physical properties, particularly the ability to favourably control the permittivity and permeabilityof the surface \cite{Jackson2011Fundamental}. Moreover, the selection criterion also depends on the fabrication availability and the cost constraint. 

From a geometrical perspective, the substrate can be in a plate shape as shown in Fig. \ref{fig:HMIMOS_structure}(a)(b), or in a microstrip line shape as depicted in Fig. \ref{fig:HMIMOS_structure}(c). The two different geometric shapes are respectively responsible for supporting the reference wave propagation in either the surface mode or the waveguide mode. The key difference between the two being that in the latter mode, waves are restricted to propagate along a confined path, i.e. the microstrip line, and not the whole surface. In addition, the substrate of an {HMIMO surface} can be composed of different materials, based on which we can categorize into individual groups mainly including the dielectric substrate and the semiconductor substrate. Details of each group will be described as follows.

\emph{\underline{Dielectric Substrate}}:  
Dielectric materials are the most widely used for constituting the substrate of an {HMIMO surface}. We enumerate three categories: printed circuit board (PCB) laminates substrate, silicon dioxide (glass) substrate, and anisotropic artificial dielectric substrate. In the first category, the commonly used PCB laminates as the substrate of {HMIMO surfaces} are commercially available, including Duroid 5880, Rogers 3010, FR4, Rogers RO4003, Rogers 3003, Rogers RT6010, Taconic TLA-5/TLA-6, etc. The PCB laminates are competitively priced products with exceptional mechanical and electrical stability, while showing a low dielectric loss which is well suited for high frequency/broadband applications. The utilization of grounded PCB that laminates to form a single/multiple layers was presented in \cite{Sievenpiper2005Holographic, Fong2010Scalar, Li2014Frequency, Pandi2015Design, Rusch2015Holographic, Yurduseven2017Dual, Yurduseven2017Design, Yurduseven2018Dynamically, Wang2018Holographic, Ramalingam2019Polarization, Kim2022Performance} for supporting various shaped sub-wavelength antenna elements, e.g., square metal/graphene patches, metallic strip gratings, and slot-shaped metamaterial elements etc.
In the second category, the silicon dioxide (glass) substrate possesses a low dielectric loss as well. In \cite{Cheng2016Real} and \cite{Ren2022Graphene},  grounded and graphene sheets/patches are implemented on silicon dioxide substrate for realizing electrically tunable THz {HMIMO surfaces}. In \cite{Galler2020High}, a glass substrate was adopted for designing a high-gain mmWave {HMIMO surface}. Such an efficient approach enables large physical antenna apertures with low fabrication tolerances and low cost material simultaneously.
In the third category, an anisotropic artificial dielectric substrate is manufactured, where a composite layer of 0.015-in thick alumina 99\% and 0.015-in thick TMM-4 materials is synthesized with acceptable requirements \cite{ElSherbiny2004Holographic}. 
Comparing the three different substrates based {HMIMO surfaces} from existing studies, we directly find that the PCB laminates substrate is more applied to frequencies below mmWave while the latter two are designed for mmWave or even THz frequencies.


\emph{\underline{Semiconductor Substrate}}:  
Semiconductor materials compose another important branch of potential substrate technologies for an {HMIMO surface}. Among various semiconductors, silicon is highly advocated because high-quality silicon wafers are both low-cost and widely available, have excellent mechanical strength and high thermal conductivity, as well as high carrier lifetime in high-resistivity silicon devices \cite{Fathy2003Silicon}. The authors in \cite{Fathy2003Silicon, Fathy2001Silicon} presented a novel direct current (DC) controlled silicon PIN diode design on a silicon substrate, exploiting lithographic technology. Such silicon PIN diodes are capable of realizing a similarly switchable functionality compared with conventional switches, thus paving the way for implementing a reconfigurable {HMIMO surface}.
The authors of \cite{Yurduseven2021Multibeam} provided a Si/GaAs laminated substrate design for an {HMIMO surface} with multibeam switchable capability. The substrate consists of a dual semiconductor layer (Si and GaAs) connected via a specified coupling slot hosted on a middle conductive layer. The choice of Si mostly exploits the appropriate wafer thickness for fabrication, while the selection of GaAs provides a possibility for reconfigurability design by further implementing Schottky diodes on this layer.

\subsubsection{Antenna Element}
The antenna elements mounted on the substrate surface form either uniform or non-uniform patterns, allowing a (reconfigurable) transformation from the reference wave in various modes to the radiated EM object waves. The antenna elements can be made of different materials and created in various shapes in discrete or (approximately) continuous forms, which will be described below. Before that, it is worth noting that the size of antenna elements and the distance between two adjacent ones is generally sub-wavelength (e.g., $\lambda/10 \sim \lambda/5$ for meta-atoms with $\lambda$ being the free space wavelength \cite{Liaskos2018New}), paving the way for dense antenna architectures that achieve full manipulation and high sampling of EM waves.

\emph{\underline{Antenna Element Material}}: 
The materials for implementing the antenna elements of an {HMIMO surface} can be metal, dielectric materials, or graphene etc., depending on their characteristics and design considerations. Metals are mostly adopted for realizing conductive antenna elements, inside of which electric currents oscillate at certain frequencies based on external EM waves. Metals are suitable for low frequencies as they behave as perfect electric conductors with neglectable losses when working at low frequencies. 
However, when transitioning to higher frequencies, the ohmic losses in metallic antenna elements, greatly degrade the radiating efficiency. Low loss and high refractive index dielectric materials are capable of addressing the loss issues faced by metals. In addition, the dielectric materials can potentially promote a smaller form factor and a wide range of bandwidth compared with their metallic counterparts \cite{Mongia1994Dielectric}. 
Notably, for its unique but excellent mechanical, electronic and optical characteristics, the newly discovered graphene \cite{Geim2007Rise} has attracted enormous attention, and has found a wide range of applications for implementing antenna elements \cite{Cheng2016Real, Ren2022Graphene}. Graphene composes of carbon atoms in the hexagonal structure, and possesses excellent properties such as low loss, electronic tunability, and strong light matter interactions. The preceding advantages enable graphene to be an excellent candidate for designing transceivers operating at THz or even optical frequencies.

\begin{table*}[!t]
	\footnotesize
	\renewcommand{\arraystretch}{1.2}
	\caption{\textsc{Antenna element configurations.}}
	\label{tab:RadiatingElement_Config}
	\centering
	\begin{tabular}{!{\vrule width0.6pt}c|c|c|c!{\vrule width0.6pt}}
	
		\Xhline{0.6pt}
		\rowcolor{yellow} \textbf{Design methodology} & \textbf{Antenna element shape} & \textbf{Aperture type} & \textbf{Surface pattern} \\ 
		
		\Xhline{0.6pt}
		\tabincell{c}{Locally maximum phase\\lines of hologram} & \tabincell{c}{Curve/line-shaped strip grating} & Discrete & A series of strip gratings \\
		
		\Xhline{0.6pt} 
		\tabincell{c}{Macroscopic surface\\ impedance based approach} & \tabincell{c}{(Slitted) square/ellipse-shaped patches\\ Circle-shaped patches with (cross) slot\\ Coffee bean\\ Grain of rice\\ Double $\pi$\\ Double anchor\\ Square/hexagon-shaped loop-wire unit\\ Cross-shaped patches\\ C-shaped antenna elements} & \tabincell{c}{Nearly\\ continuous} & \tabincell{l}{$\circ$ Textured surface pattern:\\ \quad Concentric ellipses, spiral,\\ \quad and concentric circles\\ $\circ$ A superposition of multiple\\ \quad textured surface patterns\\ $\circ$ A division of multiple\\ \quad textured surface patterns} \\
		
		\Xhline{0.6pt} 
		\tabincell{c}{Geometric polarizable\\ particle based approach} & \tabincell{c}{Slot-shaped antenna elements\\ CELC} & \tabincell{c}{Nearly\\ continuous} & Present implicit textured pattern \\
		\Xhline{0.6pt}
	\end{tabular}
\end{table*}

\begin{figure*}[t!]
	\centering
	\includegraphics[height=4.4cm, width=15.6cm]{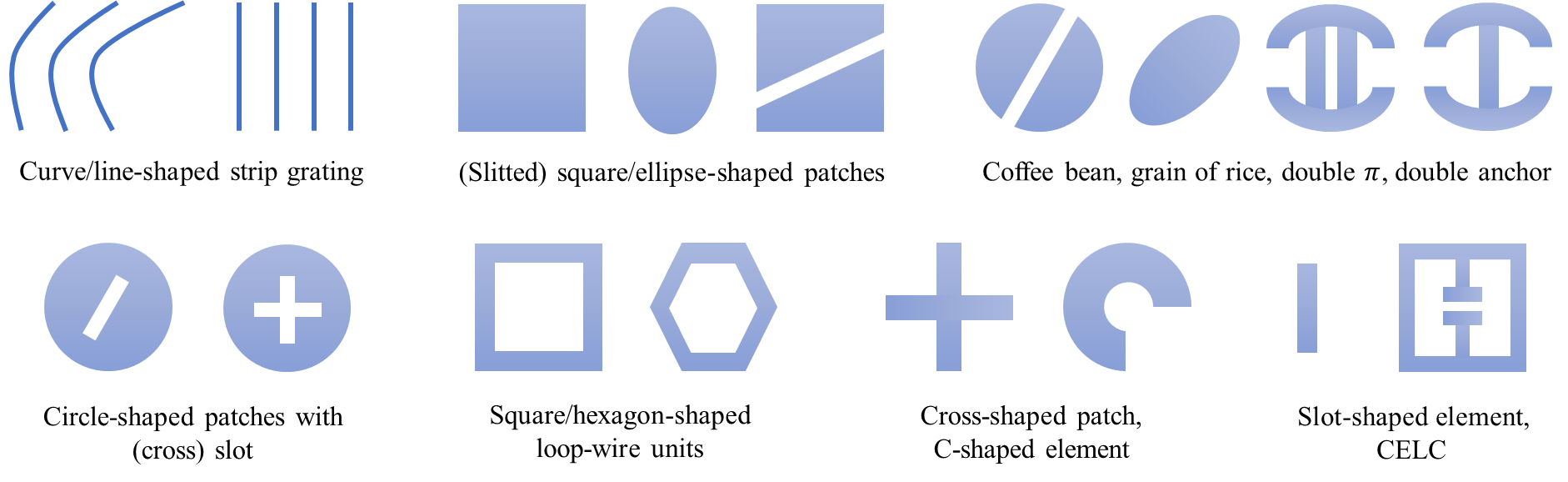}
	\caption{Various shapes of antenna elements of {HMIMO surfaces}.}
	\label{fig:ElementShapes}
\end{figure*}

\emph{\underline{Antenna Element Shape and Surface Pattern}}:
As fas as the shapes of antenna elements of an {HMIMO surface} are considered, metallic strip gratings, shown in Fig. \ref{fig:HMIMOS_structure}(a) and Fig. \ref{fig:ElementShapes}, emerge as the earliest realizations \cite{Checcacci1970Holographic, Iizuka1975Volume}, and are still employed in recent years \cite{Rusch2015Holographic}. The strip gratings are periodically located on specific positions, namely, the local maximum phase lines of the hologram (as will be further elaborated in the following subsection on the design methodologies). They can be curve-shaped or line-shaped depending on a single or an array of feeds employed. 
To enhance the flexibility, non-contacting metallic dipoles can be utilized as discrete samples of one continuous strip grating \cite{ElSherbiny2004Holographic}.
The surface pattern is formed by a series of strip gratings. As the location of one strip grating (i.e., antenna element) is only placed in a given position, EM waves of the given position can be sampled whereas the information of remaining positions is lost. As such, this kind of {HMIMO surfaces} have limited control over the radiating properties of EM waves and are thus not able to contruct a hologram with high resolution.

To mitigate the problem, it is envisioned that numerous small sized antenna elements will cover the surface for realizing dense EM wave sampling. To this aim, a multitude of sub-wavelength conductive patches were designed as antenna elements of artificial impedance surfaces \cite{Sievenpiper2005Holographic, Fong2010Scalar}. The conductive patches are relatively small compared to the wavelength of interest, allowing the scatter properties to be described using a macroscopic effective surface impedance. Moreover, the conductive patches can be customized in flexible sizes, shapes, and gaps between two adjacent ones, demonstrating a specific surface pattern, as well as allowing a realization of required macroscopic surface impedance and thereby the desired radiation. 
The surface impedance is basically guided by the holographic principle, providing a point-to-point connection between the hologram and the intended radiation as well as a point-to-point connection between the hologram and the surface pattern, i.e. the surface configuration that under the same reference wave will realize the considered hologram. 
The antenna elements can be designed based on this principle, by creating a point to point mapping between the element design and the corresponding surface pattern and impedance, which as previously explained lead to the intended EM wave radiation.
In \cite{Sievenpiper2005Holographic, Fong2010Scalar}, the authors adopted square-shaped patches and vary their gaps for obtaining required surface impedance, generating a textured surface pattern in an appearance of concentric ellipses. Employing slitted-square patches, the authors further implement tensor impedance surfaces that are capable of controlling EM wave polarization. 
In \cite{Galler2020High}, the authors utilized slitted circle-shaped patches to form a spiral surface pattern for realizing a high-gain holographic antenna. 
Alternatively, the authors of \cite{Li2014Two} designed an {HMIMO surface} with spiral surface pattern, implemented by an array of sub-wavelength metallic blocks drilled with dielectric holes.
Beyond the square-shaped patches, the antenna element shape can take various forms such as circle-shaped patch with (cross) slot, coffee bean, grain of rice, double $\pi$, double anchor \cite{Faenzi2019Metasurface}, square/hexagon-shaped loop-wire unit \cite{Li2016Polarization}, cross-shaped patch \cite{Liu2016Horn}, and C-shaped antenna element \cite{Wu2021Millimeter}, as demonstrated in Fig. \ref{fig:ElementShapes}. 
If multibeam is further supported, the conductive patches are designed to form a superposition of multiple textured surface patterns, or a division of multiple textured surface patterns in one single shared {HMIMO surface} \cite{Gonzalez2017Multibeam}. 
The previous studies follow a common point-to-point mapping between surface pattern and beam direction. Without obeying such a mapping, the authors of \cite{Wang2018Holographic} provided a novel {HMIMO surface} based on spoof surface plasmon polaritons (SPPs), which can change the propagation wavenumber of the reference wave in different directions while retaining the fixed surface pattern of concentric circles.

Following a distinct design methodology, namely, the geometric polarizable particle based approach (which will be further elaborated in the next subsection), the authors used slot-shaped antenna elements, located on holographic principle guided positions \cite{Yurduseven2017Dual}, specifically, the designed surface was fabricated to achieve a fixed multibeam radiation pattern. In another approach, researchers in \cite{Yurduseven2021Multibeam} designed an {HMIMO} surface with antenna elements located on positions that repeat periodically, the elements' response is fixed and different radiation patterns are achieved by changing the dedicated feeding port of the surface. Such antenna elements were later extended to PIN-controlled slot-shaped units for achieving reconfigurability \cite{Yurduseven2017Design, Yurduseven2018Dynamically}. Adopting more advanced complementary electric inductive-capacitive (CELC) resonator units as antenna elements, reference \cite{Smith2017Analysis} further increased the phase accuracy and the hologram range. 
It should be emphasized that the slot-shaped antenna elements \cite{Yurduseven2017Design, Yurduseven2018Dynamically, Yurduseven2021Multibeam} and the CELC based antenna elements \cite{Smith2017Analysis} are repeated periodically over the surface without presenting an explicit textured surface pattern.

To sum up, we present a list of main configurations for antenna elements, including the design methodologies (presented in the following subsection), antenna element shapes (shown in Fig. \ref{fig:ElementShapes}), aperture types, and the generated surface patterns, in Table \ref{tab:RadiatingElement_Config} for ease of reference.

\subsection{Holographic Design Methodologies of {HMIMO Surfaces}}
The previously described surface pattern of an {HMIMO surface} can be in various forms that correspond to different design methodologies. We distinguish three main holographic design methodologies, from the original locally maximum phase lines of hologram to the more recent macroscopic surface impedance based approach and geometric polarizable particle based approach. Each will be detailed as follows.

\subsubsection{Locally Maximum Phase Lines of Hologram}

The primary holographic design is achieved via the locally maximum phase lines of the hologram for its straightforward and easy interpretation. It should be first noted that the hologram generated by the reference and object waves is indeed an interference wave with a certain period. This hologram includes a certain number of locally maximum phase lines within a dedicated distance, each of which exists within one period.
In this design methodology antenna elements are located at the locally maximum phase lines of the hologram, forming a surface pattern consisted of a series of curve/line-shaped strip gratings. The resulting surface pattern allows the generation of an object wave with a specific direction. For a schematic representation, readers are encouraged to revisit Fig. \ref{fig:EM_holography}, where the maximum phase lines of the interference wave are clearly depicted. Note that, in this design methodology, the positioning of antenna elements and their response will be fixed in order to achieve a certain radiation pattern and can not be further reconfigured for creating different holograms.

To explicitly show this methodology, we assume a reference wave propagating in the $y$-direction $E_{rw} = A_r e^{-i \beta_r y}$, and an object wave steering toward a given direction, $\theta = \theta_{s}$ and $\phi = \pi/2$. Based on \eqref{EMInterferWave}, the hologram is reduced to 
\begin{equation}
	\begin{aligned}
		\hat{E}_{int} = A_{r} A_{o} e^{i \left(\beta_r + \beta_0 \sin \theta_s \right) y},
		\label{ReducedEMInterferWave}
	\end{aligned}
\end{equation}
whose wavenumber is given by $\beta_r + \beta_0 \sin \theta_s$, equivalent to wavelength of $\frac{2 \pi}{\beta_r + \beta_0 \sin \theta_s}$. 
Within each wavelength of the hologram, the antenna element is placed on the locally maximum phase line (that is, on the coordinates that the peak value of the interference wave is attained). All locally maximum phase lines of the hologram within the aperture compose the surface pattern constructed by a series of curve/line-shaped strip gratings.

From such a design methodology, one can see that the phase information between strip gratings is inevitably lost, lowering the performance of holographic mapping between reference wave and object wave. We can alternatively interpret this from a sampling perspective that the antenna elements take a small amount of specific samples of the reference wave, incurring the result of under-sampling. Moreover, as already mentioned, each surface pattern only corresponds to a specific wave radiation, such that one should change the surface pattern in order to manipulate the radiated wave. Lastly, any change in the communication frequency will also affect the radiation pattern and the main lobe will steer away from the intended direction.

\subsubsection{Macroscopic Surface Impedance Based Approach}

Deploying a large amount of sub-wavelength conductive patches over the surface, the reference wave can be densely sampled. The relatively small sizes of antenna elements and adjacent distances compared with wavelengths of both reference and object waves, enable the scatter properties to be described using macroscopic effective surface impedance. 
The macroscopic effective surface impedance is defined as the ratio of electric field $E_x$ to magnetic field $H_y$ near the surface averaged over the unit cell, which can be expressed as $Z = \int_{cell} \frac{E_x}{H_y} ds$ \cite{Fong2010Scalar}. 
It is related to substrate permittivity and thickness, as well as the period and gap between adjacent antenna elements. For instance, high impedance values are achieved with high substrate permittivities, thick substrate, large period and small gap.
Under a desired surface impedance, the antenna elements can be engineered in various shapes as shown in Table \ref{tab:RadiatingElement_Config}.

Holographic architecture realized by the macroscopic surface impedance based approach is established on collecting a large amount of specifically designed antenna elements for obtaining the expected surface impedance that is determined by the hologram created via the interference between reference and object waves. To be more specific, the surface impedance of $(x, y)$-coordinates is given by \cite{Fong2010Scalar}
\begin{equation}
	\begin{aligned}
		Z(x, y) = i \left[ X + M \Re \left( E_{obj} E_{rw}^{*} \right) \right],
		\label{ImpedanceByHologram}
	\end{aligned}
\end{equation}
where $X$ is the real-valued average impedance over the surface, and $M$ denotes the real-valued modulation depth that spans the entire available range of impedance values. It should be emphasized that the electric fields are functions of the \((x,y)\)-coordinates, thus the impedance variation over the coordinates. Equation
\eqref{ImpedanceByHologram} builds a mapping between the surface impedance and the hologram. Combining this mapping with the relation between surface impedance and geometry parameters of substrate and antenna elements, one can seamlessly correspond the hologram to geometry parameters. With a selected substrate (permittivity and thickness are fixed), the hologram is implemented as a surface pattern that is constructed by a large amount of patches with specific periods and gaps, displayed as concentric ellipses, spiral or concentric circles. These textured surface patterns correspond to different radiation properties, determining the emitted wave directions.
Holographic design via macroscopic surface impedance can be dated back to the sinusoidally-modulated reactance surfaces whose modal surface impedance is modulated sinusoidally for realizing an expected radiation \cite{Oliner1959Guided, Patel2011Printed}.

It should be emphasized that the above surface impedance is designed for radiating a single object wave. When multiple object waves are required to be radiated simultaneously, the surface impedance is designed as a division or a superposition of multiple textured surface patterns on one shared aperture. For the division case, the surface impedance design of each division follows from \eqref{ImpedanceByHologram}, indicating a corresponding object wave.
While for the superposition case, the surface impedance design is guided by the following expression
\begin{equation}
	\begin{aligned}
		Z(x, y) = i \left[ X + \frac{M}{K} \sum_{k = 1}^{K} \Re \left( E_{obj_{k}} E_{rw_{k}}^{*} \right) \right],
		\label{MultiBeamImpedanceByHologram}
	\end{aligned}
\end{equation}
where $K$ is the number of object waves. In this case, the {HMIMO surfaces} allow multi-beam radiations simultaneously, where each object wave is excited by one reference wave. 

Taking the surface impedance approach for holographic design can be boiled down to configuring the geometrical parameters of substrate and antenna elements, which impose a limitation on the realization of reconfigurable {HMIMO surfaces}, namely once the geometry parameters are designed, the radiating properties are determined accordingly. 
Tunability can be achieved by introducing special materials, such as graphene. Bridging the connection between conductivity of graphene and surface impedance, the radiation properties are tunable with the help of an external DC control, manipulating graphene's conductivity \cite{Cheng2016Real, Ren2022Graphene}. 

\subsubsection{Geometric Polarizable Particle Based Approach}

An alternative holographic design is achieved from a geometric perspective via using the polarizable particles (dipoles) based approach \cite{Johnson2014DDA, Smith2017Analysis}. Under the assumption that the element's size is small with respect to the operating free space wavelength, then the radiation field can be well approximated as that of a dipole \cite{Smith2017Analysis}.
This approach describes the generated radiations as a weighted sum of far-field patterns of all dipoles. Each antenna element corresponds to a weight that follows a specific constraint and is configured by the holographic principle.

For example, assume the far-field approximation of a radiated wave from an 1D microstrip line {HMIMO surface} (shown as one microstrip line of the aperture in Fig. \ref{fig:HMIMOS_structure}(c)). Under these settings, the radiation wave at a distance $d_o (x, y)$ can be expressed as \cite{Smith2017Analysis}
\begin{equation}
	\begin{aligned}
		&E_{rad} = \frac{A_{r} \omega^2}{4 \pi d_o (x, y)} e^{-i \beta_0 d_o (x, y)} \\
		&\times \underbrace{ \cos \theta \sum_{n=1}^{N} \alpha_{n}(\omega,x,y) e^{-i \beta_r d_r (x, y)} e^{-i \beta_0 d_r (x, y) \sin \phi} }_{AF(\theta, \phi)},
		\label{DipoleRadiationPattern}
	\end{aligned}
\end{equation}
where $\omega$ denotes the operating frequency, $\alpha_{n}(\omega,x,y)$ represents the weight of the $n$-th antenna element (located at the $(x, y)$-coordinate) at frequency $\omega$, and $AF(\theta, \phi)$ is defined as the array factor. It is emphasized that $\beta_r$ and $d_r (x, y)$ are the same as in \eqref{EMSurfaceWave}. 
Comparing the object wave in \eqref{EMObjectWave} with the radiation wave in \eqref{DipoleRadiationPattern}, we should model the array factor $AF(\theta, \phi)$ to be some constant ensuring $ E_{obj} = E_{rad}$, which means that the {HMIMO surface} radiates the desired wave to the object. As such, the weights of all antenna elements satisfying the requirement can be obtained as \cite{Smith2017Analysis}
\begin{equation}
	\begin{aligned}
		\alpha_{n}(\omega, x, y) = e^{i \beta_r d_r (x, y)} e^{j \beta_0 d_r (x, y) \sin \phi}.
		\label{Weight_Unconstrainted}
	\end{aligned}
\end{equation}
One can see from \eqref{Weight_Unconstrainted} that the weights require full control over the phase of each antenna element, which cannot be satisfied due to the inherent constraints faced by each antenna element. Each antenna element can be considered as a resonant electrical circuit scattering as a dipole, and each weight (overlooking the index $n$ and its coordinates $(x, y)$) can be expressed with the help of the Lorentzian form as \cite{Smith2017Analysis}
\begin{equation}
	\begin{aligned}
		\alpha(\omega) = \frac{F \omega^{2}}{\omega_{0}^{2} - \omega^{2} + i \omega \gamma},
		\label{Weight_Lorentzian}
	\end{aligned}
\end{equation}
where $F$ is the real-valued oscillator strength, $\omega_{0}$ denotes the resonance frequency, and $\gamma = \frac{\omega_{0}}{2 Q_{m}}$ describes the damping factor with $Q_{m}$ being the quality factor of the resonator.
The amplitude and phase of the weight are coupled through the connection $|\alpha(\omega)| = \frac{F \omega |\cos \psi|}{\gamma}$ with 
$\psi = \tan^{-1} \left( - \frac{\gamma \omega}{\omega_{0}^{2} - \omega^{2}} \right)$, 
which limits the weight range to a restricted subset compared with the independent control over amplitude and phase.
The antenna element can be tuned in two possible ways, namely by either shifting the resonance frequency or changing the damping factor. Depending on the selected tuning case, the weights can be configured in three forms: Amplitude-only, binary amplitude, and Lorentzian-constrained phase. 

First, in the amplitude-only case, the antenna element is near resonance such that $\alpha(\omega) = -i \frac{F \omega}{\gamma}$. By adjusting the oscillator strength $F$ or the damping factor $\gamma$, amplitude tuning of the weight can be achieved without changing its phase. The amplitude-only weight is deduced from \eqref{Weight_Unconstrainted} by taking its real part in the following formulation \cite{Smith2017Analysis}
\begin{equation}
	\begin{aligned}
		\alpha_{n}(\omega, x, y) = X_n + M_n \cos \left( \beta_r d_r (x, y) + \beta_0 d_r (x, y) \sin \phi \right),
		\label{Weight_AmplitudeOnly}
	\end{aligned}
\end{equation}
where $X_n$ and $M_n$ are real-valued positive variables. In this case, however, despite the intended directive beam, other beams are inevitably produced creating side lobes in the radiation pattern. 
Additionally, the binary amplitude case is applicable to antenna elements tuned between ``ON" and ``OFF" states, which can be realized by toggling the resonance frequency between a valid value within the operating frequency and a invalid value outside the operating frequency. In this case, each antenna element achieves only two amplitudes, given by \cite{Smith2017Analysis}
\begin{equation}
	\begin{aligned}
		\alpha_{n}(\omega, x, y) = X_n + M_n \Theta \cos \left( \beta_r d_r (x, y) + \beta_0 d_r (x, y) \sin \phi \right),
		\label{Weight_BinaryAmplitude}
	\end{aligned}
\end{equation}
where $\Theta \in \{0, 1\}$ enables $\alpha_{n}(\omega, x, y)$ to be an offset square wave.
Finally, the Lorentzian-constrained phase case depicts the inherent phase-amplitude dependency. The Lorentzian resonator limits the phase range to $[0, \pi]$, while the $(\pi, 2\pi]$ set is unattainable. Authors in \cite{Smith2017Analysis} proposed a simple mapping between the ideal polarizability and the constrained case, by considering the proposed mapping and the ideal weight function given in \eqref{Weight_Unconstrainted}, the constrained weight is given as, 
\begin{equation}
	\begin{aligned}
		\alpha_{n}(\omega, x, y) = \frac{i + e^{i \left( \beta_r d_r (x, y) + \beta_0 d_r (x, y) \sin \phi \right) }}{2},
		\label{Weight_LorentizianPhase}
	\end{aligned}
\end{equation}
the phase of $\alpha_{n}(\omega, x, y)$ is ensured to have a range of $[0, \pi]$, and the amplitude satisfies the constraint of $|\alpha_n(\omega, x, y)| = \left| \cos \left( \frac{\beta_r d_r (x, y) + \beta_0 d_r (x, y) \sin \phi}{2} \right) \right|$.

\subsection{Tuning Mechanisms of {HMIMO Surfaces}}
As continuous developments of {HMIMO surfaces} are realized, conventional non-tunable designs, basically being fixed in the design and fabrication process, are shifting to the paradigm of reconfigurability. The tunability enables holographic communications as an {HMIMO surface} becomes capable of achieving any required hologram by dynamically adjusting the antenna elements' response for exciting intended object waves. This characteristic is instrumental to future networks as the intended hologram will change based on the characteristics of the dynamic propagation environment.
In the following we will list the state of the art tuning mechanisms for realizing reconfigurable {HMIMO surfaces}, which are mainly based on lumped elements (PIN and varactor diodes), LCs, graphene, and photosensitive devices.



\subsubsection{Lumped Element Tuning}
PIN diodes belong to a class of tunable lumped electronic elements that can be controlled by forcing external DC bias voltages, thereby allowing a feasible solution for implementing reconfigurable {HMIMO surfaces}. 
A PIN diode is capable of presenting two states, referred to as ``ON" and ``OFF", by forward and reverse biasing via DC bias voltage, respectively. The ``ON" state can be modeled as a resistor-inductor concatenation circuit with a negligible forward resistance (effectively short-circuit), while the ``OFF" state is equivalent to a capacitor-inductor concatenation circuit presenting a high reverse resistance (effectively open-circuit) \cite{Yurduseven2017Design}. By incorporating a PIN diode in the center of each slot-shaped antenna element, the resonance frequency as well as the coupling between antenna elements can be controlled, and the emitted wave can thus be determined \cite{Yurduseven2017Design, Yurduseven2018Dynamically}. It is worth noting that the ``ON"/``OFF" states control of each PIN diode is determined by the holographic principle, i.e. controlling these states appropriately to reach an objective wave (or waves) that was captured by the surface.
Another type of tunable lumped element that can be used for achieving reconfigurability is the varactor diode. Compared with the PIN diode, it is capable of continuous phase tuning instead of the discrete ``ON" and ``OFF" states. In \cite{Boyarsky2021Electronically}, the authors presented a metasurface architecture that allows electronic beamsteering from a specifically designed layout with each antenna element controlled by two varactor diodes.
Lumped elements are one of the most prominent enablers for reconfigurable antenna elements because of their low-cost and easy implementation, opening up a wide range of applications. According to their intrinsic properties, lumped elements are mostly applied to (sub-) mmWave frequencies.



\subsubsection{Liquid Crystal Tuning}
As a special class of soft condensed matter, LCs exhibit properties between a state of liquid and solid crystals. They are capable of flowing as liquids and simultaneously exhibit anisotropy with molecules oriented in a solid crystal-like way. The LCs also demonstrate excellent tuning capabilities under external stimuli, such as electric or magnetic bias field, which controls the permittivity tensor of LCs. 
Their reconfigurable characteristics were first utilized for achieving reconfigurable {HMIMO surfaces} in \cite{Johnson2015Sidelobe, Stevenson2016Metamaterial, Kim2022Performance}. These studies design similar LC-controlled {HMIMO surfaces}, where LCs are filled in cavity boxes between slots and radiating patches to implement the reconfigurability by exciting radiating patches using external DC bias voltages. 
While the LCs are excited by external voltage, their molecules align in a specific direction, presenting a corresponding permittivity. Otherwise, the molecules are parallel to the surface resulting in a different permittivity. The varying permittivity changes the capacitance of antenna elements, consequently altering their frequency response.
The LC-controlled {HMIMO surfaces} have several advantages, including flexible integration into various structures thanks to fluidity, and low power consumption owing to the mostly capacitive operations of LCs with negligible currents. However, the main disadvantage is the relatively slow response of LCs, leading to limited applications which don't require extremely fast tunability.

\subsubsection{Graphene Tuning}
Graphene was discovered and isolated by Nobel laureates, Andre Geim and Konstantin Novoselov, in 2004 \cite{Novoselov2004Electric}. It is an atomic-scale 2D material composed of carbon atoms in hexagonal structure. It possesses superior mechanical, electronic and optical properties, namely ultra-high breaking strength, ultra-high charge carrier mobility and ultra-low resistivity for desired conductivity, as well as high transparency due to ultra-thin thickness.
Graphene's excellent properties, especially from an electronic perspective, allow dynamic control of its conductivity, thus enabling a feasible way for implementing reconfigurable antenna elements for {HMIMO surfaces} \cite{Cheng2016Real, Ren2022Graphene}. These studies design {HMIMO surfaces} by stacking the ultra-thin graphene sheet onto a  grounded silicon dioxide substrate, and then control the conductivity of graphene via DC bias voltages for achieving an arbitrary surface impedance guided by the macroscopic surface impedance approach. The main difference between the preceding studies is the adopted tuning mmethodology, specifically a pixel-by-pixel electrical control enabled by a relatively huge biasing network was proposed in \cite{Ren2022Graphene}, while a more succinct electrical control network coordinated by two groups of 1D biasing pads was introduced in \cite{Cheng2016Real}. The graphene based {HMIMO surfaces} generally work on THz frequencies and exhibit a nanosecond-scale or even faster control response, making ultra-fast tunability a reality. However, there are still challenges to be faced, for instance, a large area free-standing graphene is difficult to manufacture, which restricts its application to {HMIMO surfaces} with a large aperture area.

\begin{table*}[!t]
	\footnotesize
	\renewcommand{\arraystretch}{1.2}
	\caption{\textsc{Tuning mechanisms of HMIMO surfaces.}}
	\label{tab:TM_config}
	\centering
	\begin{tabular}{!{\vrule width0.6pt}c|c|c!{\vrule width0.6pt}}
	
		\Xhline{0.6pt}
		\rowcolor{yellow} \textbf{Tuning mechanism} & \textbf{Working frequency} & \textbf{Pros and cons} \\
		
		\Xhline{0.6pt}
		Lumped element & (Sub-) mmWave & \tabincell{c}{Pros: Low cost/power consumption, fast tunable response, easy implementation\\ Cons: Limited to a relatively low operating frequency} \\
		
		\hline 
		Liquid crystal & (Sub-) mmWave & \tabincell{c}{Pros: Flexible integration into various structures and low power consumption\\ Cons: Slow tunable response and relatively low operating frequency} \\
		
		\hline 
		Graphene & THz & \tabincell{c}{Pros: Superior electronic properties and ultra-fast tunable response\\ Cons: Difficulty in manufacturing for large surfaces} \\
		
		\hline 
		\tabincell{c}{Optical\\ (Photodiode)} & GHz - optical & \tabincell{c}{Pros: Superior electronic properties and ultra-fast tunable response\\ Cons: Difficulty in maintaining phase stability} \\
		
		\hline 
		\tabincell{c}{Electro-mechanical\\ (Piezoelectric actuator)} & mmWave & \tabincell{c}{Pros: Efficiently modify the EM properties with low loss\\ Cons: Slow tuning speed and limited device size} \\
		
		\hline 
		\tabincell{c}{Thermal\\ (Vanadium dioxide)} & mmWave & \tabincell{c}{Pros: Sufficient tuning range and integration compatibility\\ Cons: Slow tuning speed due to the time consuming process of heating and cooling} \\
		\Xhline{0.6pt}
	\end{tabular}
\end{table*}

\subsubsection{Optical Tuning}
In optical tuning, the reconfigurability can be implemented either via utilizing photosensitive semiconductors, e.g., silicon, or by driving dedicated devices, such as photodiodes. By properly illuminating/driving the corresponding photosensitive semiconductors/dedicated devices they can be optically controlled for achieving on/off inter-connections between antenna elements, thereby realizing reconfigurability. In \cite{Gonzalez2015Basic}, the authors showed a Si based checkerboard metasurface that is controlled optically by a laser source. The illuminating focal point of laser controls the electrical connection between two conductive metal patches exploiting the photoconductive properties of silicon. Different in \cite{Konkol2017High, Prather2017Optically, Carey2021Millimeter}, the authors proposed to use the dedicated photodiodes for realizing an optically controlled TCA. The photodiode is capable of achieving an optical-to-electrical transformation that can directly drive antenna elements with high power and high linearity. This scheme is mainly deployed in photonic TCA based EM holography. 
Optical tuning emerges as a promising alternative because of its inherent advantages. It is capable of obtaining an ultrafast tuning speed whilst protecting signals from EM interference. Notably, the photodiodes based scheme for TCA based {HMIMO surfaces} is proved to facilitate a full-optical processing in implementing wireless communication systems. Specifically, photodiodes assist in building optical domain processing with very large bandwidth while reducing power consumption and hardware cost compared to the conventional electrical domain processing. Combined with optical fiber, long distance deployments, where {HMIMO surfaces} and signal processing units are in different locations, can be practical. The major issue of such optical tuning systems is that the high phase stability, i.e. maintaining constant phase difference, of parallel optical signals should be strictly guaranteed which is generally difficult as optical fibers are sensitive to surrounding environments (temperature, vibration, airflow, and sound) that will greatly influence the signal phases.

\subsubsection{Others}
Besides the aforementioned tuning mechanisms, there exist various distinct approaches utilizing different principles the most significant of which are listed below. In \cite{Rabbani2020Electro}, the authors presented electro-mechanically tunable metasurfaces with high gain and beam steering capabilities at mmWave frequencies. The required tuning is obtained by controlling a piezoelectric actuator for varying the mechanical separation between antenna elements and a ground layer. It provides an extremely low loss scheme that can efficiently modify the EM properties as it directly controls the interspacing distances. However, this type of tuning approaches usually suffer from their slow tuning speed and limited device size.
In addition, a vanadium dioxide integrated reconfigurable metasurface was demonstrated in \cite{Hashemi2016Reconfigurable} for mmWave beam scanning. The vanadium dioxide belongs to the phase changing materials, which exhibit a phase transition from metal to insulator under thermal variations. Based on this useful property, \cite{Hashemi2016Reconfigurable} implemented the tunability via applying voltage to heating electrodes that control the dielectric properties of vanadium dioxide. Although a thermal tuning scheme can generally achieve a wide tuning range and is realatively easy to implement, it suffers from slow tuning speed due to the time consuming heating and cooling.


Finally, we provide a summary of various tuning mechanisms with respect to their operating frequencies as well as their pros and cons, which can be found in Table \ref{tab:TM_config}.

\subsection{Fabrication Methodologies of {HMIMO Surfaces}}

Various fabrication methodologies for implementing {HMIMO surfaces} are invented based on surface lithography technologies, mainly including photolithography, electron-beam lithography, focused-ion-beam lithography, interference lithography, self-assembly lithography and nanoimprint lithography \cite{Hsiao2017Fundamentals}.
Among these lithography technologies, photolithography is broadly applied to semiconductor integrated circuits with high throughput at microscale and nanoscale sizes. The ultimate nanostructures are shaped on substrates after experiencing several steps, such as exposure, developing, etching, depositing and lift-off processes.
Besides, electron-beam lithography is capable of drawing arbitrary patterns with several nanoscale resolution. It performs a direct-write fashion without using a photomask, creating a maskless lithography.
Alternatively, the focused-ion-beam lithography is another direct-write technology for creating fine nanostructures on a surface.
Differently, utilizing the interference of more than one coherent laser beams, interference lithography achieves the nanostructure patterning. It can be considered as a new modality of photolithography.
Furthermore, selfassembly lithography is enabled for creating various large-area nanostructures by utilizing the principle of intermolecular balance of attractive and repulsive forces for spontaneously assembling.
Lastly, nanoimprint lithography is mostly applicable to create large-area periodic sub-wavelength structures in a single step with low-cost and high throughput. 
Additionally, the commercially employed PCB laminates make it feasible to apply the off-the-shelf PCB fabrication technology directly. This results in competitive fabrication price and guarantees mechanical and electrical stability.

\subsection{Aperture Shapes of {HMIMO Surfaces}}
\label{SectionIII_Aperture_Shape}

\begin{figure*}[t!]
	\centering
	\includegraphics[height=3.4cm, width=18cm]{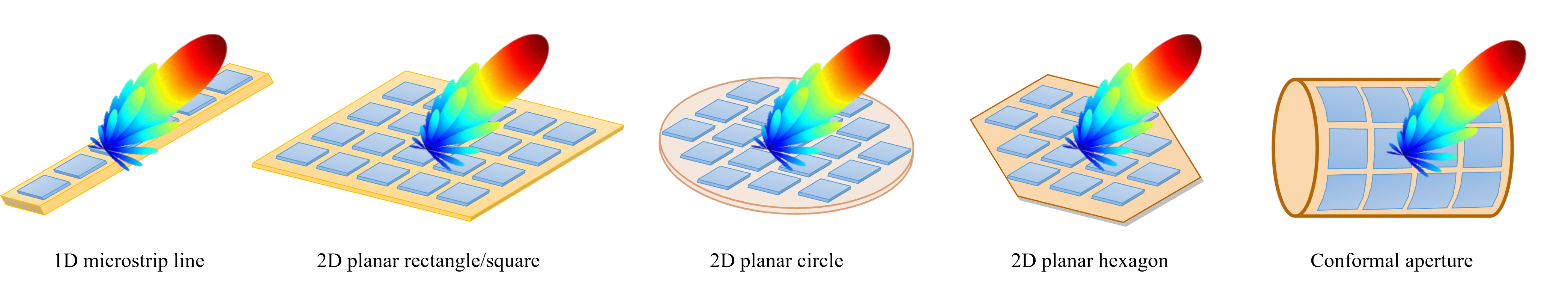}
	\caption{Typical aperture shapes of {HMIMO surfaces}.}
	\label{fig:Aperture_Shape}
\end{figure*}

{HMIMO surfaces} can be in various shapes based on the installation requirements and the design availability.
One of the simplest aperture shapes is the 1D microstrip line \cite{Yurduseven2018Dynamically, Kim2022Performance}. 
It is further emphasized that the prevalent aperture shape of {HMIMO surfaces} is 2D planar, where square/rectangle-shaped apertures \cite{Yurduseven2021Multibeam, Ramalingam2019Polarization}  and circular/hexagon-shaped apertures \cite{Movahhedi2019Multibeam, Wu2021Millimeter, Stevenson2016Metamaterial} are currently deployed. The aforementioned 1D microstrip line can be packed along with other microstrip lines onto a common panel to form a 2D planar aperture. Beyond these shapes, reference \cite{Wang2018Holographic} presented a special 2D circular aperture with 16 splitted strip branches. 
Even though the 2D planar aperture is the leading edge in shape design, it is necessary to design conformal {HMIMO surfaces} to meet special installation requirements. The authors of \cite{Li2019Wide} demonstrated a cylinder-shaped conformal lens fed by one/more sources, capable of realizing wide-angle beam steering by collimating the incident spherical wave front into a plane wave front.
Alternatively, the authors of \cite{Yoo2022Design} designed a a cylindrical conformal aperture constructed by multiple waveguide-fed 1D microstrip lines. 
Subsequently, the authors of \cite{Liu2021Conformal} presented a cylinder-shaped conformal omni-{HMIMO surface}\footnote{omni-HMIMO surfaces have reflecting and refracting capabilities thus enabling an omnidirectional manipulation of impinging EM waves. In the literature, for the specific case of RIS it can be found under the name simultaneously transmitting and reflecting (STAR) RIS.} that is capable of achieving a linear-to-circular polarization conversion. Breaking the barrier that traditional surface impedance based apertures are restricted to 2D plane or 3D cylinder, Voronoi partition was utilized in \cite{Nicholson2021Conformal} for designing an {HMIMO surface}, which can be adapted to form arbitrary conformal aperture. 
We list several typical aperture shapes in Fig. \ref{fig:Aperture_Shape}.
The non-limited aperture shapes enable great potential of {HMIMO surfaces} to be deployed in a multitude of positions both in normal or special shapes.

\subsection{Typical Functionalities of {HMIMO Surfaces}} 
{HMIMO surfaces} are capable of achieving a wide scope of functionalities for EM wave manipulations. Although they can be utilized as RIS that have been extensively investigated, we notice that RIS mainly work as passive reflectors and their working mechanisms do not directly follow the holographic operating rule. Therefore, we mainly focus on {HMIMO surfaces} working as transceivers that are guided by the holographic principle. Based on existing studies, we list two typical functionalities, namely EM wave polarization and EM wave steering. Each of the two will be described as follows. It should be emphasized that possessing one and/or more functionalities mainly depends on the design purpose and the scheme feasibility.

\subsubsection{EM Wave Polarization}
EM wave polarization indicates the oscillation orientation of a transverse wave. The polarization functionality is capable of achieving a transformation of the oscillation orientation via different designs or tuning mechanisms.
Most studies mainly focus on generating a desired wave without polarization control, such as \cite{Sievenpiper2005Holographic, Cheng2016Real, Yurduseven2018Dynamically, Ren2022Graphene} to name a few, which indicates that these {HMIMO surfaces} present a single polarization mode with a predefined oscillation orientation. Contrary to the aforementioned, authos in \cite{Fong2010Scalar} presented a framework for polarization control through creating a surface with tensor impedance properties. The control over polarization is achieved by varying the shape of antenna elements (making a specific slit on each patch) with a careful design on the geometric parameters of the slits. Later, a circularly polarized {HMIMO surface} was implemented in \cite{Minatti2011Spiral} by designing a spiral like surface pattern. Furthermore, the authors of \cite{Pandi2015Design} proposed to control over polarization using a scalar impedance surface, which greatly reduces the design complexity compared with that of tensor surfaces. The polarization control is achieved by dividing the {HMIMO surface} into different regions and changing the phase of surface impedance modulation of one region relative to the others, capable of realizing arbitrary linear and circular polarization. Alternatively, an orthogonally discrete unit-cell with four working states was proposed in \cite{Han2021Holographic}, for realizing linearly and circularly polarized waves via matching surface impedance along two orthogonal directions. It should be emphasized that the polarization control elaborated above, becomes fixed once the {HMIMO surfaces} are fabricated. 
To encompass more polarization states, the authors of \cite{Yurduseven2017Dual} demonstrated a dual-polarization {HMIMO surface} using linearly polarized, slot-shaped antenna elements located in horizontal and vertical directions. Additionally, a dual circularly polarized {HMIMO surface} was designed and experimented in \cite{Tellechea2016Dual} and \cite{Pereda2018Experimental}, respectively. Differently, in \cite{Ramalingam2019Polarization} a class of polarization-diverse {HMIMO surfaces} was introduced that enable horizontal, vertical or circular polarization via multiple excitation sources that act as switches. Subsequently, a polarization reconfigurable {HMIMO surface} was demonstrated in \cite{Li2019Design}, capable of achieving linear, and right/left-hand circular polarization via integrating an appropriately designed feed structure switched by PIN diodes into a polarization-insensitive surface \cite{Li2016Polarization}.


\begin{figure*}[t!]
	\centering
	\includegraphics[height=2.9cm, width=18cm]{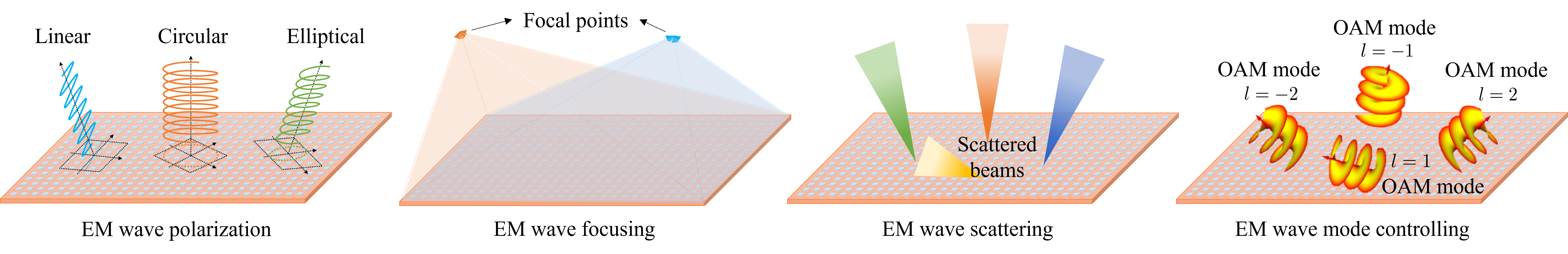}
	\caption{Typical functionalities of {HMIMO surfaces}.}
	\label{fig:Typical_functionality}
\end{figure*}

\subsubsection{EM Wave Steering}
EM wave steering defines the process of generating, either via transmission or reflection, a single or multiple beams in different directions by consrtuctively or destructively adding the phases of EM waves. As such, we encompass the EM wave focusing, which refers to beams focused at a specific point in space rather than the conventional angular beams, and scattering in the scope of steering capability. 
The initial steering capability is mainly enabled for supporting a single-beam radiation from the primary fixed configuration to the tuning capability afterwards. In this branch, the studies \cite{Checcacci1970Holographic, Iizuka1975Volume, ElSherbiny2004Holographic, Rusch2015Holographic} presented {HMIMO surfaces} with a single-beam radiation based on the ``locally maximum phase lines of hologram" design methodology. The single beam is generated based on strip grating (discrete dipole) radiations that are constructively superposed in-phase in the desired direction and destructively canceled in remaining directions. This design methodology leads to an inevitable loss of phase information that restricts the performance of beam radiation. Later, the authors of \cite{Sievenpiper2005Holographic, Fong2010Scalar, Pandi2015Design} designed single-beam radiating {HMIMO surfaces} following from the macroscopic surface impedance based approach, allowing a more accurate manipulation on EM 
waves. Recent pioneer studies inspired a multitude of research activities on single-beam radiating {HMIMO surfaces} \cite{Li2016Polarization, Galler2020High, Nicholson2020Discrete, Han2021Holographic}, and were later promoted one step forward with reconfigurable capabilities via utilizing DC controlled graphene conductivity \cite{Cheng2016Real, Ren2022Graphene} for realizing a global surface impedance control. The global surface impedance control then evolved to a local surface impedance control via using mixed-signal integrated circuits, as conceptually shown in \cite{Liu2019Intelligent}. Alternatively, based on the geometric polarizable particle based approach, different single-beam {HMIMO surfaces} were demonstrated in \cite{Smith2017Analysis} with useful comments on various radiation weights. The reconfigurability was further achieved in \cite{Yurduseven2017Design, Yurduseven2018Dynamically} facilitated by PIN diodes and in \cite{Johnson2015Sidelobe, Stevenson2016Metamaterial, Kim2022Performance} enabled by LCs. 
Beyond realizations of single-beam radiation, the steering capability for supporting multibeam radiations attracts significant interest for the promising application potential in supporting future communication systems. Several multibeam {HMIMO surfaces} with fixed beam radiations were respectively presented based on the geometric polarizable particle based approach \cite{Yurduseven2017Dual}, as well as by designing specific impedance patterns for being a transmit array \cite{Gonzalez2017Multibeam, Movahhedi2019Multibeam, Han2021Holographic} and a reflect array \cite{Karimipour2018Holographic, Karimipour2018Realization}.  Furthermore, extending the static configuration of multibeam radiations, researchers in \cite{Lv2019Holographic, Yurduseven2021Multibeam} achieved a dynamically switchable multibeam radiation, implemented by switching different feed ports.
Going one step further, a multibeam radiating {HMIMO surface} capable of achieving dynamic reconfigurability was designed in \cite{Sleasman2020Design}.


Beyond the preceding efforts, the steering capability also facilitates multibeam scanning with changes of frequencies \cite{Li2014Frequency} and promotes multiwavelength multiplexing \cite{Li2018Multiwavelength}. Moreover, specifically designed {HMIMO surfaces} are capable of respectively radiating multiple high-order Bessel beams and vortex beams that carry different OAM modes \cite{Meng2021Generation, Wu2021Millimeter, Wu2022Millimeter}.
A little bit further, it is appealing to integrate multiple functionalities using one single {HMIMO surface}. The authors of \cite{Yang2016Programmable} investigated diverse EM responses of a large-scale programmable metasurface, where they realized reconfigurable polarization conversions and dynamic steering capabilities on the same surface via using PIN diodes. We show typical functionalities of {HMIMO surfaces} in Fig. \ref{fig:Typical_functionality}.


\subsection{Representative Prototypes of {HMIMO Surfaces}}
In the evolution process, a variety of prototypes are designed and fabricated, based on which numerous functionalities and capabilities are experimented. The validations powered by prototypes will enormously promote commercial uses and wide deployments for supporting future communications. In this section, we first select representative prototypes of {HMIMO surfaces} for demonstrating the main advances in this area.
We then list recent advances of {HMIMO surfaces} aided communication prototypes that validate communication capability from an end-to-end perspective.
We believe that the existing and emergence of prototypes will promote the validations/applications to a new level, and will also play a critical role in validating future wireless communications.

\subsubsection{{HMIMO Surfaces} Prototypes}
We present prototypes of {HMIMO surfaces} by characterizing their physical parameters and functional capabilities, with which we carry out in three categories based on the design methodologies. Based on the locally maximum phase lines of hologram, the authors in \cite{Rusch2015Holographic} designed two $55.5$mm/$49.0$mm $\times$ $30$mm apertures with $20$ and $12$ strip gratings (antenna elements), respectively, for achieving a fixed direction radiation at $60$ GHz with more than $20$dBi maximum gain achieved. Utilizing the more advanced macroscopic surface impedance and geometric polarizable particle based approaches, a multitude of prototypes are presented. In \cite{Ramalingam2019Polarization}, the authors designed a $9.6 \lambda \times 8 \lambda$ ($\lambda$ free space wavelength) aperture with $3000$ metallic patches operating at $12$ GHz for demonstrating a polarization-diverse capability. This aperture achieved a $16$dBi radiation gain with $1.06$ circular-polarized axial ratio and $3.57\%$ aperture efficiency. The work \cite{Galler2020High} employed the glass package technology to enable a $35$mm $\times$ $35$mm aperture that consists of $28900$ circular pixels and operates at $150$ GHz. This structure achieved a maximum $24.7$dBi antenna gain and a $4.7^{\circ}$ $3$dB beamwidth.
Capable of supporting multibeam radiations, the authors of \cite{Lv2019Holographic} fabricated a $255.5$mm $\times$ $255.5$mm $\times$ $1.5$mm aperture covered by $73 \times 73$ units. They measured a higher than $17.2$dBi peak gain at $18$ GHz. In addition, employing a reflecting working mode, the authors of \cite{Karimipour2018Holographic, Karimipour2018Realization} designed a $18$mm $\times$ $18$mm aperture containing $2025$ units for achieving multibeam radiations with linear and circular polarization. In the linearly polarized multibeam case, they measured a $23.5$dBi peak gain of each beam, $47.2\%$ aperture efficiency, lower than $-10$dB cross-polarization and $-26$dB sidelobe at $15$ GHz. In the circularly-polarized multibeam case, $19.43$dBi peak gain of each beam, $45.8\%$ aperture efficiency, higher than $25$dB cross-polarization discrimination, below $-17$dB sidelobe, and $3$dB axial ratio bandwidths of $19.23\%$ and $18.26\%$ were achieved at $13$ GHz. Adopting a semiconductor substrate (Si/GaAs), the authors of \cite{Yurduseven2021Multibeam} presented a $56 \lambda_g \times 56 \lambda_g$ aperture with $\lambda_g$ being the wavelength of reference wave in GaAs at $94$ GHz. This prototype is capable of radiating multiple beams via switching different feed ports, with a directivity gain of $31.9$dBi and a reflection coefficient lower than $-15$dB. 
Enabled by tunable devices and materials, reconfigurable {HMIMO surfaces} are spurred. The authors used PIN diodes to dynamically control the radiation of each unit in \cite{Yurduseven2018Dynamically}. They designed a 1D microstrip line aperture that includes $79$ slot-shaped units with a unit periodicity of $\lambda_g/3.3$ and each unit $\lambda_g/2.5$ long and $\lambda_g/17.5$ wide. The antenna gain was measured as $14.6$dBi with $35\%$ aperture efficiency, $4.1^{\circ}$ half-power ($3$dB) beamwidth, and higher than $20$dB polarization purity. Alternatively, the authors of \cite{Kim2022Performance} utilized LCs to enable the tunability. An aperture with $43$ antenna elements was fabricated with a beam scanning capability of $-60^{\circ} \sim 60^{\circ}$ at $10$ GHz. With a specifically designed decoupling structure, improvements on antenna gain and sidelobe can be maximally achieved by $3.93$dB and $7.7$dB, respectively, compared with antennas without the decoupling structure.

\begin{table*}[!t]
	\footnotesize
	\renewcommand{\arraystretch}{1.2}
	\caption{\textsc{Physical aspects of HMIMO surfaces.}}
	\label{tab:physical_aspects}
	\centering
	\resizebox{\linewidth}{!}{
		\begin{tabular}{!{\vrule width0.6pt}c|c|c|c|c|c|c|c|c|c|c|c!{\vrule width0.6pt}}
			\Xhline{0.6pt}
			\rowcolor{yellow} \textbf{\tabincell{c}{Design\\methodology}} & \textbf{Ref.} & \textbf{\tabincell{c}{Beam\\mode}} & \textbf{\tabincell{c}{Polarization\\mode}} & \textbf{Tunability} & \textbf{\tabincell{c}{Tuning\\mechanism}} & \textbf{\tabincell{c}{Aperture\\shape}} & \textbf{\tabincell{c}{Substrate\\type}} & \textbf{\tabincell{c}{Reference\\wave type}} & \textbf{\tabincell{c}{Feed \& position}} & \textbf{\tabincell{c}{Antenna element}} & \textbf{\tabincell{c}{Operating\\frequency}} \\
			\hline
			\multirow{5}{*}{\tabincell{c}{Locally\\ maximum\\ phase lines\\ of hologram}} & \cite{Iizuka1975Volume} & \multirow{5}{*}{\tabincell{c}{Single}} & \multirow{5}{*}{\tabincell{c}{Single}} & \ding{55} & \ding{55} & \multirow{5}{*}{\tabincell{c}{2D planar\\(square)}} & \multirow{5}{*}{\tabincell{c}{Dielectric}} & \multirow{5}{*}{\tabincell{c}{Surface\\ wave}} & \tabincell{c}{A pyramidal horn\\ Edge-fed} & \tabincell{c}{Curve-shaped\\ metal strips} & \tabincell{c}{11.7 ---\\ 12.3 GHz} \\
			\cline{2-2} \cline{5-6} \cline{10-12}
			& \cite{ElSherbiny2004Holographic} &  &  & \ding{55} & \ding{55} &  &  &  & \tabincell{c}{WR28 waveguide\\ Edge-fed} & \tabincell{c}{Curve-shaped\\ metallic dipoles} & 35 GHz \\
			\cline{2-2} \cline{5-6} \cline{10-12}
			& \cite{Rusch2015Holographic} &  &  & \ding{55} & \ding{55} &  &  &  & \tabincell{c}{AUF, Planar YUF\\ Surface-fed} & \tabincell{c}{Curve-shaped metal\\ strip gratings} & 60 GHz \\
			\Xhline{0.6pt} 
			\multirow{31}{*}{\tabincell{c}{Macroscopic\\ surface\\ impedance\\ based\\ approach}} & \cite{Sievenpiper2005Holographic} & \multirow{15}{*}{\tabincell{c}{Single}} & \multirow{9}{*}{\tabincell{c}{Single}} & \ding{55} & \ding{55} & \tabincell{c}{2D square,\\ 3D cylinder} & \multirow{15}{*}{\tabincell{c}{Dielectric}}  & \multirow{15}{*}{\tabincell{c}{Surface\\ wave}} & \multirow{7}{*}{\tabincell{c}{\tabincell{c}{A monopole\\ Bottom-fed}}} & \tabincell{c}{Square-shaped\\ metal patches} & 17 GHz \\
			\cline{2-2} \cline{5-7} \cline{11-12}
			& \cite{Fong2010Scalar} &  &  & \ding{55} & \ding{55} & \multirow{11}{*}{\tabincell{c}{2D planar\\(square)}} &  &  &  & \tabincell{c}{Slitted square-shaped\\ metal patches} & 10 GHz \\
			\cline{2-2} \cline{5-6} \cline{11-12}
			& \cite{Pandi2015Design} &  &  & \ding{55} & \ding{55} &  &  &  &  & \multirow{3}{*}{\tabincell{c}{Square-shaped\\ metal patches}} & \tabincell{c}{\\ 12 GHz} \\
			\cline{2-2} \cline{5-6} \cline{12-12}
			& \cite{Nicholson2020Discrete} &  &  & \ding{55} & \ding{55} &  &  &  &  &  & \tabincell{c}{\ding{55}\\ \quad} \\
			\cline{2-2} \cline{5-6} \cline{10-12}
			& \cite{Galler2020High} &  &  & \ding{55} & \ding{55} &  &  &  & \tabincell{c}{A TGV port\\ Bottom-fed} & \tabincell{c}{Slitted circular-shaped\\ metal patches} & 150 GHz \\
			\cline{2-2} \cline{4-6} \cline{10-12}
			& \cite{Ramalingam2019Polarization} &  & \tabincell{c}{Polarization-\\diverse} & \ding{55} & \ding{55} &  &  &  & \tabincell{c}{Two monopoles\\ Bottom-fed} & \tabincell{c}{Square-shaped\\ metal patches} & 12 GHz \\
			\cline{2-2} \cline{4-6} \cline{10-12}
			& \cite{Cheng2016Real} &  & \multirow{3}{*}{\tabincell{c}{Single}} & \ding{51} & Graphene &  &  &  & \tabincell{c}{A Dipole array\\ Edge-fed} & \tabincell{c}{Graphene strips} & 2 THz \\
			\cline{2-2} \cline{5-7} \cline{10-12}
			& \cite{Ren2022Graphene} &  &  & \ding{51} & Graphene & \tabincell{c}{2D square,\\ 3D cylinder} &  &  & \tabincell{c}{A monopole\\ Bottom-fed} & \tabincell{c}{Square-shaped\\ graphene patches} & 1 THz \\
			\cline{2-12}
			& \cite{Han2021Holographic} & \multirow{19}{*}{\tabincell{c}{Multiple}} & \multirow{19}{*}{\tabincell{c}{Single}} & \ding{55} & \ding{55} & \multirow{3}{*}{\tabincell{c}{2D planar\\(square)}} & \multirow{19}{*}{\tabincell{c}{Dielectric}} & \multirow{9}{*}{\tabincell{c}{Surface\\ wave}} & \multirow{3}{*}{\tabincell{c}{\tabincell{c}{A monopole\\ Bottom-fed}}} & \tabincell{c}{Square-shaped cells\\ (four metal patches)} & 11 GHz \\
			\cline{2-2} \cline{5-6} \cline{11-12}
			& \cite{Li2014Frequency} &  &  & \ding{55} & \ding{55} &  &  &  &  & \tabincell{c}{Square-shaped\\ metal patches} & \tabincell{c}{16 ---\\ 19GHz} \\
			\cline{2-2} \cline{5-7} \cline{10-12}
			& \cite{Gonzalez2017Multibeam} &  &  & \ding{55} & \ding{55} & \multirow{3}{*}{\tabincell{c}{2D planar\\(circular)}} &  &  & \tabincell{c}{Four monopoles\\ Bottom-fed} & \tabincell{c}{Ellipse-shaped\\ metal patches} & \ding{55} \\
			\cline{2-2} \cline{5-6} \cline{10-12}
			& \cite{Movahhedi2019Multibeam} &  &  & \ding{55} & \ding{55} &  &  &  & \tabincell{c}{An AUF\\ Surface-fed} & \multirow{3}{*}{\tabincell{c}{\tabincell{c}{Square-shaped\\ metal patches}}} & \tabincell{c}{12 ---\\ 18 GHz} \\
			\cline{2-2} \cline{5-7} \cline{10-10} \cline{12-12}
			& \cite{Han2021Holographic} &  &  & \ding{55} & \ding{55} & \multirow{11}{*}{\tabincell{c}{2D planar\\(square)}} &  &  & \tabincell{c}{A monopoles\\ Bottom-fed} &  & 11 GHz \\
			\cline{2-2} \cline{5-6} \cline{9-12}
			& \cite{Karimipour2018Holographic}&  &  & \ding{55} & \ding{55} &  &  & \multirow{3}{*}{\tabincell{c}{Space\\ wave}} & \multirow{3}{*}{\tabincell{c}{\tabincell{c}{A horn feed\\ External-fed}}} & \tabincell{c}{Square-shaped\\ metal patches} & 15 GHz \\
			\cline{2-2} \cline{5-6} \cline{11-12}
			& \cite{Karimipour2018Realization}&  &  & \ding{55} & \ding{55} &  &  &  &  & \tabincell{c}{Slitted square-shaped\\ metal patches} & 13 GHz \\
			\cline{2-2} \cline{5-6} \cline{9-12}
			& \cite{Lv2019Holographic} &  &  & \tabincell{c}{\ding{55}\\Switchable} & \ding{55} &  &  & \multirow{5}{*}{\tabincell{c}{Surface\\ wave}} & \tabincell{c}{Four monopoles\\ Bottom-fed} & \multirow{3}{*}{\tabincell{c}{Square-shaped\\ metal patches}} & 18 GHz \\
			\cline{2-2} \cline{5-6} \cline{10-10} \cline{12-12}
			& \cite{Li2018Multiwavelength}&  &  & \ding{55} & \ding{55} &  &  &  & \tabincell{c}{Two monopoles\\ Bottom-fed} &  & \tabincell{c}{17 GHz,\\ 20 GHz} \\
			\cline{2-2} \cline{5-6} \cline{10-12}
			& \cite{Meng2021Generation}&  &  & \ding{55} & \ding{55} &  &  &  & \tabincell{c}{Two dipoles\\ Bottom-fed} & \tabincell{c}{Slitted square-shaped\\ metal patches} & \tabincell{c}{30 GHz,\\ 30.5 GHz} \\
			\Xhline{0.6pt} 
			\multirow{15}{*}{\tabincell{c}{Geometric\\ polarizable\\ particle\\ based\\ approach}} & \cite{Kim2022Performance} & \multirow{9}{*}{\tabincell{c}{Single}} & \multirow{9}{*}{\tabincell{c}{Single}} & \ding{51} & LC & \tabincell{c}{1D micro-\\strip line} & \multirow{9}{*}{\tabincell{c}{Dielectric}} & \multirow{9}{*}{\tabincell{c}{Waveguide}} & \multirow{5}{*}{\tabincell{c}{\tabincell{c}{A coaxial feed\\ Edge-fed}}} & \tabincell{c}{Square-shaped\\ metal patches} & 10 GHz \\
			\cline{2-2} \cline{5-7} \cline{11-12}
			& \cite{Johnson2015Sidelobe}&  &  & \ding{51} & LC & \tabincell{c}{2D planar\\(square)} &  &  &  & \multirow{3}{*}{\tabincell{c}{\tabincell{c}{Square-shaped\\ meta-atom}}} & 30 GHz \\
			\cline{2-2} \cline{5-7} \cline{12-12}
			& \cite{Stevenson2016Metamaterial}&  &  & \ding{51} & LC & \tabincell{c}{2D planar\\(hexagon)} &  &  &  &  & \tabincell{c}{Ku band,\\ Ka bands} \\
			\cline{2-2} \cline{5-7} \cline{10-12}
			& \cite{Yurduseven2017Design} &  &  & \ding{51} & PIN diode & \tabincell{c}{2D planar\\(square)} &  &  & \tabincell{c}{A coaxial feed\\ Bottom-fed} & \multirow{3}{*}{\tabincell{c}{\tabincell{c}{Slot-shaped\\ antenna elements}}} & 20 GHz \\
			\cline{2-2} \cline{5-7} \cline{10-10} \cline{12-12}
			& \cite{Yurduseven2018Dynamically} &  &  & \ding{51} & PIN diode & \tabincell{c}{1D micro-\\strip line} &  &  & \tabincell{c}{A coaxial feed\\ Edge-fed} &  & 20 GHz \\
			\cline{2-12} 
			& \cite{Yurduseven2017Dual} & \multirow{5}{*}{\tabincell{c}{Multiple}} & Dual & \ding{55} & \ding{55} & \multirow{5}{*}{\tabincell{c}{2D planar\\(square)}} & Dielectric & \multirow{3}{*}{\tabincell{c}{Waveguide}} & \tabincell{c}{A coaxial feed\\ Bottom-fed} & \multirow{3}{*}{\tabincell{c}{\tabincell{c}{Slot-shaped\\ antenna elements}}} & 25 GHz \\
			\cline{2-2} \cline{4-6} \cline{8-8} \cline{10-10} \cline{12-12} 
			& \cite{Yurduseven2021Multibeam} &  & \multirow{3}{*}{\tabincell{c}{Single}} & \tabincell{c}{\ding{55}\\Switchable} & \ding{55} &  & \tabincell{c}{Semi-\\conductor} &  & \tabincell{c}{Multiple SIWs\\ Edge-fed} &  & 94 GHz \\
			\cline{2-2} \cline{5-6} \cline{8-8} \cline{9-12}
			& \cite{Boyarsky2021Electronically} &  &  & \ding{51} & Varactor &  & Dielectric & \tabincell{c}{Surface\\ wave} & \tabincell{c}{Multiple SIWs\\ Edge-fed} & CELC & 10 GHz \\
			\Xhline{0.6pt} 
			\multirow{3}{*}{\tabincell{c}{Others\\ (Spoof SPPs)\\ (OAM)}} & \cite{Wang2018Holographic} & Single & \multirow{3}{*}{\tabincell{c}{Single}} & \ding{55} & \ding{55} & \multirow{3}{*}{\tabincell{c}{2D planar\\(circular)}} & \multirow{3}{*}{\tabincell{c}{Dielectric}} & Waveguide & \tabincell{c}{A coaxial feed\\ Bottom-fed} & \tabincell{c}{Periodic (rotated)\\ metal strip lines} & 14 GHz \\
			\cline{2-3} \cline{5-6} \cline{9-12}
			& \cite{Wu2021Millimeter}& Multiple &  & \ding{55} & \ding{55} &  &  & \tabincell{c}{Surface\\ wave} & \tabincell{c}{A monopole\\ Bottom-fed} & \tabincell{c}{C-shaped\\ metallic units} & 60 GHz \\
			\Xhline{0.6pt} 
		\end{tabular}}
\end{table*}

Beyond the representative prototypes mentioned above, several initial commercial products ($39$ GHz CPE beamformer,  $28$ GHz RAN beamformer,  $28$ GHz repeater beamformer, and $14$ GHz A2G beamformer) were released by Pivotalcommware company for different business users (from original equipment manufacturers to network operators) with distinct requirements, such as radio access, signal relay, and air-to-ground broadband communications. Take the $28$ GHz RAN beamformer for instance, the product is capable of achieving a wide angle beam steering of $-60^{\circ} \sim 60^{\circ}$ in both azimuth and elevation directions, as well as a fast beam switching of $100$ns execution rate and $4 \mu s$ update rate. It also supports horizontal and vertical polarization, and achieves a high beamforming gain of $25$dBi with $6^{\circ}$ half-power beamwidth in both azimuth and elevation directions \cite{Pivotalcommware2021RANBeamformer}.
These products highlight the potential of {HMIMO surfaces} in promoting communication systems with a tremendous reduction in cost, size, weight and power consumption.

\subsubsection{{HMIMO Surfaces} Aided Communication Prototypes}
By integrating {HMIMO surfaces} into communication systems, one can evaluate communication performance and verify new capabilities through {HMIMO surfaces} aided communication prototypes, which is critical for meeting future wireless networks' requirements. 
The authors of \cite{Tang2020Wireless} first integrated a $256$-element aperture as RF chain-free transmitter and space-down-conversion receiver into a communication system to form an end-to-end MIMO prototype working at $4.35$ GHz. Using a 16 quadrature amplitude modulation (QAM), an experimental $2 \times 2$ MIMO-16QAM transmission with $20$ megabit-per-second data rate is achieved, validating the great potential to enable the cost-effective and energy-efficient systems. Beyond such an early work, the same team built various communication system prototypes by fully exploiting the functionalities of {HMIMO surfaces}. They demonstrated various end-to-end communication prototypes, including binary frequency-shift keying/phase-shift keying/QAM transmitters, pattern modulation system, multichannel direct transmission system, and space-/frequency-division multiplexing system \cite{Cheng2022Reconfigurable}. 
Besides, the authors in \cite{Dai2020Reconfigurable} built an {HMIMO surface} aided wireless communication prototype using a $256$-element reflecting platform. The prototype was constructed by modular hardware (hosts, USRPs and an RIS) and flexible software (source and channel coding, orthogonal frequency-division multiplexing (OFDM) modulation, etc.) to validate wireless transceiver capabilities. They measured $21.7$dBi and $19.1$dBi antenna gains at $2.3$ GHz and $28.5$ GHz, respectively. 
Afterwards, \cite{Pei2021RIS} adopted a $1100$-element reflecting platform and presented a prototype of {HMIMO surface} aided wireless communications working at $5.8$ GHz. The indoor and outdoor field trials showed that more than $26$dB power gains were achieved. Live-streamed $1080$p videos can be smoothly played under a $32$ megabit-per-second transmission rate over a distance of $500$m.
In addition, the authors of \cite{Fara2022Prototype,Mustafa_ArbitraryBeam_6G2022} integrated a $14 \times 14$-element reflecting platform into an ambient backscattering communication system, in which they demonstrated a significant improvement on system performance.
Moreover, the authors of \cite{Araghi2022Reconfigurable} used a $2430$-element reflecting platform to form a real-world test-bed. The experiments were performed across two rooms with LoS channel between transmitter and receiver, where the experimental results showed that at least $15$dB signal power enhancement was observed. 
Employing a $640$-element omni-{HMIMO surface} that is capable of signal reflection and refraction, the authors of \cite{Zhang2022Intelligent} built a communication system prototype based on host computer and USRPs. The experimental results validated the full-dimensional communication capability of generating effective beams toward both sides of the aperture. 
From a more macroscopic perspective, an end-to-end interplay between user devices, {HMIMO surfaces}, and the programmable-wireless-environment's control system was investigated in \cite{Liaskos2022Software}.  


In the end, we select representative studies and provide a summary table with respect to all mentioned physical aspects of {HMIMO surfaces} for ease of reference. We advise readers to refer to Table \ref{tab:physical_aspects} for details of each study.

\section{Theoretical Foundations of HMIMO Communications}
\label{SectionIV}
The enormous physical advances of {HMIMO surfaces} increase the possibility of turning HMIMO communications into reality. 
However, the corresponding theoretical foundations of HMIMO communications are under development and many aspects remain to be unveiled. In particular, the channel model of HMIMO systems faces an underlying shift due to the dense packing of nearly infinite small antenna elements and the large area deployments of {HMIMO surfaces}. The DoF (or number of communication modes) of HMIMO systems and the corresponding system capacity are fundamentally transformed. One of the most distinctive transformations is the shifting from the conventional digital domain to the EM-domain. The new domain opens new possibilities for wireless communications and also introduces new issues deserved to be extensively studied, such as EM wave sampling and EM information theory. They will not only unveil the fundamental limits of HMIMO communications and facilitate the ultimate performance analysis on HMIMO systems, but also lay the foundation to develop critical enabling technologies. It is quite necessary to present recent advances on these areas. To this aim, we summarize in this section the existing theoretical studies on HMIMO communications with a particular emphasis on channel modeling, DoF, system capacity, EM wave sampling and EM information theory. We analyze each of the contents subsequently.

\begin{figure}[t!]
	\centering
	\includegraphics[height=3.0cm, width=8.6cm]{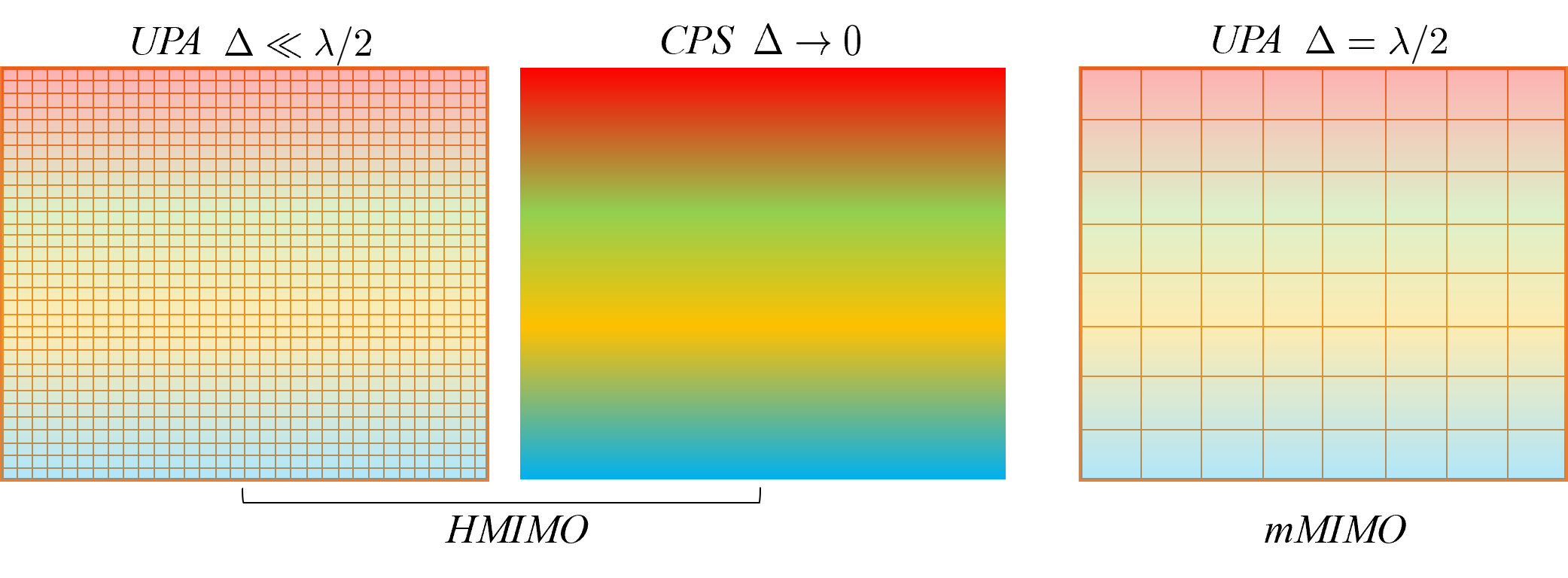}
	\caption{The UPA and CPS representation for {HMIMO surfaces}, with a comparison with mMIMO antenna arrays.}
	\label{fig:UPA_CPS}
\end{figure}

Before we formally present these contents, it is necessary to summarize a generalized representation of {HMIMO surfaces} from the previously delivered physical counterparts, such that the theoretical analysis of HMIMO can be facilitated. 
According to the physical aspects of {HMIMO surfaces} presented previously, they can be mathematically modeled as uniform planar array (UPA) apertures with small antenna element spacing, or spatially continuous planar surface (CPS) apertures. The former currently prevails and follows a mMIMO antenna array fashion, while the latter considers the ultimate limit of a densely packed aperture as the antenna element spacing tends to zero, allowing the modeling of HMIMO from a continuous EM perspective. It is note worthy that the UPA form of HMIMO is different from that of mMIMO with respect to the antenna element spacing. UPA for HMIMO has an antenna element spacing (denoted by $\Delta$) much smaller than half a wavelength (denoted by $\lambda$), i.e., $\Delta \ll \lambda/2$. Differently, UPA for mMIMO generally meet the half a wavelength requirement, namely, $\Delta = \lambda/2$. 
Without loss of generality, we take the rectangular shape of apertures as our indicative example, and demonstrate the UPA and CPS for HMIMO, as well as the differences with mMIMO in Fig. \ref{fig:UPA_CPS}. 
In addition, we consider the uniform linear array (ULA) as a 1D special case of UPA. 
Particularly, we emphasize that the ULA representation for HMIMO is different from the conventional ULA based mMIMO in terms of the antenna element spacing, namely, $\Delta \ll \lambda/2$ for HMIMO and $\Delta = \lambda/2$ for mMIMO.

\subsection{HMIMO Channel Modeling}

\begin{figure}[t!]
	\centering
	\includegraphics[height=6.8cm, width=8.6cm]{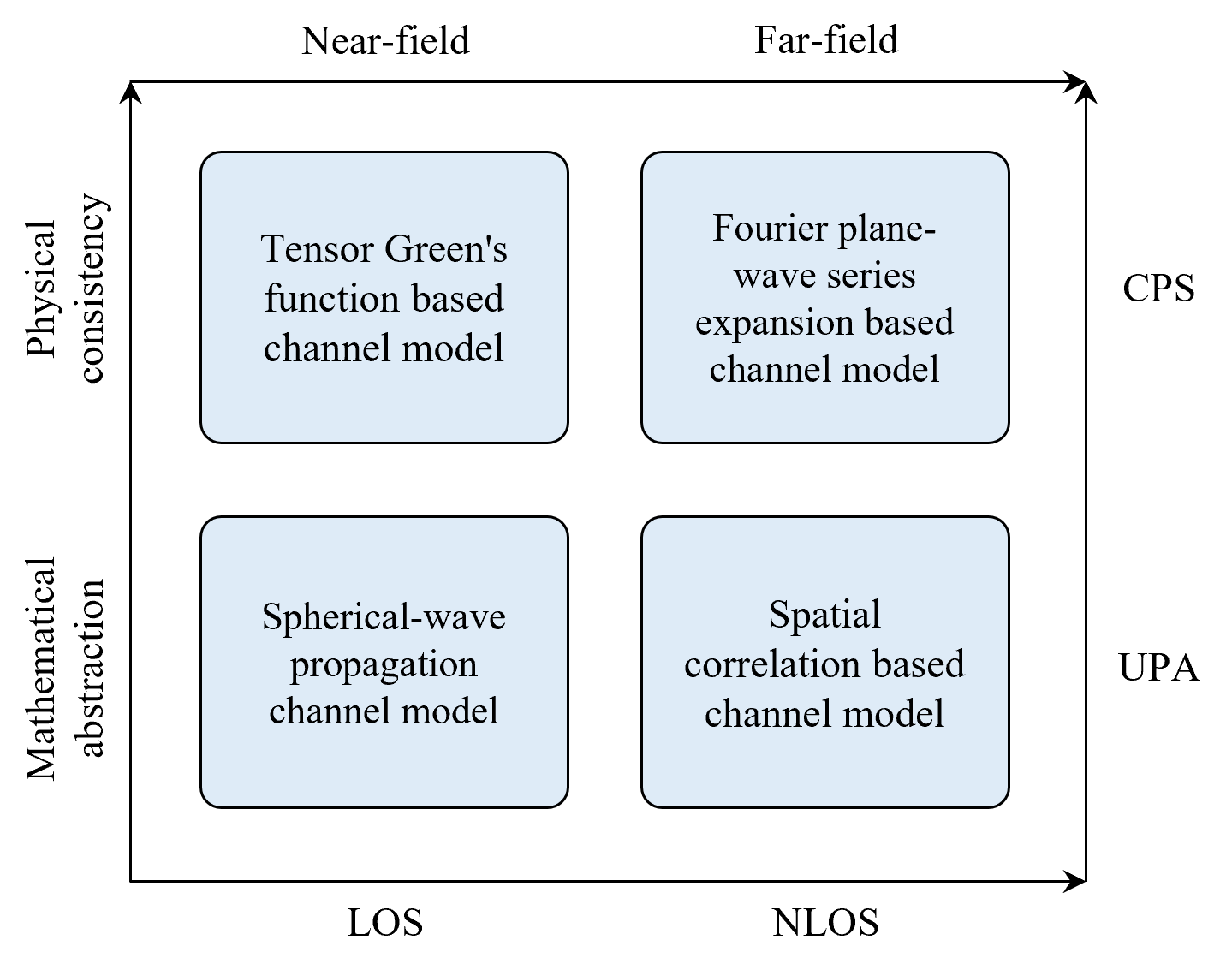}
	\caption{A panoramic demonstration of the four different channel modeling technical routes.}
	\label{fig:ChannelModelSummary}
\end{figure}

Channel modeling for HMIMO mainly faces two fundamental changes, which are brought by the dense feature and the large characteristic of {HMIMO surfaces}. Particularly, with densely distributed antenna elements, the channel modeling of HMIMO must address the strong spatial correlations and mutual coupling of nearly continuous apertures. Moreover, as the aperture size becomes large, the communication distances fall within the Fresnel zone, thereby leading to the near-field regime of communications. As such, the conventional far-field plane-wave assumption is no longer valid. 
Conventional channel models, such as Rayleigh fading \cite{Tse2005Fundamentals}, correlated Rayleigh fading \cite{Bjornson2018Massive}, and cluster-based geometric model \cite{Heath2016Overview}, built for far-field mMIMO communications, are not applicable to depict HMIMO channels. 
Therefore, it is quite necessary and critical for HMIMO communications to capture the essence of realistic physical channels by building efficient and easy-to-handle channel models. 
Since HMIMO channel modeling is an emerging research topic, it is still under development with a certain number of different technical routes. 
In the following, we embrace these representative channel modeling approaches of HMIMO, including the spherical-wave propagation channel model, the tensor Green's function based channel model, the spatial correlation channel model and the Fourier plane-wave series expansion based channel model, with respect to scenarios under LoS and NLoS environments. 
We emphasize that each technical route addresses only a few (not all) of the near-field, far-field, LoS and NLoS. They also correspond to different aperture representations (UPA and CPS) and follow different modeling ideas (mathematical abstraction and physical consistency). We summarize all these aspects in Fig. \ref{fig:ChannelModelSummary} \footnote{The spatial correlation based channel model is mostly applied to far-field scenarios, while it is noted that few studies exist for near-field scenarios by combining with the spherical-wave propagation channel model. Furthermore, the tensor Green's function based channel model is an inherent representation for CPS based HMIMO, but it can be used for modeling the UPA based HMIMO as well. Similarly, by taking the limit of the spherical-wave propagation channel model, it is possible to use it for CPS based HMIMO. 
This figure is a summary of existing works. It is still possible to break the boundaries of each technical route in future studies.}.

\subsubsection{LoS Channel Modeling} 
\begin{figure*}[t!]
	\centering
        \includegraphics[height=4cm, width=12.6cm]{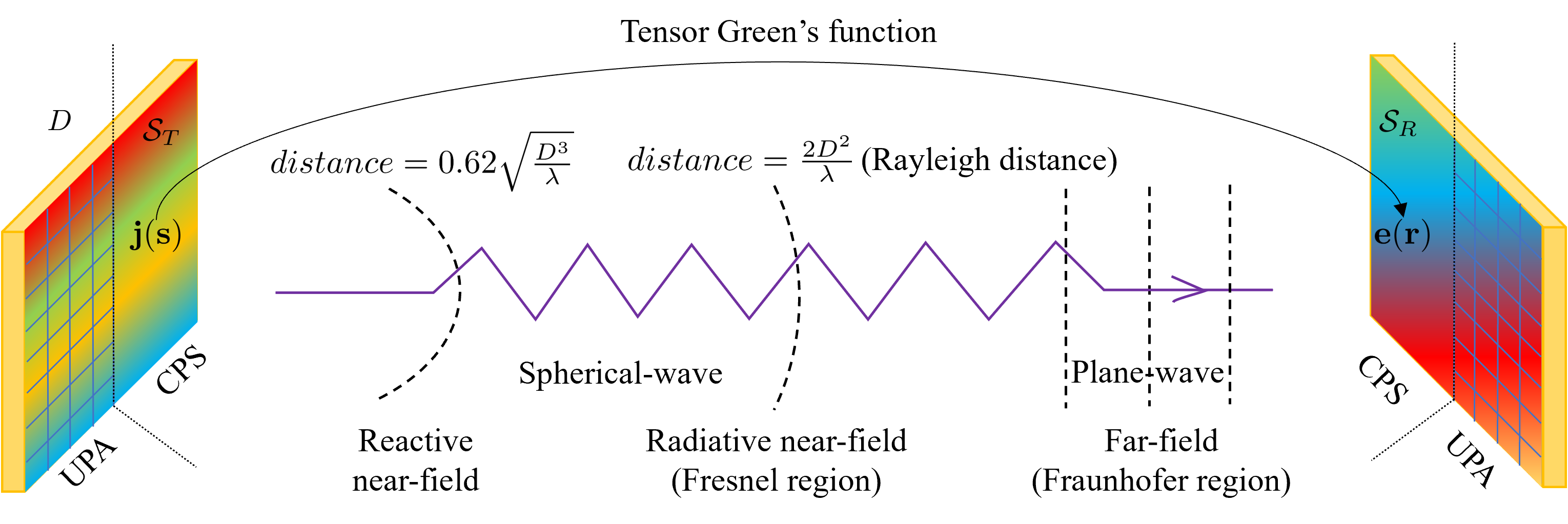}
	\caption{A completed demonstration of LoS channel between two {HMIMO surfaces}.}
	\label{fig:LOSChannelModel}
\end{figure*}

Future HMIMO communications are expected to operate in an LoS dominated environment. This is mainly due to twofold, the first being that higher frequencies are mostly adopted, migrating from microwave and mmWave to THz or even optical frequencies, and the other important factor is induced by the extremely large antenna aperture of {HMIMO surfaces}, shifting from far-field NLoS communications to near-field LoS dominated communications. It is, therefore, critical to model the LoS channel for HMIMO systems, especially in near-field regions. 
We next emphasize two typical paradigms for near-field LoS channel modeling by depicting the spherical-wave propagation effects and by the tensor Green's function, respectively. 
We present a completed demonstration of LoS communications between two {HMIMO surfaces} in Fig. \ref{fig:LOSChannelModel}, encompassing the wave propagation in an LoS channel, different communication regions and their distinguished distances, as well as the UPA and CPS based apertures. The UPA based aperture corresponds to the spherical-wave propagation channel model, and the CPS based aperture is in line with the tensor Green's function based channel, which will be detailed as follows. 

\emph{\underline{Spherical-Wave Propagation Channel Model}}: This modeling paradigm describes the near-field LoS channel based on a mathematical abstraction of the actual wave propagation effects. For an HMIMO system, the distance between each transmit and each receive element is different, such that the path amplitude and phase variations corresponding to the distance are distinct to each antenna element. This wave propagation phenomenon can be mathematically expressed as
\begin{align}
    \label{eq:SWPM}
    h_{mn} = A \left( d_{mn} \right) e^{- i \frac{2 \pi}{\lambda} d_{mn}} 
\end{align}
for each $mn$-pair, where $A \left( d_{mn} \right)$ indicates the path amplitude, depends on the distance $d_{mn}$, which is fundamentally disparate from the conventional plane-wave propagation model for far-field channel models, whose path amplitudes are insensitive to specific antenna elements. Specifically, $A \left( d_{mn} \right)$ is inversely proportional to the distance, and can be found following \cite{Ellingson2021Path}. In addition, an in-depth EM-domain analysis is performed in \cite{Dovelos2021Electromagnetic} for THz near-field channel modeling. 
In the following, we list studies on this spherical-wave propagation model.

As a simple starting point, the near-field spherical wavefront propagation model and the performance analysis for ULA based XL-MIMO systems were presented in \cite{Lu2021How,Li2022Near}. It is emphasized that these works focus more on the large aspect of an ULA antenna array while assuming that the antenna element spacing keeps the half a wavelength configuration. 
This channel model was later extended to the UPA based XL-MIMO in \cite{Lu2022Communicating} by further considering the projected aperture effect to the channel model. The authors applied this channel model for deriving a closed-form receive SNR with maximum-ratio combining (MRC) beamforming, as well as a new distinguishing criterion between the near- and far-field regimes leading to a new limiting distance as a complement of the classic Rayleigh distance, where their results generalized existing studies and revealed useful insights. 

Furthermore, in \cite{Wang2022HIHO} a fundamental spherical-wave propagation channel model was developed for a point-to-point HMIMO system, considering a practical scenario that the propagation direction is not perpendicular to the transmit and receive {HMIMO surfaces}. 
This channel model coincides with that proposed in \cite{Lu2022Communicating} when the point-to-point HMIMO system reduces to a single-input, multipe-output setup. What is more, this established channel model is consistent with that of classic mMIMO systems in LoS channels, paving the way for a convenient and fair comparison between HMIMO and mMIMO systems. The developed channel model was later employed to unveil the power gain and spectral efficiency through an EM-domain system modeling and analysis.

\emph{\underline{Tensor Green's Function Based Channel Model}}: 
This modeling paradigm is established on a more fundamental wave propagation principle from electromagnetism, where transmissions are viewed from a physically-consistent EM perspective. In this regard, a transmit current distribution is generated for triggering a receive electric field after passing through the wireless channel. In the free-space transmission condition, the transmission relation, satisfying the wave equation from Maxwell equations, is expressed as \cite{Chew1999Waves}
\begin{align}
    \label{eq:EF-GF-CD}
    {\bm{e}} \left( {{{\bm{r}}}} \right) = \int_{V} {{\bm{G}} \left( {{{\bm{r}}},{{\bm{s}}}} \right)} {\bm{j}} \left( {{{\bm{s}}}} \right) {\rm{d}}{{\bm{s}}},
\end{align}
where the integral region $V$ can be a volume or a surface; ${\bm{e}} \left( {{{\bm{r}}}} \right)$ and ${\bm{j}} \left( {{{\bm{s}}}} \right)$ are electric field and current distributions, respectively, relevant to their locations $\bm{r}$ and $\bm{s}$; ${{\bm{G}}\left( {{{\bm{r}}},{{\bm{s}}}} \right)}$ denotes the tensor (or dyadic) Green's function, given by \cite{Poon2005Degrees}
\begin{align}
	\nonumber
	&{\bm{G}}\left( {{{\bm{r}}},{{\bm{s}}}} \right)
	= \frac{i \eta}{{2 \lambda d}} \left[ \left( {1 + \frac{i}{{2 \pi d / \lambda}} - \frac{1}{{(2 \pi d / \lambda)^2}}} \right) {{\bm{I}}_3} \right. \\
	\label{eq:GreenFunc}
	& \qquad \quad  \left. + \left( {\frac{3}{{(2 \pi d / \lambda)^2}} - \frac{{3i}}{{2 \pi d / \lambda}} - 1} \right) \frac{ \bm{d} \bm{d}^{T} } {{{d^2}}} \right] 
	e^{i 2 \pi d / \lambda} 
\end{align}
with $\eta$ being the free-space impedance, ${{\bm{I}}_3}$ being a $3 \times 3$ identity matrix, $\bm{d} = \bm{r} - \bm{s}$ and $d = \|\bm{d}\|_{2}$. 
The integral region is over the whole area of the transmit {HMIMO surface}. 
On this basis, the LoS channel, connecting the transmit current distribution and the receive electric field, can be depicted using the tensor Green's function. Relevant studies are listed in the following.

The authors of \cite{Wei2022Tri} investigated the channel modeling of a near-field HMIMO system, considering parallel placements of {HMIMO surfaces}, therein they emphasized in the triple-polarization representation of the channel model and validated its feasibility. Based upon their proposed channel model, the authors suggested two precoding schemes for mitigating the cross-polarization and inter-use interference. Later, up-to-date studies \cite{Gong2023Generalized, Gong2023HMIMO} were carried out considering a more general and practical scenario, where {HMIMO surfaces} can be deployed with arbitrary surface placements, and established the generalized EM-domain channel models via the tensor Green's function with reasonable and moderate assumptions. The authors showed the effectiveness of their proposed channel models. They further employed their model to unveil the capacity limit of a point-to-point HMIMO system. 

Compared with the spherical-wave propagation channel modeling, the tensor Green's function based EM-domain channel modeling strictly obeys the physical principle of wave propagation. In addition, the latter captures three states of polarization, capable of depicting any arbitrarily polarized wave, and it seems that inside cross polarizations can potentially enhance system performance over uni-polarization systems, especially in near-field regions.

\subsubsection{NLoS Channel Modeling} 
Although the LoS near-field scenario is the leading trend for future HMIMO communications, it is still important to study the NLoS channel modeling, possibly existing in most far-field scenarios as well as few near-field scenarios where scattering of waves occurs. In NLoS channel modeling, different technical routes exist, in which we mainly introduce two typical routes prevalent recently, and present them as follows.


\begin{figure*}[t!]
	\centering
	\includegraphics[height=4.0cm, width=18cm]{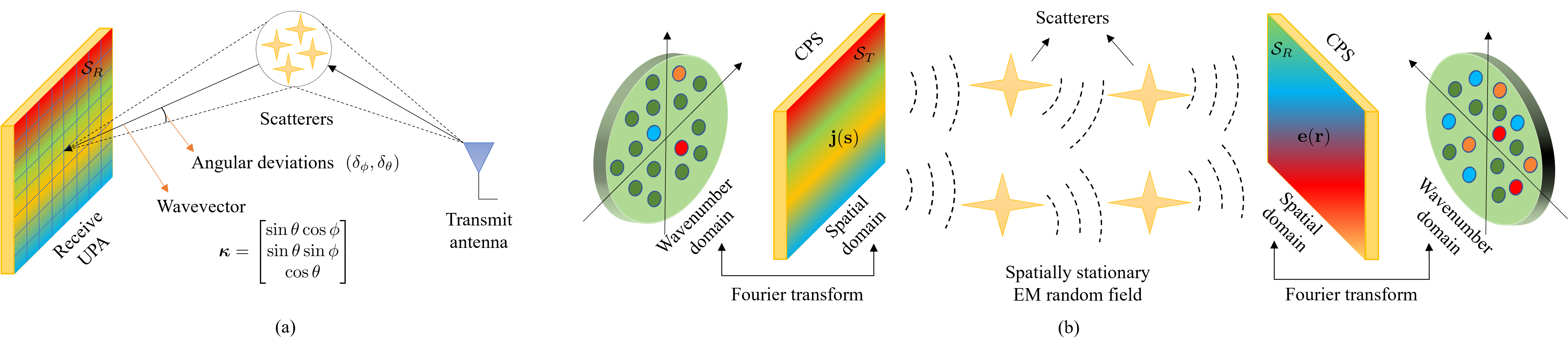}
	\caption{NLoS channel modeling schematics for HMIMO: (a) Spatial correlation based channel modeling, and (b) Fourier plane-wave series expansion based channel modeling.}
	\label{fig:NLOSChannelModel}
\end{figure*}

\emph{\underline{Spatial Correlation Based Channel Model}}: 
In this technical route, channel models are built from a spatial perspective, where channels are generally characterized by the spatial correlation matrix, which is a second-order channel statistic. 
For instance, the classic Kronecker channel model is characterized by the second-order channel statistics, and many wireless applications \cite{Gong2020RF, Gong2019Improved, Gong2019Compressive, Gong2020Compressive} are facilitated based on the second-order statistics.  
For an uplink HMIMO system, shown in Fig. \ref{fig:NLOSChannelModel}(a), the channel can be denoted by $\bm{h} \sim \mathcal{CN}(\bm{0}, \bm{R})$, where $\bm{R} = \mathbb{E}\left\{ \bm{h} \bm{h}^{H} \right\}$, detailed as \cite{Demir2022Channel}
\begin{align}
    \nonumber
    \bm{R} &= \beta \iint f(\phi + \delta_{\phi}, \theta + \delta_{\theta}) \bm{a}(\phi + \delta_{\phi}, \theta + \delta_{\theta}) \\
    &\qquad \qquad \qquad \qquad \qquad \bm{a}^{H}(\phi + \delta_{\phi}, \theta + \delta_{\theta}) 
    {\rm{d}} \delta_{\phi} {\rm{d}} \delta_{\theta},
\end{align}
with $\beta$ being the average channel gain; $f(\phi, \theta)$ being the normalized spatial scattering function, describing the angular multipath distribution and the directivity gain of antennas; $\bm{a}(\phi, \theta)$ being the array response vector; $\phi$ and $\theta$ being the azimuth and elevation angle, respectively; $\delta_{\phi}$ and $\delta_{\theta}$ being the angular deviations corresponding to $\phi$ and $\theta$, respectively, due to the distributed scatters. Generally, $\delta_{\phi}$ and $\delta_{\theta}$ are assumed to follow a certain distribution, such as Gaussian, Laplace, uniform, and von
Mises distributions \cite{Abdi2002A}, for depicting different scattering environments. 

Based on this framework, a far-field channel model was developed in \cite{Bjornson2020Rayleigh} for an RIS-aided system in an isotropic scattering environment, where the multiple paths are uniformly distributed. A closed-form expression of the correlation matrix was derived based on the ${\rm{sinc}}(\cdot)$ function.
Going one step further, considering a non-isotropic scattering environment and using directive antennas modeled after a cosine directivity pattern, \cite{Demir2022Channel} established a spatial correlation channel model for a UPA based holographic mMIMO system, using the Gaussian distribution, and derived a tight approximation.

Considering both spatial and temporal correlations, \cite{Sun2021Small} investigated a joint spatial-temporal correlation model in an isotropic scattering environment. This model is characterized by a four-dimensional (4D) sinc function, which can be simplified to the spatial-only correlation at a certain time instant that was later adopted for DoF analysis of an RIS-aided communication system. Later, the authors of \cite{Sun2022Characteristics} extended the spatial correlation analysis for a quasi-CPS\footnote{ With the term quasi-CPS we refer to apertures with inter-element spacing less than half a wavelength, but not at the asymptotical continuous limit.} aperture with a particular consideration of the mutual coupling effect.

Furthermore, the near-field spatial correlation of a ULA based XL-MIMO was examined in \cite{Dong2022Near}, where the non-uniform spherical wave characteristics are taken into consideration. This is a primary attempt to construct the near-field NLoS spatial correlation channel model for a ULA based system, paving the way for future UPA/CPS based channel modeling. Therein, the near-field spatial correlation, determined by the scattered power distribution, is relevant to both the incident angles and the distances from the scatterers. The authors derived an integral expression of spatial correlation to demonstrate the distance's influence. The results revealed a non-stationarity of the spatial correlation.

\emph{\underline{Fourier Plane-Wave Series Expansion Based Channel Model}}: 
This technical route mainly follows the principle that the channel response measured in far-field regions between two CPS apertures can be characterized as a spatially-stationary EM random field, as demonstrated in Fig. \ref{fig:NLOSChannelModel}(b). It can be statistically represented with respect to plane waves depicted by the 4D Fourier representation. Each plane wave corresponds to a propagation direction that is characterized by the wave vector, i.e., a vector encompassing three components corresponding to $x$, $y$ and $z$ orientations of the wavenumber $2 \pi / \lambda$, respectively.
Similar to the Fourier transform between time and frequency domains, the relation between spatial and wavenumber domains is enabled by the Fourier transform as well, such that the channel response $h(\bm{r},\bm{s})$ between location $\bm{s}$ and $\bm{r}$ is 
expressed as \cite{Pizzo2020Spatially,Pizzo2022Spatial,Pizzo2022Fourier}
\begin{align}
    \nonumber
    h(\bm{r},\bm{s}) &= \frac{1}{(2\pi)^2} \iiiint\limits_{\mathcal{R}(\frac{2 \pi}{\lambda}) \times \mathcal{R}(\frac{2 \pi}{\lambda})} a_r(\bm{k},\bm{r}) H_a(k_x,k_y,\kappa_x,\kappa_y) \\
    \label{4D1}
    & \times a_s(\bm{\kappa},\bm{s}) \cdot {\rm{d}} k_x {\rm{d}} k_y {\rm{d}} \kappa_x {\rm{d}} \kappa_y,
\end{align}
where $a_s(\bm{\kappa},\bm{s}) = e^{-i \bm{\kappa}^{T} \bm{s}} = e^{-i \left( \kappa_x s_x + \kappa_y s_y + \kappa_z s_z \right)}$ reveals the transmit response, realizing the transformation from the excitation current at location $\bm{s} = [s_x, s_y, s_z]^{T}$ to the transmit wave propagation direction $\bm{\kappa}/\|\bm{\kappa}\|$, with $\bm{\kappa} = [\kappa_x, \kappa_y, \kappa_z]^{T}$ being the transmit wave vector and $\kappa_z = \sqrt{(\frac{2 \pi}{\lambda})^2 - \kappa_x^2 - \kappa_y^2}$; $a_r(\bm{k},\bm{r}) = e^{i \bm{k}^{T} \bm{r}} = e^{i \left( k_x r_x + k_y r_y + k_z r_z \right)}$ denotes the receive response, mapping the propagation direction $\bm{k}/\|\bm{k}\|$ of the receive wave to the measured current at point $\bm{r} = [r_x, r_y, r_z]^{T}$ with $\bm{k} = [k_x, k_y, k_z]^{T}$ being the receive wave vector and $k_z = \sqrt{(\frac{2 \pi}{\lambda})^2 - k_x^2 - k_y^2}$; $H_a(k_x,k_y,\kappa_x,\kappa_y)$ represents the wavenumber domain channel response; The integral regions are limited to $\mathcal{R}(\frac{2 \pi}{\lambda}) = \left\{ (u_x, u_y) \in \mathbb{R}^{2}: u_x^2 + u_y^2 \leq (\frac{2 \pi}{\lambda})^2 \right\}, u \in \{ k, \kappa \}$.
 Therefore, modeling the spatial domain channel can be equivalent to an alternative modeling of the wavenumber domain channel, which is given by \cite{Pizzo2020Spatially,Pizzo2022Spatial,Pizzo2022Fourier}
\begin{equation}
\begin{aligned}
    H_a(k_x,k_y,\kappa_x,\kappa_y) = \sqrt{S(k_x,k_y,\kappa_x,\kappa_y)} W(k_x,k_y,\kappa_x,\kappa_y),
    \label{4D2}
\end{aligned}
\end{equation}
where the wavenumber domain channel can be represented by the channel spectral density $S(k_x,k_y,\kappa_x,\kappa_y)$ which is relevant to the scattering environment that describes an NLoS channel, and the antenna arrangement; $W(k_x,k_y,\kappa_x,\kappa_y)$ involves the random characteristics of the channel. 
The wavenumber domain channel generally has a sparse structure, namely, a finite number of non-zero dominating coefficients. Based on sampling theory, the finite integration area can be sampled uniformly to approximate the wavenmuber domain channel, as demonstrated in Fig. \ref{fig:NLOSChannelModel}(b). 
The channel approximation accuracy depends on the amount of sampling points. One can obtain a more accurate channel representation via generating more samples, leading to an increase in computational complexity.
Based on the approximate wavenumber domain channel, the spatial domain channel can be obtained by taking the Fourier transformation. A multiple-user extension of the channel model was performed in \cite{WeiLi2022Multi-user}. 


For ease of reference, we list representative studies of channel modeling for HMIMO communications in a summary table, as demonstrated in Table \ref{tab:ChannelModeling}.

\begin{table*}[!t]
    \tiny
    \renewcommand{\arraystretch}{1.2}
    \caption{\textsc{Channel modeling for HMIMO communications.}}
    \label{tab:ChannelModeling}
    \centering
    \resizebox{\linewidth}{!}{
        \begin{tabular}{!{\vrule width0.6pt}c|c|c|c|c|l!{\vrule width0.6pt}}
            \Xhline{0.6pt}
            \rowcolor{yellow} \textbf{\tabincell{c}{Channel\\ modeling}} & \textbf{Ref.} & \textbf{\tabincell{c}{System model}} & \textbf{\tabincell{c}{Aperture\\ type}} & \textbf{\tabincell{c}{Channel\\ type}} & \qquad \qquad \qquad \qquad \qquad \qquad \qquad \textbf{\tabincell{c}{Main contributions}}\\

            \Xhline{0.6pt}
            \multirow{4}{*}{\tabincell{c}{Spherical-wave\\ propagation\\ channel model}} & \tabincell{c}{\cite{Lu2022Communicating}} & \tabincell{c}{Multiple UEs\\ BS: XL-MIMO array\\ UE: Single-antenna} & \tabincell{c}{UPA, {CPS}} & 
            \tabincell{c}{LoS\\ (Near-field)} & \tabincell{l}{Investigate channel modeling and performance analysis of XL-MIMO communications\\ based on the generic spherical wave propagation model.} \\
            
            \cline{2-6}
            & \tabincell{c}{\cite{Wang2022HIHO}} & \tabincell{c}{Single UE\\ BS: {HMIMO surface}\\ UE: {HMIMO surface}} & \tabincell{c}{{CPS}, UPA} & 
            \tabincell{c}{LoS\\ (Near-field)} & \tabincell{l}{Develop a simple channel model for LoS holographic-input holographic-output system,\\ which is consistent with that for conventional LoS mMIMO systems.} \\

            \Xhline{0.6pt}
            \multirow{4}{*}{\tabincell{c}{Tensor Green's\\ function based\\ channel model}} & \tabincell{c}{\cite{Wei2022Tri}} & \tabincell{c}{Downlink multiple UEs\\ BS: {HMIMO surface}\\ UE: {HMIMO surface}} & \tabincell{c}{{CPS}, UPA} & 
            \tabincell{c}{LoS\\ (Near-field)} & \tabincell{l}{Establish a near-field LoS channel model based on the tensor Green's function\\ considering an utilization of triple polarization, and propose two precoding schemes.} \\
            
            \cline{2-6}
            & \tabincell{c}{\cite{Gong2023Generalized}\\ \cite{Gong2023HMIMO}} & \tabincell{c}{Single UE\\ BS: {HMIMO surface}\\ UE: {HMIMO surface}} & \tabincell{c}{{CPS}, UPA} & 
            \tabincell{c}{LoS\\ (Near-field)} & \tabincell{l}{Prvide channel modeling and capacity limit of a near-field LoS channel for HMIMO\\ communications with arbitrary surface placement based on the tensor Green's function.} \\

            \Xhline{0.6pt}
            \multirow{6}{*}{\tabincell{c}{Spatial corre-\\lation based\\ channel model}} & \cite{Demir2022Channel} & \tabincell{c}{Single UE\\ BS: {HMIMO surface}\\ UE: Single-antenna} & \tabincell{c}{UPA} & 
            \tabincell{c}{NLoS\\ (Far-field)} & \tabincell{l}{Provide a channel model for holographic mMIMO with the consideration of non-\\isotropic scattering and directive antennas} \\
            
            \cline{2-6}
            & \tabincell{c}{\cite{Sun2021Small}\\\cite{Sun2022Characteristics}} & \tabincell{c}{Single UE\\ BS: /, UE: RIS} & \tabincell{c}{UPA,\\ quasi-{CPS}} & 
            \tabincell{c}{NLoS\\ (Near-/far-field)} & \tabincell{l}{Derive a 4D sinc function depicted joint small-scale spatial-temporal correlation model\\ in isotropic scattering and extend it to an extra consideration of mutual coupling effect.} \\

            \cline{2-6}
            & \cite{Dong2022Near} & \tabincell{c}{Single UE\\ BS: Single-antenna\\ UE: XL-MIMO array} & \tabincell{c}{ULA} & 
            \tabincell{c}{NLoS\\ (Near-field)} & \tabincell{l}{Present an analysis on near-field spatial correlation with an emphasis on one-ring\\ scatter distribution, including the far-field spatial correlation as a special case.} \\

            \Xhline{0.6pt}
            \multirow{7}{*}{\tabincell{c}{Fourier plane-\\wave series\\ expansion based\\ channel model}} & \tabincell{c}{\cite{Marzetta2018Spatially}\\ \cite{Pizzo2020Spatially}} & \tabincell{c}{Single UE\\ BS: {HMIMO surface}\\ UE: {HMIMO surface}} & \tabincell{c}{{CPS}, UPA} & \tabincell{c}{NLoS\\ (Far-field)} & \tabincell{l}{Describe the 3D small-scale fading as a Fourier plane-wave spectral representation and\\ present a discrete representation for EM wave via Fourier plane-wave series expansion.} \\
            
            \cline{2-6} 
            & \tabincell{c}{\cite{Pizzo2020Holographic}\\ \cite{Pizzo2022Fourier}} & \tabincell{c}{Single UE\\ BS: {HMIMO surface}\\ UE: {HMIMO surface}} & \tabincell{c}{{CPS}, UPA} & 
            \tabincell{c}{NLoS\\ (Far-field)} & \tabincell{l}{Present a 4D plane-wave representation of channel response in arbitrary scattering and\\ provide a low-rank semi-unitarily equivalent approximation of the spatial EM channel.} \\
            
            \cline{2-6}
            & \cite{WeiLi2022Multi-user} & \tabincell{c}{Downlink multiple UEs\\ BS: {HMIMO surface}\\ UEs: {HMIMO surface}} & \tabincell{c}{{CPS}, UPA} & 
            \tabincell{c}{NLoS\\ (Far-field)} & \tabincell{l}{Extend the 4D plane-wave representation of channel response in arbitrary scattering to\\ the multi-user scenario, where each UE is equipped with an {HMIMO surface}.} \\

            \Xhline{0.6pt}
        \end{tabular}}
\end{table*}

\subsection{HMIMO Performance Analysis}
\subsubsection{DoF}
The DoF represents the number of communication modes of EM waves which reveals the number of independent data streams can be transmitted simultaneously by the wireless propagation media. It is capable of indicating the optimal communication of an HMIMO system and should be determined to fully understand the limitations. Since the mathematical modeling of HMIMO transceivers shifts from the conventional discrete antenna array to the continuous EM surface, and the far-field region tends to be near-field, the DoF of HMIMO systems varies accordingly which needs to be examined. In the following contents, we list up-to-date research for DoF of HMIMO communications, show their considered scenarios, and present their main results.


Pioneer studies were conducted in \cite{Hu2017LIS,Hu2018LIS} for investigating the DoF of an LIS-based HMIMO system, where an LIS communicates with multiple single-antenna UEs in near-field LoS channels. The authors applied the spherical-wave propagation channel model in their analysis. Accordingly, they derived that the DoF of such a system tends to be $2/\lambda$ per meter for UEs with a 1D line layout, and $\pi/\lambda^2$ per square meter for UEs designed in a 2D plane and a 3D volume layouts, where $\lambda$ represents the wavelength.

Later, the studies \cite{Dardari2020Communicating1,Dardari2020Communicating2} explored the DoF of a point-to-point LIS-based HMIMO system, where an LIS communicates with a small RIS in arbitrary configurations, with a particular emphasis on the near-field communication. They contributed the optimal communication between the LIS and the small RIS as an EM-level eigenfunction problem which is conventionally solved by time consuming and less instructive numerical computations, and then presented an approximate but accurate closed-form expression of the DoF that facilitates the computations and unveils underlying optimal communications. The obtained results indicated that the DoF of LoS channel is determined only by geometric factors normalized to the wavelength, and can be higher than $1$. 
Exploiting the DoF for achieving optimal communications, references \cite{Decarli2021Beamspace,Decarli2021Communication} proposed a beamspace modeling, describing generic orientations and near-field operating conditions, for defining the communication modes and determining its DoF, as well as deriving the closed-form basis set between transceivers that guarantees a near-optimal communication without complicated weights' configurations for precoding/combing and exhaustive numerical computations.

Conversely, the authors of \cite{Pizzo2020Degrees} put an emphasis on a far-field isotropic scattering environment and analyzed the spatial DoF of a point-to-point HMIMO system. They examined the DoF for 1D linear, 2D planar and 3D volumetric apertures, respectively, based on the newly derived 4D Fourier plane-wave series expansion of the channel response. The results provided a guideline of discrete antenna elements spacing to achieve the DoF, which is $2 L_x / \lambda$ for 1D linear apertures, $\pi L_x L_y / \lambda^2$ for 2D planar apertures, and $2 \pi L_x L_y / \lambda^2$ for 3D volumetric apertures, where $L_x$ and $L_y$ are lengths of the surface. The work was then extended to a more general study on the relation between DoF and Nyquist sampling under arbitrary propagation conditions in \cite{Pizzo2022Nyquist}. This was performed by modeling the EM wave propagation to a corresponding linear system, for which multidimensional sampling theorem and Fourier theory are applied for analysis. The study showed that the DoF per unit of area is the Nyquist samples per square meter for large antenna surfaces.
With further consideration in the presence of evanescent waves, reference \cite{RanJi2022Extra} studied the spatial DoF for near-field HMIMO communications using the Fourier plane-wave series expansion. It mainly focused on the isotropic scattering environment while is capable of being extended to the non-isotropic case. The study revealed that the evanescent waves can be further exploited to add extra spatial DoF and increase the system capacity.

According to \cite{Sun2021Small}, the authors explored the DoF through a specially developed joint spatial-temporal correlation model for isotropic scattering environment. Therein, they noticed that the spatial DoF decreases with an increase of the number of antennas of HMIMO transceivers, which seems counter-intuitive. An additional explanation of this anomalous phenomenon was presented in \cite{Sun2022Characteristics} via utilizing the power spectrum of the spatial correlation function. With a particular emphasis on the existence of mutual coupling, this study presented an analysis on the effective spatial correlation as well as the eigenvalue structures of the spatial correlation matrix in terms of different antenna element intervals, by leveraging a specially designed metric that indicates the inter-element coupling strength. The corresponding results revealed the connection between eigenvalues and evanescent waves which are potentially beneficial for near-field communications.

\subsubsection{Capacity}

For emerging HMIMO communication systems, the intrinsic system capacity is critical to investigate their potential. Since HMIMO exhibits strong mutual coupling effect due to numerous of antenna elements and behaves differently in near-field and far-field regions, therefore, the fundamental limits have yet to be uncovered and novel performance evaluation methods should be developed. In this regard, we overview the contemporary studies on capacities of HMIMO communications for demonstrating the recent progresses of this area. To be specific, we classify the advances based on different channel types, namely, capacity of LoS channel and capacity of channel with presence of NLoS. The details are expanded as follows.

\emph{\underline{Capacity of LoS Channel}}: 
In this category, the LoS channel between HMIMO transceivers does not include any scattering that affects the wave propagation. 
Under this condition, references \cite{Hu2017LIS,Hu2018LIS} first presented pioneer studies on the system capacity of HMIMO communications implemented by LIS. Therein, the studies focused on the uplink system capacity for multiple single-antenna UEs experiencing LoS channels, where the capacities for UEs on line/plane shapes or in a cube shape were derived. The reference provided an asymptotic limit on the normalized capacity as the terminal density increases and the wavelength approaches zero, which was derived as $\frac{P}{2N_0}$ where $P$ denotes the transmit power per volume unit and $N_0$ is the noise spatial power spectral density. 


The authors of \cite{Williams2020Communication} studied the receive power for LIS-based-enabled HMIMO downlink transmission communicating with single UE in an LoS environment. The investigation was proceeded by presenting a new mathematical communication model that captures system the impedance for both isotropic and planar antenna elements. The work was later extended to multiple single-antenna UEs in \cite{Williams2021Multiuser}, where radiated and received powers were characterized by expressions derived using a circuital description of the LIS-based HMIMO system accounting for super-directivity and mutual coupling. With specially designed matched filter (MF) and weighted MMSE transmitters, the authors verified the variation of sum-rate in terms of antenna element spacing, ohmic losses, and mutual coupling among UEs.

In \cite{Jesus2020Near}, the authors assessed the uplink spectral efficiency of a large trasmitting metasurface communicating with two single-antenna UEs, with an emphasis on the near-field communication. They validated the uplink spectral efficiency using MRC and minimum mean squared error (MMSE) combining schemes, and showed that MMSE combining is superior to MR combining in achieving high spectral efficiency by mitigating the interference. The results indicated that channel estimation errors degrade MMSE fast. The study also demonstrated the impact of polarization mismatch on the spectral efficiency.


With a particular focus on the XL-MIMO system, the authors of \cite{Lu2022Communicating} studied the receive SNR of both UPA and CPS BS enabled multi-user system under a LoS near-field environment depicted based on the spherical-wave propagation model. They derived a close-form expression of the receive SNR for MRC or maximum ratio transmission (MRT) beamforming, and the results showcased that the receive SNR increases with the number of transmit antennas with diminishing return instead of the conventional linear scaling law. The authors also interpreted the receive SNR with respect to vertical and horizontal angular spans. Their results encompass the conventional uniform plane wave assumption as a special case, and facilitate a new uniform-power distance to the classical Rayleigh distance counterpart.

In another recent work \cite{Wang2022HIHO} the system capacity of an LoS point-to-point HMIMO system was investigated from an EM-domain system modeling, where the spherical-wave propagation channel model is employed. The authors presented comparison results of HMIMO over conventional MIMO, indicating that the considered HMIMO system is capable of achieving up to $\pi^{2}$ times higher power gain then the conventional LoS MIMO system with a same surface area. This extra power induced by HMIMO systems can interpreted to up to $3.30$ bits/s/Hz spectral efficiency gain.

From a more fundamental EM perspective, where a tensor Green's function based channel model was applied, the authors of \cite{Gong2023HMIMO} analyzed the near-field capacity limit of point-to-point HMIMO systems in an LoS propagation environment. A tight upper bound was derived based upon an EM-domain analysis framework, where it revealed that the capacity limit grows logarithmically with the product of transmit element area, receive element area, and the combined effects of $1/{{d}_{mn}^2}$, $1/{{d}_{mn}^4}$, and $1/{{d}_{mn}^6}$ over all transmit and receive antenna elements, where $d_{mn}$ indicates the distance between each transmit and receive element.

\begin{table*}[!t]
    \tiny
    \renewcommand{\arraystretch}{1.2}
    \caption{\textsc{Performance analysis of HMIMO communications.}}
    \label{tab:PerformanceAnalysis}
    \centering
    \resizebox{\linewidth}{!}{
        \begin{tabular}{!{\vrule width0.6pt}c|c|c|c|c|l!{\vrule width0.6pt}}
            \Xhline{0.6pt}
            \rowcolor{yellow} \textbf{\tabincell{c}{Performance\\analysis}} & \textbf{Ref.} & \textbf{\tabincell{c}{System model}} & \textbf{\tabincell{c}{Aperture type}} & \textbf{\tabincell{c}{Channel type}} & \qquad \qquad \qquad \qquad \qquad \qquad \textbf{\tabincell{c}{Main contributions}}\\
            \Xhline{0.6pt}
            \multirow{15}{*}{\tabincell{c}{DoF}} & \tabincell{c}{\cite{Hu2017LIS}\\ \cite{Hu2018LIS}} & \tabincell{c}{Uplink multiple UEs\\ BS: LIS, UE: Single-antenna} & \tabincell{c}{{CPS}} & 
            \tabincell{c}{LoS\\ (Near-/far-field)} & \tabincell{l}{Derive the DoF, per area unit of deployed surface, for 1D linear, 2D planar,\\ and 3D cubic terminal deployments.} \\
            
            \cline{2-6}
            & \tabincell{c}{\cite{Dardari2020Communicating1}\\ \cite{Dardari2020Communicating2}} & \tabincell{c}{Single UE\\ BS: LIS, UE: LIS} & \tabincell{c}{{CPS}} & 
            \tabincell{c}{LoS\\ (Near-field)} & \tabincell{l}{Derive the DoF as well as show it is only determined by geometric factors and\\ is larger than $1$ in near-field region.} \\
            
            \cline{2-6}
            & \tabincell{c}{\cite{Decarli2021Beamspace}\\ \cite{Decarli2021Communication}} & \tabincell{c}{Single UE\\ BS: LIS, UE: Small RIS} & \tabincell{c}{{CPS}, ULA} & 
            \tabincell{c}{LoS\\ (Near-field)} & \tabincell{l}{Propose a beamspace modeling that defines the communication modes and\\ determines the DoF.} \\
            
            \cline{2-6}
            & \tabincell{c}{\cite{Pizzo2020Degrees}} & \tabincell{c}{Single UE\\ BS: {HMIMO surface}\\ UE: {HMIMO surface}} & \tabincell{c}{{CPS}, ULA,\\ UPA, volume} & 
            \tabincell{c}{NLoS\\ (Far-field)} & \tabincell{l}{Investigate the DoF of HMIMO systems based on the 4D Fourier plane-wave\\ series expansion of the HMIMO channel.} \\
            
            \cline{2-6}
            & \tabincell{c}{\cite{Pizzo2022Nyquist}} & \tabincell{c}{Single UE\\ BS: {HMIMO surface}\\ UE: {HMIMO surface}} & \tabincell{c}{{CPS}, UPA} & 
            \tabincell{c}{/} & \tabincell{l}{Study the relation between DoF of HMIMO system and Nyquist sampling.} \\
            
            \cline{2-6}
            & \tabincell{c}{\cite{RanJi2022Extra}} & \tabincell{c}{Single UE\\ BS: {HMIMO surface}\\ UE: {HMIMO surface}} & \tabincell{c}{{CPS}, UPA} & 
            \tabincell{c}{NLoS\\ (Near-field)} & \tabincell{l}{Reveal that evanescent waves can be exploited to increase extra spatial DoF.} \\
            
            \cline{2-6}
            & \tabincell{c}{\cite{Sun2021Small}\\\cite{Sun2022Characteristics}} & \tabincell{c}{Single UE\\ BS: /, UE: RIS} & \tabincell{c}{UPA,\\ quasi-{CPS}} & 
            \tabincell{c}{NLoS\\ (Near-/far-field)} & \tabincell{l}{Study the DoF with and without the presence of mutual coupling.} \\
            
            \Xhline{0.6pt}
            \multirow{30}{*}{\tabincell{c}{Capacity}} & \tabincell{c}{\cite{Hu2017LIS}\\ \cite{Hu2018LIS}} & \tabincell{c}{Uplink multiple UEs\\ BS: LIS, UE: Single-antenna} & \tabincell{c}{{CPS}} & 
            \tabincell{c}{LoS\\ (Near-/far-field)} & \tabincell{l}{Derive capacity for 1D, 2D and 3D terminal deployments, as well as present\\ the asymptotic limit of normalized capacity.} \\
            
            \cline{2-6}
            & \tabincell{c}{\cite{Williams2020Communication}} & \tabincell{c}{Single UE\\ BS: LIS, UE: Single-antenna} & \tabincell{c}{{CPS}, UPA} & 
            \tabincell{c}{LoS\\ (Near-field)} & \tabincell{l}{Analyze the received power based on a designed mathematical communication model\\ that depicts the mutual coupling of antenna elements.} \\
            
            \cline{2-6}
            & \tabincell{c}{\cite{Williams2021Multiuser}} & \tabincell{c}{Downlink multiple UEs\\ BS: LIS, UE: Single-antenna} & \tabincell{c}{{CPS}, UPA} & 
            \tabincell{c}{LoS\\ (Near-field)} & \tabincell{l}{Examine received power by a circuital description of the system, and evaluate sum-\\rate with respect to antenna spacing, ohmic losses, and inter-UE mutual coupling.} \\
            
            \cline{2-6}
            & \tabincell{c}{\cite{Jesus2020Near}} & \tabincell{c}{Uplink two UEs\\ BS: LIS, UE: Single-antenna} & \tabincell{c}{UPA} & 
            \tabincell{c}{LoS\\ (Near-/far-field)} & \tabincell{l}{Assess uplink spectral efficiency using MRC and MMSE combing schemes in\\ the near-field region.} \\
            
            \cline{2-6}
            & \tabincell{c}{\cite{Lu2022Communicating}} & \tabincell{c}{Multiple UEs\\ BS: XL-MIMO arrays\\ UE: Single-antenna} & \tabincell{c}{UPA, {CPS}} & 
            \tabincell{c}{LoS\\ (Near-field)} & \tabincell{l}{Investigate the received SNR of XL-MIMO systems based on the generic spherical-\\ wave propagation model.} \\

            \cline{2-6}
            & \tabincell{c}{\cite{Wang2022HIHO}} & \tabincell{c}{Single UE\\ BS: {HMIMO surface}\\ UE: {HMIMO surface}} & \tabincell{c}{{CPS}, UPA} & 
            \tabincell{c}{LoS\\ (Near-/far-field)} & \tabincell{l}{Compare LoS holographic-input holographic-output system with conventional\\ LoS MIMO systems in terms of capacity and power gain.} \\
            
            \cline{2-6}
            & \tabincell{c}{\cite{Gong2023HMIMO}} & \tabincell{c}{Single UE\\ BS: {HMIMO surface}\\ UE: {HMIMO surface}} & \tabincell{c}{{CPS}, UPA} & 
            \tabincell{c}{LoS\\ (Near-field)} & \tabincell{l}{Study the capacity limit of HMIMO systems based on the EM-domain tensor\\ Green's function based channel model.} \\
            
            \cline{2-6}
            & \tabincell{c}{\cite{Alegria2019Achievable}} & \tabincell{c}{Single UE\\ BS: LIS, UE: single-antenna} & \tabincell{c}{{CPS}, UPA} & 
            \tabincell{c}{LoS\\ (Near-/far-field)} & \tabincell{l}{Analyze the effect of hardware impairments on the achievable rate using a\\ simplified receiver structure.} \\
            
            \cline{2-6}
            & \tabincell{c}{\cite{Hu2022Holographic}} & \tabincell{c}{Downlink multiple UEs\\ BS: {HMIMO surface}, UE: Single-antenna} & \tabincell{c}{UPA} & 
            \tabincell{c}{LoS\\ (Far-field)} & \tabincell{l}{Study the effect of quantization on the sum-rate and present a lower bound in\\ terms of quantization.} \\
            
            \cline{2-6} 
            & \tabincell{c}{\cite{Pizzo2020Holographic}\\ \cite{Pizzo2022Fourier}} & \tabincell{c}{Single UE\\ BS: {HMIMO surface}\\ UE: {HMIMO surface}} & \tabincell{c}{{CPS}, UPA} & 
            \tabincell{c}{NLoS\\ (Far-field)} & \tabincell{l}{Evaluate the capacity of a point-to-point HMIMO system based on the built\\ 4D Fourier plane wave representation of the channel model.} \\
            \cline{2-6}
            & \cite{WeiLi2022Multi-user} & \tabincell{c}{Downlink multiple UEs\\ BS: {HMIMO surface}\\ UE: {HMIMO surface}} & \tabincell{c}{{CPS}, UPA} & 
            \tabincell{c}{NLoS\\ (Far-field)} & \tabincell{l}{Study system capacity using MRT and ZF procoding schemes based on the 4D\\ Fourier plane wave representation of HMIMO channel for multiple UEs.} \\
            
            \cline{2-6}
            & \tabincell{c}{\cite{Shlezinger2019Dynamic}\\\cite{Wang2019Dynamic}} & \tabincell{c}{Uplink/Downlink multiple UEs\\ BS: DMA, UE: Single-antenna} & \tabincell{c}{UPA} & 
            \tabincell{c}{NLoS} & \tabincell{l}{Study uplink and downlink capacities of DMA-based HMIMO system based on\\ a specially designed mathematical model of DMA.} \\
            
            \cline{2-6}
            & \tabincell{c}{\cite{Jung2020Performance}} & \tabincell{c}{Uplink multiple UEs\\ BS: LIS, UE: Single-antenna} & \tabincell{c}{UPA} & 
            \tabincell{c}{NLoS\\ (Near-/far-field)} & \tabincell{l}{Asymptotically analyze uplink data rate under channel estimation errors,\\ interference channels, and hardware impairments.} \\
            
            \cline{2-6}
            & \tabincell{c}{\cite{Jung2021Performance}} & \tabincell{c}{Uplink multicell multiple UEs\\ BS: LIS, UE: Single-antenna} & \tabincell{c}{UPA} & 
            \tabincell{c}{NLoS\\ (Near-/far-field)} & \tabincell{l}{Evaluate the spectral efficiency of a multi-LIS multicell system in the presence of\\ pilot contamination, and derive an asymptotic bound.} \\

            \Xhline{0.6pt}
        \end{tabular}}
\end{table*}

In addition to previous theoretical analyses on system capacity, \cite{Alegria2019Achievable} studied the achievable rate of a receiving LIS system with the presence of correlated hardware impairments. The correlation was modeled by means of distance between considered points on the LIS, based on which the closed-form expression of the achievable rate was derived. Moreover, a research on the impact of quantization of amplitude controlled HMIMO surfaces on the sum-rate of a downlink HMIMO surface-assisted multi-user system was conducted in \cite{Hu2022Holographic}. 
They presented a lower bound of the sum-rate in terms of quantization, and unveiled the required minimum quantized bits accordingly.

\emph{\underline{Capacity with presence of NLoS Channel}}: 
In this category, the channel between HMIMO transceivers includes scatterers that influence the wave propagation. 
To present an exact depiction from an EM perspective, references \cite{Pizzo2020Holographic,Pizzo2022Fourier} investigated the system capacity of a point-to-point HMIMO system based on the 4D Fourier plane wave representation of HMIMO channels with arbitrary spatially-stationary scattering. Particularly, instead of using the conventional spatial domain channel model, the wavenumber domain channel was established to capture the essence of the physical channel and used to evaluate the system capacity for rectangular volumetric arrays. 
On this basis, the authors of \cite{WeiLi2022Multi-user} extended the Fourier plan wave representation of HMIMO channels to the scenario including multiple UEs with each equipped by an {HMIMO surface}, based on which they investigated the system capacity using the MRT
and zero-forcing (ZF) precoding schemes, respectively. 
The study revealed that large spectral efficiency can be achieved by packing more antenna elements on HMIMO transceivers. Moreover, as spaces among antennas are reduced, strong mutual coupling deteriorates the spectral efficiency under a fixed number of antenna elements.

By identifying the multi-path channel model using a linear filter characterized by multiple channel taps, the authors of \cite{Shlezinger2019Dynamic} and \cite{Wang2019Dynamic} studied the uplink and the downlink system capacities of an HMIMO system, respectively, in which a dynamic metasurface (DMA) implemented BS communicates with multiple single-antenna UEs. The authors derived the uplink/downlink capacities based on a specifically designed mathematical model of the DMA. They unveiled the fundamental limits under different weight configurations of DMA.


Moreover, the authors of \cite{Jung2020Performance} analyzed the uplink data rate for an LIS-based HMIMO system in an asymptotic manner. They assumed an LoS desired channel interfered by spatially correlated Rician fading channels. In particular, the study put an emphasis on a practical scenario where channel estimation errors, interference channels, and hardware impairments are considered. Under these settings, the theoretical bound was presented, which demonstrates that the noise, interference from estimation error, hardware impairments, as well as the NLoS paths can be eliminated with the increase of antennas. Later, \cite{Jung2021Performance} evaluated the spectral efficiency of a multi-LIS system in the presence of pilot contamination. The authors derived its theoretical bound asymptotically, which can be utilized for determining pilot training length and number of UEs. The study suggested that the system spectral efficiency is limited by the pilot contamination as well as both inter and intra-LIS LoS interference no matter how large the number of antennas become.

Lastly, to provide a panoramic view of performance analysis, we sum up related studies to a summary table, as demonstrated in Table \ref{tab:PerformanceAnalysis}.

\subsection{EM Wave Sampling}
Since EM waves modulated by HMIMO are continuous in space, they have to be sampled and discretized for digital processing, which is related to the sampling of spatial EM waves aiming at retaining the maximum EM information with the minimum samples.  

Preliminary explorations of EM wave sampling were carried out in \cite{Bucci1987On}\cite{Bucci1998Representation}, investigating the non-redundant representation of spatially continuous EM waves using a limited number of samples. 
A recent study on EM wave sampling for an LIS-based HMIMO system was conducted in \cite{Hu2018LIS}. For a LoS uplink LIS-based BS serving multiple single-antenna UEs, the authors showed that the hexagonal sampling lattice is capable of optimally minimizing the surface area while retaining one independent signal dimension for each spent antenna.
The authors in \cite{Pizzo2022Nyquist} further investigated the EM propagation characteristics in different scenarios with planar {HMIMO surfaces}. The optimal Nyquist sampling theorem in the spatial domain and the spatial DoF were derived accordingly. It is concluded that sampling at the Nyquist rate allows to fully capture DoF of EM wave with minimum number of samples.  Specifically, the authors demonstrated the redundancy of conventional half a wavelength interval sampling approach. Based on this observation, they extended the proposed spatial domain Nyquist sampling to the non-isotropic scattering environment, and made a preliminary design of the Nyquist sampling matrix for the complex environment to derive sampling efficiency.  Employing prior knowledge of the scattering conditions, they derived an elongated hexagonal sampling structure for achieving an efficient representation. The results revealed that a reduction of $13\%$ samples per square meter is realized compared to half a wavelength sampling for isotropic propagation, and more sample reduction is expected for non-isotropic propagation.


\subsection{EM Information Theory}

\begin{figure*}[t!]
	\centering
	\includegraphics[height=4.6cm, width=17.6cm]{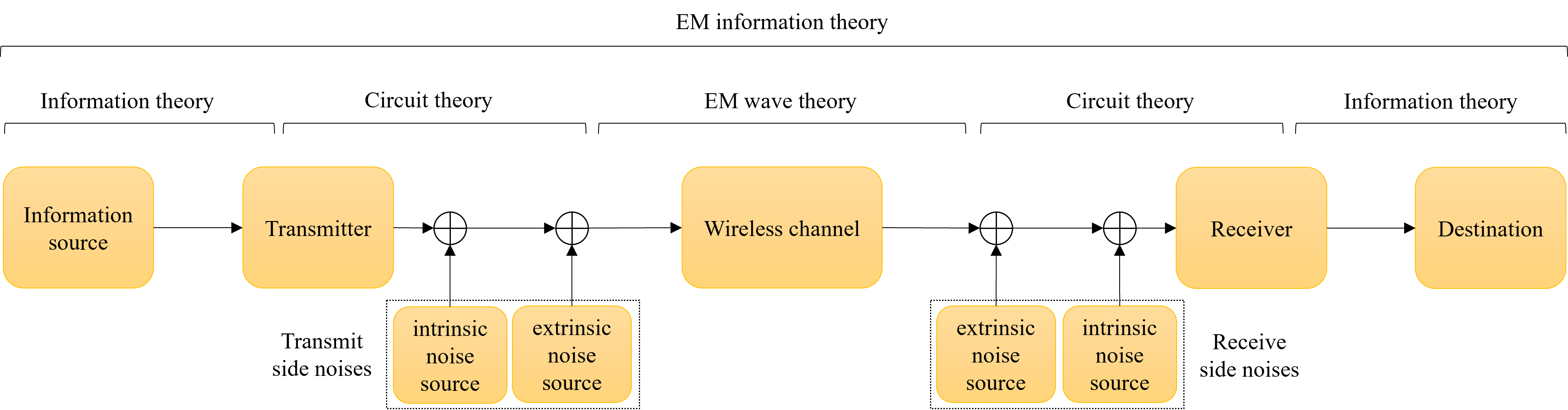}
	\caption{Schematic diagram of EM information theory of a general communication system, a seamless blend of information theory, circuit theory and EM wave theory.}
	\label{fig:EMInformationTheory}
\end{figure*}

It is widely known that Shannon's information theory \cite{Shannon1948Mathematical} has become the most important pillar for guiding and supporting various developments of wireless communications. It was built as a mathematical abstraction of the actual wireless communication process. The communication channel, treated by Shannon's information theory, is merely the medium used to transmit the signal between transmitter and receiver, which is mathematically described as a conditional probability distribution. This mathematical abstraction brings huge benefits in tackling various problems in wireless communications, while it, however, neglects the physical effects of actual signal transmissions, which was emphasized by Dennis Gabor \cite{Gabor1953Communication}. Therefore, as a natural consequence, the question that arises is if extra incremental capacity can be obtained using a more physical consistency framework? Exploring the answer is becoming more and more urgent in the pursue of extreme requirements of 6G. 
In addition, with the emergence of advanced technologies, such as metamaterials, metasurfaces, and HMIMO, etc., it is possible to proceed signal processing in the EM-domain, which stands for another strong motivation. EM information theory is studied under such a circumstance, which is an interdisciplinary framework to evaluate the fundamental limits of wireless communications with a fusion of EM theory and information theory. It is envisioned as the next milestone for guiding wireless analyses and designs.

Beyond the conventional analysis and design framework built based upon Shannon's information theory, EM wave theory and circuit theory are two extra frameworks expected to be incorporated into the EM information theory. These frameworks belong to the EM theory, and are becoming more and more important and effective in analyzing newly emerged wireless systems, such as HMIMO. EM information theory is envisioned as a seamless blend of the conventional information theory, EM wave theory, and circuit theory, as demonstrated in Fig. \ref{fig:EMInformationTheory}. Particularly, information theory mostly analyzes the information source, destination, and is less relevant to the transmitter and receiver. The circuit theory is more applied to the antenna and the RF circuit parts of the transmitter and receiver. It can be used to design the impedance matching, model the extrinsic and intrinsic noise that are generated from background radiations and RF devices, respectively. The remaining EM wave theory is more suitable for analyzing the wave propagation effects in wireless channels. 
Since the conventional information theory is well studied and is more familiar to us, in the subsequence, we will omit it and mainly introduce the latter two typical frameworks.

\subsubsection{EM Wave Theory}

Since the medium of wireless transmissions are EM waves, it is inevitable to follow the physical principle of EM wave propagation. Particularly, EM waves can be fully described via using Maxwell's equations, whose final forms were proposed by James Clerk Maxwell in 1865 \cite{Maxwell1865Viii}. Maxwell's equations, bridging the connection between varying electric field and varying magnetic field, are provided in differential forms as \cite{Chew1999Waves}
\begin{align}
    \nabla \times \bm{E}(\bm{r}, t) &= - \frac{\partial }{\partial t} \bm{B}(\bm{r}, t), \\
    \nabla \times \bm{H}(\bm{r}, t) &= - \frac{\partial }{\partial t} \bm{D}(\bm{r}, t) + \bm{J}(\bm{r}, t), \\
    \nabla \cdot \bm{B}(\bm{r}, t) &= 0, \\
    \nabla \cdot \bm{D}(\bm{r}, t) &= \varrho (\bm{r}, t),
\end{align}
where $\bm{E}$ is the electric field, $\bm{H}$ is the magnetic field, $\bm{D}$ is the electric flux, $\bm{B}$ is the magnetic flux, $\bm{J}(\bm{r}, t)$ is the current density and $\varrho(\bm{r}, t)$ is the charge density; $\nabla \times$ denotes the curl operator, $\nabla \cdot$ indicates the divergence operator; $\bm{r}$ denotes the location and $t$ reveals the time. These equations can be further simplified if the field is time harmonic, namely, the field is a sinusoidal function of time. As EM waves are vector fields, the vector wave equation of a time harmonic field in a homogeneous isotropic medium can be derived from Maxwell's equations as \cite{Chew1999Waves}
\begin{align}
    \nabla \times \nabla \times \bm{E}(\bm{r}) - k^{2} \bm{E}(\bm{r}) = i \epsilon \mu \bm{J}(\bm{r}),
\end{align}
where $k$ is the wavenumber, $\epsilon$ and $\mu$ are the permittivity and permeability of the medium. 
The vector wave equation can be solved using the tensor Green's function, such that \cite{Chew1999Waves}
\begin{align}
    \label{eq:ECR}
    \bm{E}(\bm{r}) = \int_{V} \bm{G}( \bm{r}, \bm{s} ) \bm{J}(\bm{s}) {\rm{d}} \bm{s},
\end{align}
where $V$ denotes the integral region, $\bm{G}( \bm{r}, \bm{s} )$ is the tensor Green's function, given by 
\begin{align}
    {\bm{G}}\left( {{{\bm{r}}},{{\bm{s}}}} \right) = i \epsilon \mu \frac{{{e^{i{k} \left \| {{{\bm{r}}} - {{\bm{s}}}} \right \|_{2}}}}}{{4\pi \left \| {{{\bm{r}}} - {{\bm{s}}}} \right \|_{2}}} \left( {{{\bm{I}}_3} + \frac{1}{{k^2}} {{\nabla}} {{{\nabla}}}} \right), 
\end{align}
with $\bm{I}_{3}$ being a $3 \times 3$ identity matrix. It is noteworthy that the tensor Green's function is the solution of the vector wave equation for a point source. 

The current and electric field relation in \eqref{eq:ECR} reveals that the source current $\bm{J}$ at location $\bm{s}$ can excite a certain electric field $\bm{E}$ at location $\bm{r}$. It is natural to correspond this current-electric field relation to the conventional transmit-receive relation of wireless transmissions, while with a more physically-consistent form. The wireless channel is thus depicted based upon the tensor Green's function. 

Empowered by such an EM wave theory inspired model, a few studies were performed with their emphasis on different aspects, such as channel modeling \cite{Pizzo2022Fourier,Gong2023HMIMO}, DoF analysis \cite{Poon2005Degrees,Dardari2020Communicating2}, capacity limit analysis \cite{Gong2023HMIMO}, as well as the transmit and receive pattern design \cite{Sanguinetti2021Wavenumber,Zhang2022Pattern_1}. As early attempts, these works are expected to inspire more future studies.

\subsubsection{Circuit Theory}
\label{circuit_theoy}
Another important analysis framework for wireless communications, especially for HMIMO communications, is using the circuit theory, where communication systems are depicted using a multi-port network, including multiple transmit ports and multiple receive ports, where each two-port, corresponding to a circuit voltage and a circuit current, indicates a single signal. The circuit-theoretic multi-port network framework facilitates analyses on physical factors, such as impedance matching, antenna mutual coupling, and distinct noise sources. It is capable of mitigating the gap between physics of EM waves and the mathematical abstraction of information theory \cite{Ivrlac2010Circuit,Ivrlac2014Multiport}, such that physically-consistent analyses and designs can be guaranteed.

\begin{figure}[t!]
	\centering
	\includegraphics[height=4.2cm, width=8cm]{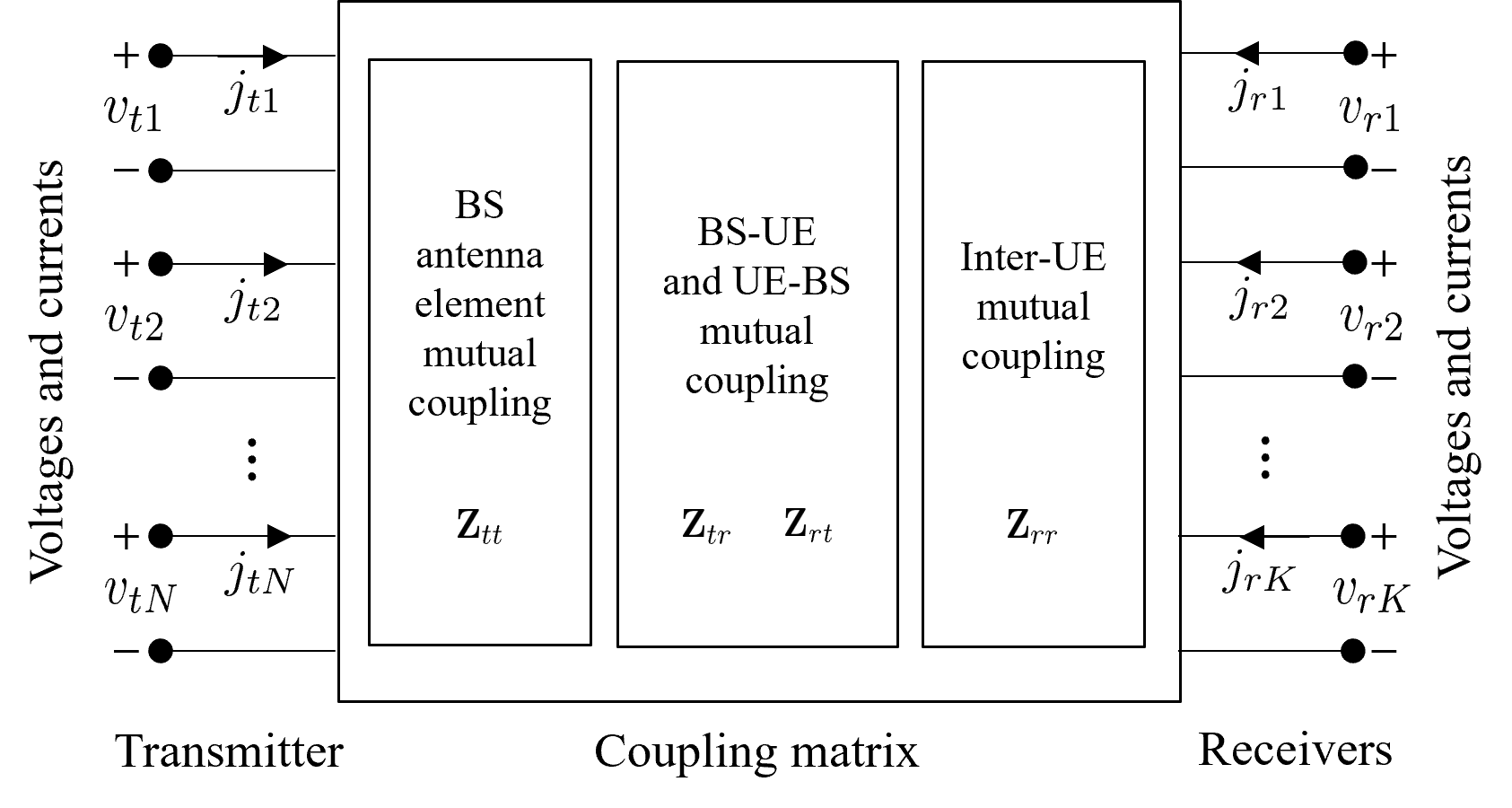}
	\caption{The circuit theory based multi-port network.}
	\label{fig:CircuitTheory}
\end{figure}

In a multi-port network, the communication model is generally built based on connections between the voltages and the currents of ports, as shown in Fig. \ref{fig:CircuitTheory} for a communication system with one BS and multiple UEs. Specifically, the communication model is given by \cite{Ivrlac2010Circuit, Ivrlac2014Multiport}
\begin{align}
    \begin{bmatrix}
    \bm{v}_{t} \\
    \bm{v}_{r}
    \end{bmatrix}=
    \begin{bmatrix}
    \bm{Z}_{tt} & \bm{Z}_{tr} \\
    \bm{Z}_{rt} & \bm{Z}_{rr}
    \end{bmatrix}
    \begin{bmatrix}
    \bm{j}_{t} \\
    \bm{j}_{r}
    \end{bmatrix},
\end{align}
where $\bm{v}_{t}$, $\bm{v}_{r}$ and $\bm{j}_{t}$, $\bm{j}_{r}$ are circuit voltages and circuit currents corresponding to the transmitter and multiple receivers, respectively; $\bm{Z}_{tt}$ and $\bm{Z}_{rr}$ are self-impedance matrices corresponding to the transmitter and the receiver, respectively; $\bm{Z}_{tr}$ and $\bm{Z}_{rt}$ denote the mutual impedance matrices between the transmitter (receiver) and the receiver (transmitter). As demonstrated by Fig. \ref{fig:CircuitTheory}, $\bm{Z}_{tt}$ depicts the BS antenna element mutual coupling effects; $\bm{Z}_{rr}$ captures the inter-UE mutual coupling effects; $\bm{Z}_{tr}$ and $\bm{Z}_{rt}$ describe the BS-UE and UE-BS mutual coupling effects, respectively.

The mutual impedance matrices $\bm{Z}_{tr}$ and $\bm{Z}_{rt}$, as well as the off-diagonal elements of self-impedance matrices $\bm{Z}_{tt}$ and $\bm{Z}_{rr}$ in the free-space condition can be obtained following an EM-domain analysis using \eqref{eq:ECR}. In existing works, the diagonal elements of self-impedance matrices $\bm{Z}_{tt}$ and $\bm{Z}_{rr}$ in the free-space regime are obtained in three ways: 1) Obtaining self-impedance matrices via the energy conservation law, where the transmit power equals the radiated power of electric fields assuming lossless antennas \cite{Ivrlac2010Circuit}. 2) Taking the limit of mutual impedance when $d \to 0$, where $d$ is the distance between two antenna elements \cite{Williams2021Multiuser}. 3) Using a closed-form expression of the ``Chu's antenna" to determine the self-impedance \cite{Chu1948Physical, Shyianov2022Achievable}. 
It is noteworthy that, in far-field regions, $\bm{Z}_{tr}$ can be neglected, i.e., $\bm{Z}_{tr} \approx \bm{0}$, because the effects at the transmitting side caused by currents at the receiving side are negligible in far-field regions, forming the unilateral approximation relation \cite{Ivrlac2010Circuit}. However, this mutual impedance cannot be ignored if near-field regions are considered \cite{Akrout2022Achievable}.

\begin{figure}[t!]
	\centering
	\includegraphics[height=3.0cm, width=8cm]{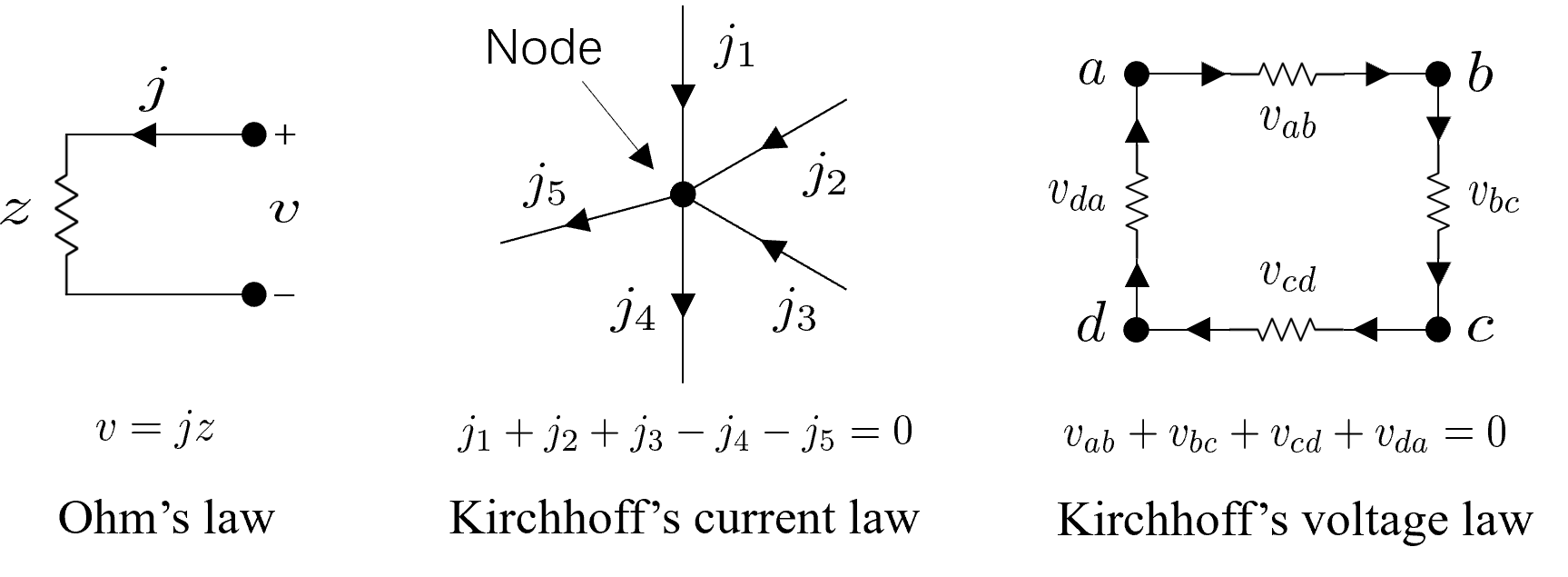}
	\caption{Three basic laws in circuit theory.}
	\label{fig:ThreeLaws}
\end{figure}

The circuit-theoretic multi-port network framework depends on three basic laws in assisting its analysis, as shown in Fig. \ref{fig:ThreeLaws}. The first one is Ohm's law, which states that the voltage between two ports equals the product of the conductor impedance and the current flowing through the conductor. It bridges the connections of voltage, impedance, and current. 
Another frequently used one is Kirchhoff’s current law, which describes that the currents entering a common node equals the currents leaving this node. It reveals that the currents are conserved within a common node. 
The last one is Kirchhoff’s voltage law, which indicates that the sum of the voltage drops around a closed circuit loop equals zero. It manifests that the voltages are conserved within a closed circuit loop.

Exploiting the useful circuit theory based multi-port network, several recent works \cite{Williams2022Electromagnetic, Saab2022Optimizing, Akrout2022Achievable, Akrout2022Super, Yang2022Channel, Han2023Using} were carried out with different emphases, such as communication models, near-field communications, channel modeling, and channel capacities.

\section{Enabling Technologies of HMIMO Communications}
\label{SectionV}

Since {HMIMO surfaces} only provide a basic physical entity in implementing HMIMO communications, it is necessary to develop corresponding physical-layer enabling technologies to drive {HMIMO surfaces} for approaching the fundamental limits of HMIMO communications. To this purpose, critical physical-layer technologies, such as channel estimation, and beamforming in holographic settings, are of great significance. However, designing these enabling technologies encounters fundamental differences compared to those designs in conventional mMIMO communication systems. The differences mainly come from the revolution of HMIMO over mMIMO in following aspects. 

Firstly, the hardware structures and working mechanisms of {HMIMO surfaces}, as shown in Section \ref{subsec:HTR}, Section \ref{SectionIII} and Section \ref{subsec:CCT}, are totally different from that of mMIMO antenna arrays. This will inevitably cause a distinct mathematical model for depicting the working process of {HMIMO surfaces}, and will also introduce different practical constraints. Different mathematical models and practical constraints will significantly affect the designs for enabling technologies. Secondly, the approximately continuous aperture, covered by infinite small antenna elements, opens up the possibility to directly manipulate EM waves by generating any arbitrary current distribution in a holographic mannrt. It will enable the conventional digital signal processing to be processed in EM-domain. As such, conventional signal processing for mMIMO will suffer from an ultra-high computational complexity due to the massive number of antenna elements. In addition, mutual coupling, mitigated in mMIMO arrays, emerges due to the ultra-dense positioning of antenna elements. Conventional transceiver designs will be invalid, thereby mutual coupling should be considered in enabling technologies designs for HMIMO communications. Lastly, the low power consumption and low-cost characteristics of {HMIMO surfaces} facilitate fabrication of large area continuous apertures, allowing transceivers to communicate in the near-field region. Not only the angle information but also the distance information between transceivers can be exploited to assist near-field communications, which clearly differentiates from the conventional far-field assumption that is the norm for enabling technologies of mMIMO communication systems. 

In summary, the new features bring significant differences in designing enabling technologies for HMIMO communications, compared with previously designed physical-layer technologies for mMIMO communications. This poses many challenges, while at the same time, also motivates new technology designs. In the following of this section, we focus on newly emerged enabling technologies of HMIMO communications, with a particular emphasis on HMIMO channel estimation, and HMIMO beamforming/beam focusing.

\subsection{HMIMO Channel Estimation}

Similar to channel estimation in conventional mMIMO communication systems, channel information is also necessary in HMIMO communication systems for accurate decoding and recovery of the transmitted signal at the receiving end. At the same time, HMIMO channel estimation is more complicated than the traditional discrete uncorrelated antenna array scenario because in HMIMO, effects such as inter-element coupling have to be carefully addressed. Conventional estimation methods based on pilot transmission lead to unacceptably large overhead and the far-field assumption usually assumed in wireless systems till now, does not necessarily apply to HMIMO due to the increasing size of the surfaces. In addition, the new features introduced in HMIMO, such as the EM-domain and near-field processing, inevitably give rise to new communication and channel models that are critical to asses and perform channel estimation on. So far, it is infeasible to develop channel estimation approaches accounting for all new features due to the lack of a practical design framework and mature communication/channel models. However, as more attention is attracted to this area, we delightedly observe that pioneer studies emerge with a special focus on near-field channel estimations, hybrid near- and far-field channel estimations, and near-field beam training via different approaches, such as compressed sensing (CS) and deep learning. In what follows, we present these state-of-the-art works as a basis for inspiring future, more practical and completed studies.\footnote{It should be noted that, when referring to the term XL-MIMO in the literature, cases with single antenna and multiantenna receivers are both included, as long as the BS consists of an extremely large-scale antenna array. Furthermore, part of the works studied the ULA based system with an emphasis on the `large' characteristic while ignoring the `dense' feature of HMIMO, in which the antenna element spacing is either the half a wavelength or not specified at all. These works are included because they can act as foundations to extend their ideas to the HMIMO regime, by incorporating both the `large' and `dense' features.}

\subsubsection{Near-Field Channel Estimation}

CS algorithms exploiting channel sparsity \cite{He2021Wireless} in the angular domain rely on far-field channel models. On the other hand, for HMIMO apertures which are large in physical size, the near-field region, where the channel is also related to the distance, appears to be dominant. In \cite{Cui2021Near-FieldChannelEstimation,Cui2022Channel}, the angle domain is replaced by the polar domain, in which both far- and near-field are sparse. A polar-domain simultaneous orthogonal
matching pursuit (OMP) algorithm is proposed for the near-field channel estimation. 
In \cite{Zhang2022Near1}, a joint dictionary learning and sparse recovery empowered channel estimation approach was suggested for near-field communications. The authors employed the spherical-wave propagation channel model to depict the near-field, and proposed an angular-domain sparse dictionary model, parameterized by the distance information, where the latter is included as an unknown variable in the dictionary. Their modeling approach reduced the heavy storage burden and high dictionary coherence caused by the polar-domain channel estimation approaches, whilst showcasing superiority in accurately retrieving channel information over polar-domain approaches.

In an attempt to realize low-complexity channel estimation for near-field extremely large RIS systems, in \cite{Yang2023Channel} the authors utilized the spherical-wave propagation channel model, and proposed a two-phase estimation procedure, where angular-domain parameters were first estimated and the cascaded angular- and polar-domain parameters were assessed afterwards. The aforementioned phases are CS based, thus decreasing the pilot use and computational complexity.

In a distinct approach, a model-based deep learning framework for near-field channel estimation of XL-MIMO systems was showcased in \cite{Zhang2023Near}. The spherical-wave propagation channel model was applied in building a spatial gridding based sparse dictionary, such that the channel estimation problem was attributed to a CS problem that was solved using a learning iterative shrinkage and thresholding algorithm. To improve the estimation accuracy and reduce the computation complexity, a sparsifying dictionary learning version of the latter algorithm was proposed by embedding the dictionary as a neural network layer into the layers of the algorithm.

In a recent work \cite{Elbir2023Near}, the near-field wideband THz channel estimation in the presence of beam-split effects was considered. A model-based channel estimation approach was first proposed, established on the OMP algorithm and a dedicated beam-split aware dictionary that exploited angular and distance deviations due to the near-field beam-split phenomenon. A model-free approach was then suggested, employing a federated learning scheme to reduce complexity and mitigate training overhead.

In order to tackle the non-stationarity of the HMIMO channel, a thorough research was carried out in \cite{han_channel_2020}, implementing a subarray-wise CS-based channel estimation method and a scatterer-wise method to explore the near-field non-stationary channel. The multipath channel is modeled with the last-hop scatterers under a spherical wavefront, and the large aperture array is divided into multiple subarrays. Numerical results demonstrate that the subarray-wise method can achieve an excellent mean square error performance with low complexity, whereas the scatterer-wise method can accurately position the scatterers and determine the non-stationary channel.

In \cite{Near_Field_Mixed_NLoS_LoS_Est}, in an XL-MIMO system the estimation problem is decomposed into two sub-problems, the LoS and NLoS component evaluation. The estimation of the LoS term reduces to geometric parameter estimation, i.e. the distance of the first antenna at the receiver from the first antenna at the transmitter, relative angle between receiver and transmitter, and the angle of departure. This is realized by searching a collection with coarse on-grid parameters, and then refined by iterative optimization. As far as the NLoS part is considered, OMP- based estimation is established, exploiting their polar-domain sparsity. Lastly, the CRLB of the proposed scheme is derived.

What is more, in \cite{yang2023nearfield}, the channel estimation problem of a near-field wideband MIMO XL-RIS aided system is investigated. The BS-RIS channel is assumed to be a LoS dominated channel that is previously known, so only the RIS-user channel is evaluated. Firstly, a crucial component of the developed framework is designing the wideband spherical domain dictionaries. Subsequently, a multi-frequency parallelizable subspace recovery framework is put forward for solving the wideband channel estimation problem, using the designed dictionaries. This framework converts the 2D-CS problem into multiple sparse vector recovery problems, with multi-frequency joint processing.

\subsubsection{Hybrid-Field Channel Estimation}

In addition to the merely near-field channel estimation, a few studies are emerged in investigating the hybrid near- and far-field channel estimation, which is a potential case in future HMIMO communication systems.
Specifically, researchers in \cite{Wei2022Channel} proposed a hybrid near- and far-field channel for XL-MIMO, where some scatterers are in the far-field, while others lie in the near-field. The hybrid-field channel estimation scheme deployed, individually exploits the channel sparsity of its near- and far-field components in the angular and polar domains respectively. 

{Furthermore, in paper \cite{Yu2022Hybrid} a deep learning framework was developed for channel estimation in the preceding context of hybrid near- and far-field channel. The deep learning model was implemented via a fixed point networks, where a closed-form linear estimator and a non-linear neural network estimator are incorporated inside. The authors demonstrated the significant performance gain of their approach over state-of-the-art in accuracy and convergence range. Likewise, in \cite{Nayir2022Hybrid} the hybrid near- and far-field channel estimation problem was investigated by proposing an OMP cascaded convolutional autoencoder neural network.}

\subsubsection{Near-Field Beam Training}
In the envisioned 6G communications, the integration of THz communications and large aperture antennas will occupy most of the communication systems. Due to severe path loss induced on the transmitted signals in the aforementioned frequency bands, communications will heavily rely on the LoS path. In addition, the numerous antenna elements make full channel knowledge acquisition nearly prohibitive. To that end, beam training arises as a fast and efficient solution to estimating only the LoS path by retrieving some channel parameters, i.e. direction and distance estimation. 

 Indicatively, in \cite{DL_Beam_Training_XL_MIMO}, an XL-MIMO system was considered where both the transmitter and the receiver only have one RF chain, a deep learning framework was proposed for localization in the near-field region. The authors assumed that a predefined codebook is applied for beamforming, searching for the optimal codeword to align the beam to the LoS path and thus achieve the largest data rate.
Motivated by the hierearchical codebook, they decomposed the problem of direction and distance estimation using two mapping functions as input to the neural networks.

Moreover, in \cite{Near_Field_Rainbow} along the lines of near-field wideband channel estimation, a distinct approach was proposed, established on the beam split phenomenon by employing true time delay beamforming and only one RF chain. More in detail, a wideband XL-MIMO scenario is considered where the LoS path is dominating. The BS is equipped with a ULA utilizing time delaying beamforming, the proposed scheme searches the
optimal angle in a frequency division manner, and the optimal distance in a time division manner. As a result, the estimation overhead is significantly reduced, since only the time slots for distance estimation are necessary.

\begin{table*}[!t]
	\tiny
	\renewcommand{\arraystretch}{1.2}
	\caption{\textsc{HMIMO channel estimation.}}
	\label{tab:channel_estimation}
	\centering
	\resizebox{\linewidth}{!}{
	\begin{tabular}{!{\vrule width0.6pt}c|c|c|c|c|l!{\vrule width0.6pt}}
		\Xhline{0.6pt}
		\rowcolor{yellow} \textbf{Ref.} & \textbf{\tabincell{c}{Uplink/\\downlink}} & \textbf{\tabincell{c}{Aperture\\type}} & \textbf{\tabincell{c}{Channel\\type}} & \textbf{\tabincell{c}{Estimation approach}} & \qquad \qquad \qquad \qquad \qquad \qquad \qquad  \textbf{\tabincell{c}{Main contributions}} \\

		\hline
		\tabincell{c}{\cite{Cui2021Near-FieldChannelEstimation}\\ \cite{Cui2022Channel}} & Uplink & \tabincell{c}{ULA} & \tabincell{c}{NLoS\\ (Near-field)} & \tabincell{c}{CS-based method} &  \tabincell{l}{Near-field channel estimation by exploiting the polar-domain sparsity.}\\

        \hline
		\cite{Zhang2022Near1} & Uplink & \tabincell{c}{ULA} & \tabincell{c}{NLoS\\ (Near-field)} & \tabincell{c}{CS-based method} &  \tabincell{l}{A joint dictionary learning and sparse recovery empowered channel estimation\\ approach was proposed to relieve high storage burden and dictionary coherence\\ of polar-domain approaches.}\\
		
        \hline
		\cite{Yang2023Channel} & Uplink & \tabincell{c}{ULA} & \tabincell{c}{NLoS\\ (Near-field)} & \tabincell{c}{CS-based methods} &  \tabincell{l}{A two-phase estimation procedure was proposed, where two CS based frameworks\\ and estimation algorithms were presented.}\\

        \hline
		\cite{Zhang2023Near} & Uplink & \tabincell{c}{ULA} & \tabincell{c}{NLoS\\ (Near-field)} & \tabincell{c}{Deep learning-\\ based method} &  \tabincell{l}{A model-based deep learning approach for XL-MIMO systems was proposed,\\ where a sparsifying dictionary learning approach was proposed.}\\

        \hline
		\cite{Elbir2023Near} & Uplink & \tabincell{c}{ULA} & \tabincell{c}{NLoS\\ (Near-field)\\ (Wideband)} & \tabincell{c}{CS-based and\\ federated learning-\\ based methods} &  \tabincell{l}{An OMP enabled model-based approach and a federated\\ learning empowered model-free approach were suggested for wideband THz\\ channel estimation with beam-split effects.}\\

        \hline
		\cite{han_channel_2020} & Uplink & \tabincell{c}{ULA} & \tabincell{c}{NLoS\\ (Near-field)\\ (Non-stationary)} & \tabincell{c}{Subarray-/scatterer-wise\\ CS-based methods} &  \tabincell{l}{The multipath channel is modeled with the last-hop scatterers under a spherical\\ wavefront and the large aperture array is divided into multiple subarrays.}\\

        \hline
		\cite{Near_Field_Mixed_NLoS_LoS_Est} & Downlink & \tabincell{c}{ULA\\ } & \tabincell{c}{NLoS\\ (Near-field)} & \tabincell{c}{CS based method} &  \tabincell{l}{The channel estimation problem is divided into two sub-problems, estimating the LoS\\ component via searching a collection with coarse on-grid parameters, and evaluating\\ the NLoS part by OMP- based estimation exploiting polar-domain sparsity field.}\\

        \hline
		\cite{yang2023nearfield} & Uplink & \tabincell{c}{UPA\\} & \tabincell{c}{NLoS\\ (Near-field)\\(Wideband)} & \tabincell{c}{CS based method} &  \tabincell{l}{A multi-frequency parallelizable subspace recovery framework is put forward for\\ solving the wideband channel estimation problem, using the designed spherical\\ domain dictionaries.}\\

        \hline
		\cite{Wei2022Channel} & Downlink & \tabincell{c}{ULA} & \tabincell{c}{NLoS\\ (Hybrid-field)} & \tabincell{c}{CS-based method} &  \tabincell{l}{The hybrid-field channel estimation scheme deployed, individually exploits channel\\ sparsity of its far-and near-field components in the angular and polar domains.}\\

        \hline
		\cite{Yu2022Hybrid} & Uplink & \tabincell{c}{UPA} & \tabincell{c}{NLoS\\ (Hybrid-field)} & \tabincell{c}{Deep learning-\\based method} &  \tabincell{l}{A deep learning based channel estimator was developed, where the deep learning\\ model was implemented via using a fixed point networks.}\\

        \hline
		\cite{Nayir2022Hybrid} & Downlink & \tabincell{c}{ULA} & \tabincell{c}{NLoS\\ (Hybrid-field)} & \tabincell{c}{Deep learning-\\based method} &  \tabincell{l}{An OMP cascaded convolutional autoencoder neural network\\ was developed for hybrid near- and far-field channel estimation.}\\

        \hline
		\cite{DL_Beam_Training_XL_MIMO} & Uplink & \tabincell{c}{ULA\\ } & \tabincell{c}{NLoS\\ (Near-field)} & \tabincell{c}{Deep learning-\\based method} &  \tabincell{l}{Utilizing the hierarchical beam training codebook, they decompose the problem of\\ LoS path estimation in direction and distance estimation \\using two mapping functions as input to the neural networks.}\\
  
        \hline
		\cite{Near_Field_Rainbow} & Uplink & \tabincell{c}{ULA} & \tabincell{c}{LoS\\ (Near-/far-field)\\ (Wideband)} & \tabincell{c}{Beam split\\ phenomenon\\
        based method} &  \tabincell{l}{The beam split phenomenon in wideband channels is exploited for assisting multiple\\ directions searching in one time slot, where the direction estimation is carried out\\ in a frequency division manner, while the distance in a time division scheme.}\\

        \hline
		\cite{Demir2022Channel} & Uplink & \tabincell{c}{UPA} & \tabincell{c}{NLoS\\ (Far-field)} & \tabincell{c}{Array geometric\\ information-aided\\ subspace method} &  \tabincell{l}{Provide a spatial channel correlation model and suggest an array geometric\\ information aided subspace channel estimation approach.}\\

        \hline
		\cite{Ghermezcheshmeh2022Parametric} & Uplink & \tabincell{c}{UPA\\ {CPS}} & \tabincell{c}{LoS\\ (Near-/far-field)} & \tabincell{c}{Parametric model\\ based method} &  \tabincell{l}{Channel estimation scheme based on a parametric physical channel model for LoS\\ dominated communications in mmWave and THz wave bands.}\\

         \hline
		\cite{tadele2023channel} & Downlink & \tabincell{c}{UPA\\ } & \tabincell{c}{NLoS\\ (Far-field)\\(Wideband)} & \tabincell{c}{Circuit theory model\\ Linear MMSE method} &  \tabincell{l}{Based on an EM-compliant circuit theory based channel model, a channel estimation\\ framework is showcased utilizing the antennas' scattering parameters, where a mutual\\ coupling aware linear MMSE algorithm is proposed.}\\

		\Xhline{0.6pt} 
	\end{tabular}}
\end{table*}

\subsubsection{Other Channel Estimation Schemes}

In reducing the pilot training overhead, the authors in \cite{Demir2022Channel} introduced a spatially correlated channel model for NLoS environments, and argued that with the large number of closely deployed antenna elements, the rank of the spatial correlation matrix needed for MMSE channel estimation, decreases. Instead, a reduced-subspace least-squares estimator was developed by exploiting the array geometry to identify
a subspace of reduced rank that covers the eigenspace of any spatial correlation matrix. The proposed estimator outperforms
the least-squares estimator, without using any user-specific channel statistics. The concept of reduced-subspace least-squares estimator was further employed for channel estimation of RIS-aided communications in \cite{Demir2022Exploiting}, such that pilots with a much shorter length can be realized to reduce the training overhead.

In \cite{Ghermezcheshmeh2022Parametric}, the authors proposed a channel estimation scheme based on a parametric physical channel model for LoS dominated communication in mmWave and THz wave bands. The proposed channel
estimation scheme exploits the specific structure of the radiated beams generated by the continuous surface to estimate the channel parameters in a LoS dominated channel model. It considers both the far- and near-field regions by partitioning the continuous aperture into tiles for which the far-field assumption holds. It is numerically demonstrated that
the proposed estimation scheme significantly outperforms other benchmark schemes under poor scattering.

Towards realizing a coupling-aware channel estimation framework, in \cite{tadele2023channel}, an HMIMO architecture with tightly packed antennas was considered. To that end, an EM-compliant circuit theory channel model is taken into account, incorporating the mutual coupling effects of both transmitting and receiving antennas. Using the available information on the mutual coupling matrices of each antenna, a novel framework for channel estimation is demonstrated, which superiors in comparison with the standard linear MMSE method without coupling awareness. The proposed algorithm is further extended in a wideband OFDM communication setting as well, resulting in the same improvements.

Lastly, we list relevant studies of HMIMO channel estimation in Table \ref{tab:channel_estimation} for ease of reference.

\subsection{HMIMO Beamforming and Beam Focusing}

The approximately continuous aperture of {HMIMO surfaces}, where the spacing of antenna elements tends to be much less than half a wavelength, can achieve both higher spatial resolution and stronger EM wave gain compared with conventional discrete antenna arrays. In addition, the continuous aperture is envisioned as a large surface. As a result, the opportunity of near-field communications emerges, in which an extra distance dimension is introduced for assisting communications on the basis of the conventional angle dimension. With those fundamental changes, 
the traditional beamforming technologies are facing a transformation from the conventional angle-dependent manner to the joint distance-angle fashion. Alternatively, conventional beamforming, only achieving the angular level of energy focus, namely, the transmitted energy is focused to a certain transmitting angle, will potentially shift to the near-field HMIMO beam focusing, capable of realizing precise location energy focus exploiting both distance and angle information. Recent advances on HMIMO beamforming and beam focusing will be further presented in the current subsesction with a macroscopic classification as follows.

\subsubsection{DMA Input-Output Response Based Work}
\begin{figure}[t!]
	\centering
	\includegraphics[height=5.6cm, width=7cm]{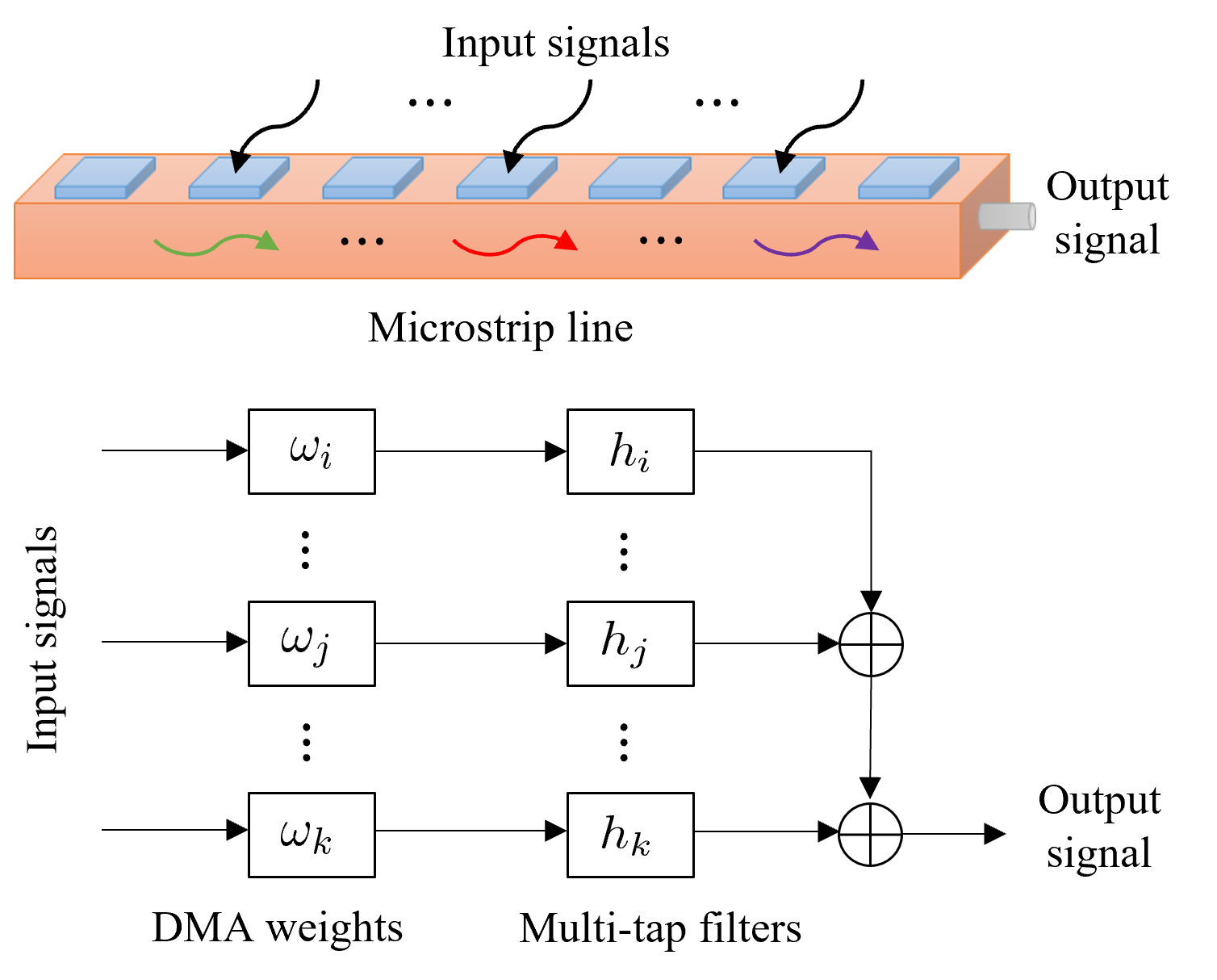}
	\caption{Modeling of the input-output response of a microstrip line of DMA (at the receiving side).}
	\label{fig:DMA_IO_Model}
\end{figure}

In this group, most of the past studies are performed on the basis of the input-output response model of a DMA, one of the typical {HMIMO surfaces} that can be utilized for realizing DMA-based HMIMO communication systems. Fig. \ref{fig:DMA_IO_Model} presents a demonstration on the receive modeling of the input-output response of a microstrip line of DMA. Each antenna element corresponds to a tunable weight, and the input signal corresponding to each antenna element propagates along the microstrip line together with other signals withing the waveguide. This propagation is modeled via a linear multi-tap filter. Specifically, the weight of the $\ell$-th element, following the Lorentzian-constrained phase model, is expressed as \cite{Shlezinger2019Dynamic}
\begin{align}
    \omega_\ell \in \left\{ \frac{i+e^{i \phi}}{2} \; | \; \phi \in [0, 2 \pi] \right\}, \;\;\; \forall \ell,
\end{align}
where $\phi$ indicates the tunable phase. 
The signal propagation inside the microstrip line is given by 
\begin{align}
    h_\ell = e^{- d_\ell \left( \alpha + i \beta \right)}, \;\;\; \forall \ell,
\end{align}
where $d_\ell$ is the distance between the $\ell$-th element and the output port of the microstrip line; $\alpha$ and $\beta$ are the waveguide attenuation coefficient and the wavenumber, respectively. DMAs typically consist of multiple microstrip lines \(N_m\), that each have \(N_e\) metamaterial elements, thus the whole architecture consists of \(N\triangleq N_m N_e\) elements. That being said, we formulate the input output relationship between the input signal \(\mathbf{y}\), and the signal at the ouput ports of the microstrips \(\mathbf{z}\) by accounting for all the microstrip lines included. Letting \(\omega_{i,\ell}\) denote the weight of the \(\ell\)-th element of the \(i\)-th microstrip, the \(N_m \times N\) analog weight matrix is given as,   
\begin{align}\label{eq:DMA_weights}
    [\mathbf{W}]_{\left(j,(i-1)N_e+\ell\right)}\triangleq \begin{cases} \omega_{i,\ell},&i=j\\0,&i\neq j\end{cases},
\end{align}
while the \(N \times N\) diagonal matrix \(\mathbf{H}\) accounts for the microstrip line propagation and is expressed as,
\begin{align}\label{eq:Waveguide_Propagation_Matrix}
    [\mathbf{H}]_{\left((i-1)N_e+\ell,(i-1)N_e+\ell\right)}\triangleq h_{i,\ell}.
\end{align}
Exploiting the preceding definitions, the input-output response is formulated as
\begin{equation}\label{eq:DMA_in_out}
    \mathbf{z}=\mathbf{W}\mathbf{H}\mathbf{y}.
\end{equation}
Using the established input-output response model, HMIMO beamforming/beam focusing can be attained by properly configuring the DMA weights.

To this aim, \cite{Shlezinger2019Dynamic} and \cite{Wang2019Dynamic} first formulated a mathematical model for DMA-based mMIMO systems and designed efficient alternating optimization (AO) algorithms to dynamically configure the DMA weights for the achievable sum-rate maximization in the uplink and downlink case, respectively. The AO algorithms were designed by approaching an optimally derived unconstrained configuration, based on which DMA weights with practical constraints can be obtained. Later, both \cite{wang_dynamic_2020} and \cite{wang_dynamic_2021} extended the use of DMAs in wideband setups, such as OFDM systems, using low resolution analog-to-digital converters (ADCs) by jointly optimizing the DMA weights along with the dynamic range of the ADCs and the digital processing, under a given bit constraint. 
Exploring a different optimization objective, \cite{xu_dynamic_2021} studied the energy efficiency maximization problem in single-cell multi-user mMIMO systems, by jointly optimizing the transmit precoding matrices of the UEs and the DMA weights of the BS, based on Dinkelbach’s transform and AO framework. This effort was later extended in \cite{You2022Energy} for further exploiting both instantaneous and statistical channel state information (CSI) in designing the joint transmit precoding and DMA weights.

With a particular focus on the beam focusing in near-field scenario, a mathematical model for DMA-based near-field multi-user MIMO systems was proposed in \cite{zhang_beam_2021}, incorporating both the feasible processing of DMAs as well as the propagation of the transmitted EM waves in near-field wireless communications. Then, the joint optimization of the DMA weights and the digital precoding vector was considered, in order to maximize the sum-rate when operating in near-field, while accounting for the specific Lorentzian-form response of metamaterial elements. 
Moreover, exploiting the great potential of beam focusing in near-field wireless power transfer (WPT), the authors in \cite{Zhang2022Near} presented a mathematical model for DMA-based radiating near-field WPT systems, characterized the weighted sum-harvested energy maximization problem of the considered system, and proposed an efficient solution to jointly design the DMA weights and digital precoding vector. 
Lastly, in \cite{Xu2022Near-Field}, an algorithmic framework was proposed to design the DMA beamforming for the near-field wideband multi-user uplink scenario based on a spherical wave channel model that incorporates both the near-field and dual-wideband effects. The DMA-based beam combining problem was first formulated, and then a sum-mean-square-error minimization algorithmic framework was developed, adopting the AO methodology.

\begin{figure}[t!]
	\centering
	\includegraphics[height=5.2cm, width=8cm]{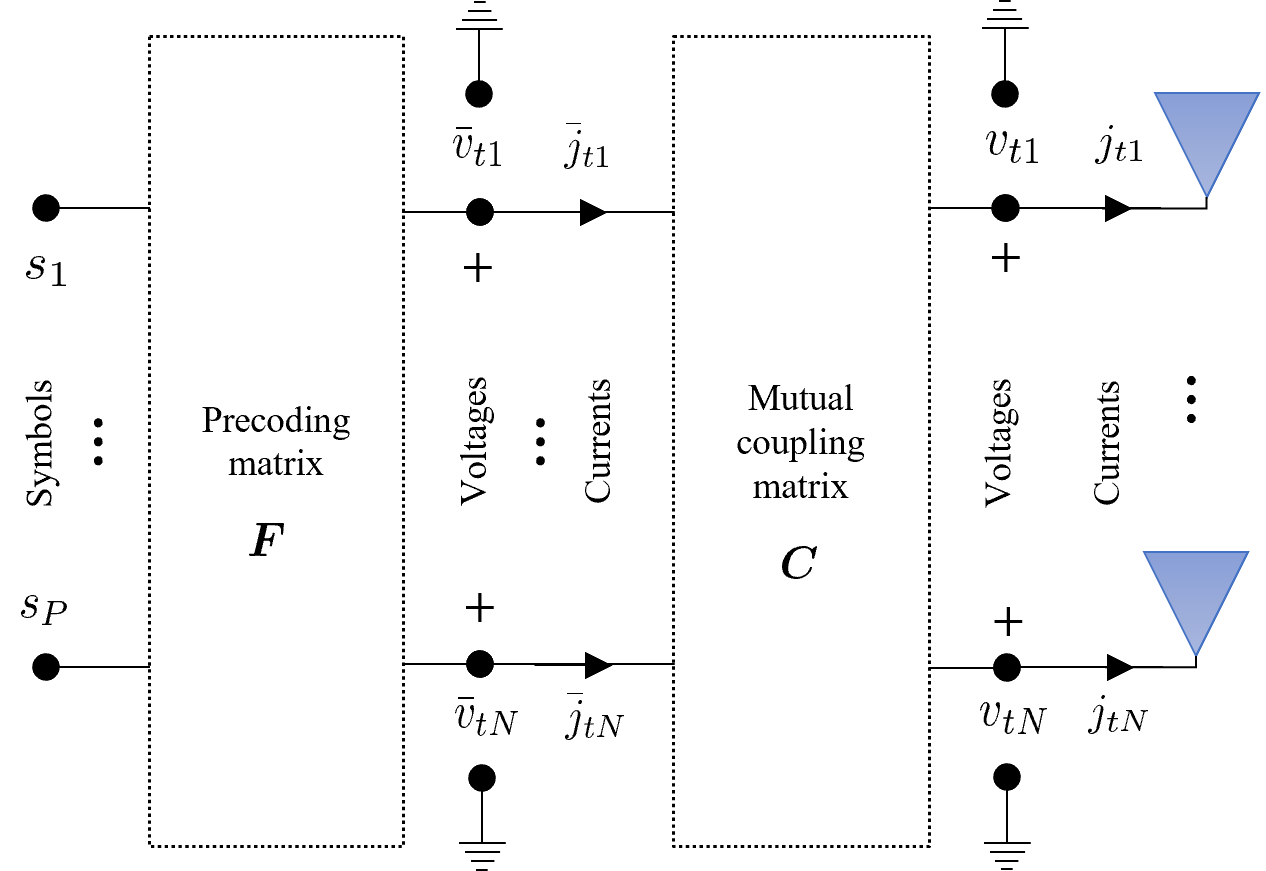}
	\caption{Mutual coupling based transmit beamforming scheme, depicted by the circuit model.}
	\label{fig:MC_Circuit_Model}
\end{figure}


\subsubsection{Mutual Coupling Based Work}
This group mainly describes a specific classification of recent studies that focus on characterizing mutual coupling mathematically, examining its influence, and further exploiting it for achieving super-directivity based on coupling-aware transceiver designs. Fig. \ref{fig:MC_Circuit_Model} showcases a mutual coupling based transmit beamforming scheme that is depicted by the circuit model via voltages and currents. The mutual coupling effects among transmit antenna elements are fully captured by the mutual coupling matrix, which is generally formed by the self and mutual impedances of antenna elements, as demonstrated in the circuit theory of EM information theory in \ref{circuit_theoy}. 
More in detail, the symbols are precoded in parallel through a parallel input parallel output multiport network, and in continuance pass through the mutual coupling block to form the transmitted voltage and current signals, which can be expressed as 
\begin{align}
    \bm{x} = \bm{C} \bm{F} \bm{s},
\end{align}
where $\bm{s}$ is the symbol vector, $\bm{F}$ denotes the precoding matrix, $\bm{C}$ is the mutual coupling matrix, and $\bm{x} \in \{ \bm{v}, \bm{j} \}$ is the transmitted voltage or current vector.
By designing a coupling-aware precoding matrix, the symbols can be optimally transmitted.

To the best of the authors' knowledge, the single user downlink LIS communication scenario was first investigated in \cite{Williams2020Communication}, where two models are studied for the LIS: one based on isotropic antenna elements depicted as spheres and the other being a collection of planar antenna elements. For both models, the expression for the mutual coupling was derived and the precoding vector from the coupling-aware MF was provided. The coupling-aware precoding showed that super-directivity can be potentially achieved as the antenna elements are more densely spaced. Later, \cite{Williams2021Multiuser} investigated the LIS-based multi-user MIMO communication scenario taking into account mutual coupling, super-directivity and near-field effects. The LIS design was based on the infinitesimal dipoles and two coupling-aware transmission schemes were proposed: MF and Weighted MMSE under practical limitations, such as the limited radiated power and existence of ohmic losses. The authors showed that those practical limitations cannot be neglected in achieving super-directivity. 
Recent works were carried out in this direction \cite{Han2022Coupling, Han2023Superdirective}, where a coupling-aware beamforming scheme was proposed for achieving super-directivity in a nearly continuous surface. They further presented how to obtain the coupling matrix which is used for coupling-aware beamforming. The authors validated the effectiveness of their approaches through full-wave EM simulations and real-world experiments, and demonstrated that superdirectivity can be achieved. 
In \cite{Wang2022Beamforming}, the beamforming performance of HMIMO was studied  with the consideration of mutual coupling and arbitrary radiation power patterns. The authors proposed to obtain the mutual coupling matrix by following the energy conservation law. The derived result was then used as a means of analyzing the beamforming gain, where numerical results showcased the superiority of coupling-aware beamforming in comparison to typical beamforming.

\subsubsection{Holographic Principle Based Work}
\begin{figure}[t!]
	\centering
	\includegraphics[height=6.6cm, width=6cm]{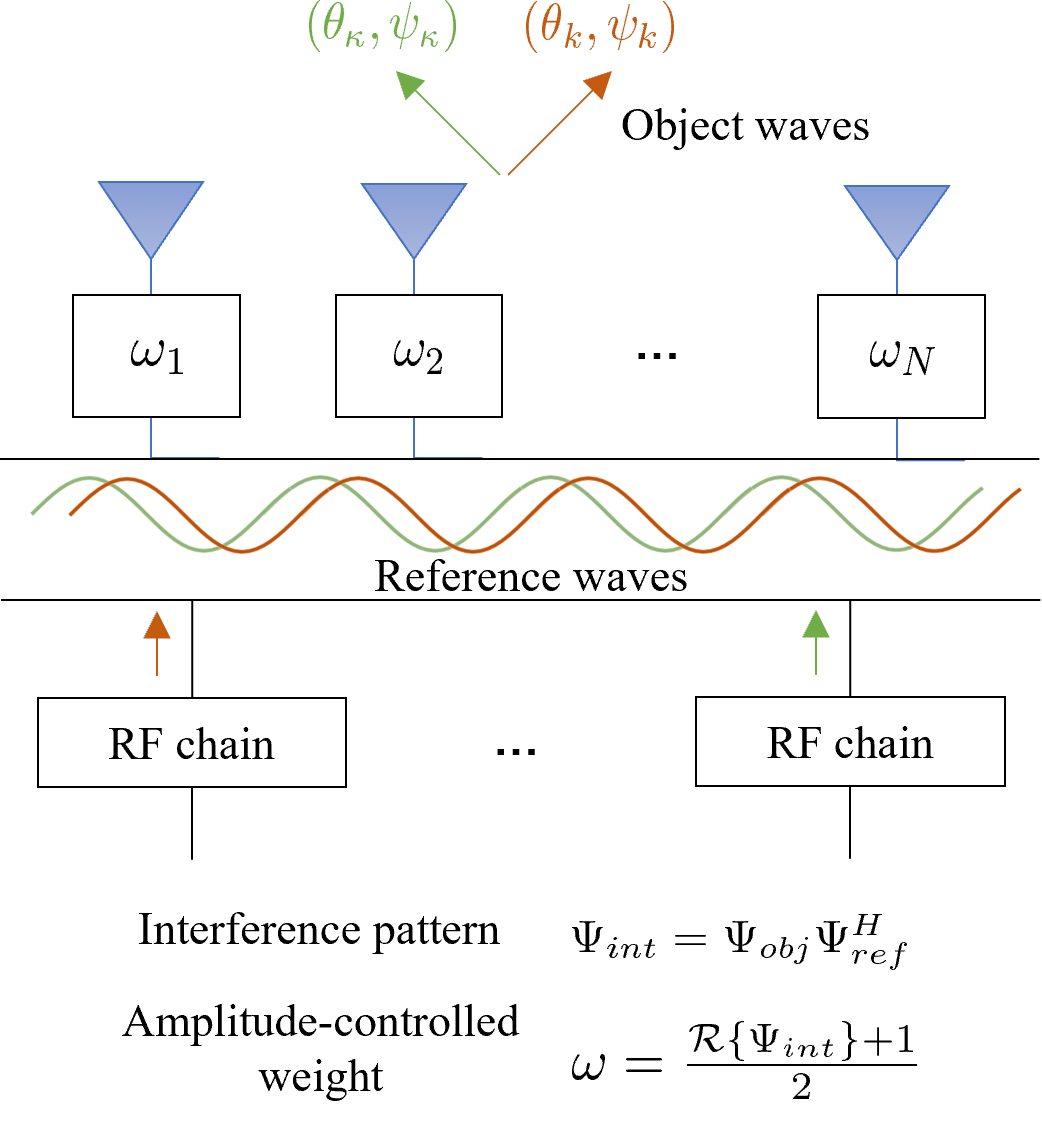}
	\caption{Holographic principle guided transmissions (at the transmitting side).}
	\label{fig:HoloPrinciple_Model}
\end{figure}

This group encompasses recent studies of HMIMO beamforming under the guidance of holographic principle as demonstrated previously in \emph{Holographic LWA based EM Holography} in Section \ref{subsec:HTR}. As shown in Fig. \ref{fig:HoloPrinciple_Model}, reference waves, loaded by RF chains, propagate along the substrate and thus excite each antenna element with a specific weight to finally form intended object waves. The weights are amplitude-controlled and tuned based on the interference pattern. 
Specifically, the reference wave generated by an RF chain, and the $(\theta_{k}, \psi_{k})$-direction object wave are respectively expressed as \cite{di_reconfigurable_2021}
\begin{align}
    \Psi_{ref} (\bm{r}_{n}^{m}) &= e^{ -i \bm{a}_{r} \bm{r}_{n}^{m} }, \\
    \Psi_{obj} (\bm{r}_{n}, \theta_{k}, \psi_{k}) &= e^{ -i \bm{a}_{o} (\theta_{k}, \psi_{k}) \bm{r}_{n} },
\end{align}
where $\bm{r}_{n}^{m}$ denotes the distance vector from the $m$-th RF chain port to the $n$-th element; $\bm{r}_{n}$ represents the location vector of the $n$-th element; 
$\bm{a}_{r}$ indicates the propagation wave vector of the reference wave; $\bm{a}_{o} (\theta_{k}, \psi_{k})$ is the propagation wave vector of the object wave. As such, the interference pattern between the reference wave and the object wave is obtained as $\Psi_{int} (\bm{r}_{n}, \bm{r}_{n}^{m}, \theta_{k}, \psi_{k}) = \Psi_{obj} (\bm{r}_{n}, \theta_{k}, \psi_{k}) \Psi_{ref}^{H} (\bm{r}_{n}^{m})$. Employing an amplitude controlling approach to construct the interference pattern, the normalized configured weight of the $n$-th element is provided by \cite{di_reconfigurable_2021}
\begin{align}
    \omega (\bm{r}_{n}, \bm{r}_{n}^{m}, \theta_{k}, \psi_{k}) = \frac{\mathcal{R} \left\{ \Psi_{int} (\bm{r}_{n}, \bm{r}_{n}^{m}, \theta_{k}, \psi_{k}) \right\} + 1}{2},
\end{align}
where $\mathcal{R}\{ \cdot \}$ denotes the real part of a complex number. The configuration of these weights is also termed as holographic beamforming, is realised in the analog domain and follows the holographic principle for achieving intended object waves.
Based on this model, representative studies are listed as follows.

In \cite{di_reconfigurable_2021}, an {HMIMO surface}-assisted wideband OFDM downlink single-user scenario with frequency selective channels was considered, focusing on the achievable data rate maximization problem by jointly optimizing the digital and holographic beamforming, respectively, according to an amplitude control optimization algorithm. An interference pattern multiplexing based scheme was developed to mitigate the beam squint effect arising from the frequency selectivity. Subsequently,
\cite{Deng2021ReconfigurableBeamforming} introduced an iterative joint optimization algorithm for digital beamforming at the BS and holographic beamforming at the {HMIMO surface} in a downlink multi-user scenario, aiming to maximize the achievable sum-rate. In order to handle the resulting complex-domain optimization problem subject to unconventional real-domain amplitude constraints, coupled with the superposition of the radiation waves from different radiation elements as well as the overall coupling between the radiation elements simultaneously with the propagating surface, the closed-form optimal holographic beamforming scheme was also derived. By simulation results, it was shown that the {HMIMO surface} was capable of accurate beam-steering with low side-lobes. In addition, \cite{Deng2022Reconfigurable} proposed a joint beamforming optimization technique to maximize the sum-rate in an {HMIMO surface}-assisted downlink multi-user communication system. Particularly, an AO algorithm was developed solving the digital beamforming sub-problem at the BS by ZF beamforming with power allocation, the holographic beamforming sub-problem at the {HMIMO surface} via fractional programming, and the receive combining sub-problem at the UEs by a coordinate ascent approach. 
Lastly, in \cite{deng_hdma_2022} a new type of space-division multiple access was proposed, called holographic-pattern division multiple access (HDMA) along with a multi-user holographic beamforming scheme for HDMA. Theoretical analysis resulted in an optimal holographic beam pattern through which the sum-rate with simple ZF precoding can achieve the asymptotic capacity of the HDMA system.

In addition, \cite{Deng2022Holographic} extended the work to satellite communications. An uplink {HMIMO surface}-aided communication system comprising of one UE and multiple low-earth-orbit (LEO) satellites was considered, and a sum-rate maximization problem was formulated. ZF digital beamforming was applied at the UE and holographic beamforming was optimized at the {HMIMO surface} via dynamic programming. The satellites’ positions were predicted according to the temporal variation law. Simulation results showed the superiority of the {HMIMO surface}-aided system compared to the traditional phased antennas in terms of the achievable sum-rate as well as the cost, and proved the technique's robustness against tracking errors of the satellites’ positions. Furthermore, the authors of \cite{Zhang2022Holographic} formulated an integrated sensing and communication problem in an {HMIMO surface}-assisted scenario for simultaneously detecting multiple radar targets and serving multiple UEs. The beamforming gains towards the directions of the targets were maximized, whilst minimizing the cross-correlation among these directions, with an iterative algorithm. Theoretical analysis for the maximum beamforming gain lower bound was conducted, and simulation results revealed the superiority of the holographic beamforming scheme over traditional MIMO systems.

\subsubsection{EM Level Model Based Work}
\begin{figure}[t!]
	\centering
	\includegraphics[height=6.6cm, width=6.4cm]{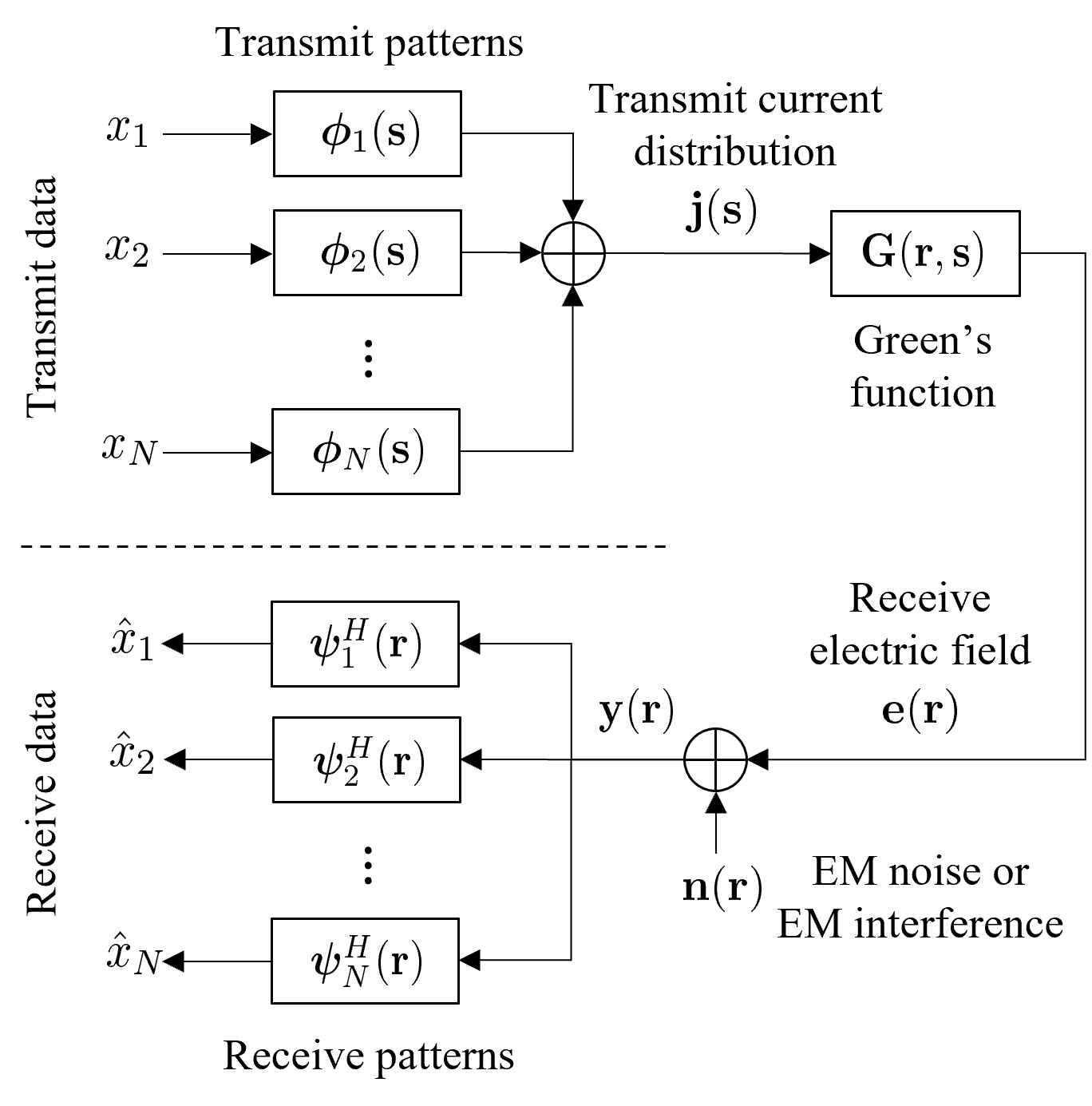}
	\caption{EM level modeling of a point-to-point HMIMO communication system.}
	\label{fig:EM_Level_Model}
\end{figure}

Beyond above-mentioned groups of mathematical model spurred studies, another branch, aiming to model the communication system in an EM level, emerges. Fig. \ref{fig:EM_Level_Model} exhibits an EM level modeling of point-to-point HMIMO communications, where the transmit data are carried by dedicated transmit patterns, and the transmit current distribution is generated accordingly {as a combination of all weighted transmit patterns, namely,} \cite{Dardari2020Communicating2, Sanguinetti2021Wavenumber, Gong2023HMIMO}
\begin{align}
    \bm{j}(\bm{s}) = \sum_{n=1}^{N} x_{n} \bm{\phi}_{n} (\bm{s}),
\end{align}
where $\bm{s}$ indicates the location of a transmit point that belongs to the transmit surface $A_{T}$; $\bm{\phi}_{n}$ is the $n$-th transmit pattern for the transmit surface. The transmit current excites an electric field $\bm{\bm{r}}$ that can be measured at the receiver's end. Together with the EM noise or EM interference, the received signal can be expressed by using the tensor Green's function as \cite{Chew1999Waves}
\begin{align}
    \bm{y}(\bm{r}) = \int_{A_{T}} \bm{G}(\bm{r}, \bm{s}) \bm{j}(\bm{s}) {\rm{d}} \bm{s} + \bm{n}(\bm{r}).
\end{align}
At the receiver's side, the acquired signal is further decoded to obtain the data by means of receive patterns \cite{Dardari2020Communicating2, Sanguinetti2021Wavenumber, Gong2023HMIMO}, 
\begin{align}
    \hat{x}_{n} = \int_{A_{R}} \bm{\psi}_{n}^{H}(\bm{r}) \bm{y}(\bm{r}) {\rm{d}} \bm{r},
\end{align}
where $\bm{\psi}_{n}(\bm{r})$ denotes the $n$-th receive pattern for the receiving surface $A_{R}$.
Based on this EM level model, it is possible to design dedicated patterns, capable of generating any current distributions for CPS based {HMIMO surfaces}, for maximizing the communication performance from an EM perspective. The patterns can also be interpreted as basis functions, as described in \cite{Dardari2021Holographic,Gong2023HMIMO}. Based on this EM level model, emerging studies are listed as follows.

Firstly, \cite{Sanguinetti2021Wavenumber} built an EM level transmission model for a point-to-point HMIMO communication system, and analyzed the optimal pattern achieved in specific conditions. Furthermore, to provide a simple and practical pattern design, a suboptimal wavenumber-division multiplexing (WDM) scheme was introduced for LoS HMIMO, inspired by OFDM, assuming that the spatially-continuous transmit currents and received fields are represented using the Fourier basis functions. The orthogonality among the communication modes (in the wavenumber domain) is achieved with WDM only when the size of the receiver grows infinitely large, due to the unbounded support of the channel response in the spatial domain. 
Later, in \cite{Zhang2022Pattern,Zhang2022Pattern_1}, a pattern design for CPS enabled multi-user communication system is investigated to maximize the information modulated on EM waves. The authors modeled the system in EM level, and proposed a pattern-division multiplexing method accordingly. Specifically, the authors first construct a communication architecture between CPSs from an EM wave theory perspective, and then project the orthogonal continuous current functions carrying information at transceiver side onto a Fourier orthogonal basis, thus transforming the design of orthogonal continuous current functions into a design problem of their projection length on a finite orthogonal basis, thus enabling optimization of the problem and realizing a significant capacity improvement over the existing WDM scheme.
In addition, another recent study \cite{Wei2022Tri} suggested two precoding schemes for a multi-user HMIMO communication system, where near-field and triple polarization were considered. The two precoding schemes are respectively responsible for cross-polarization elimination and inter-user interference mitigation. This study was performed based on a specially established near-field channel model by means of the dyadic Green’s function.

\begin{table*}[!t]
	\tiny
	\renewcommand{\arraystretch}{1.2}
	\caption{\textsc{HMIMO beamforming and beam focusing.}}
	\label{tab:beamforming}
	\centering
	\resizebox{\linewidth}{!}{
	\begin{tabular}{!{\vrule width0.6pt}c|c|c|c|c|c|l!{\vrule width0.6pt}}
		\Xhline{0.6pt}
		\rowcolor{yellow} \textbf{\tabincell{c}{Category}} & \textbf{Ref.} & \textbf{\tabincell{c}{Uplink/\\downlink}} & \textbf{\tabincell{c}{Multiple\\UEs}} & \textbf{\tabincell{c}{Channel\\ type}} & \textbf{\tabincell{c}{Optimization\\ approach}} & \qquad \qquad \qquad \qquad \qquad \qquad \qquad \qquad \qquad \textbf{\tabincell{c}{Main contribution}} \\ 
		
		\Xhline{0.6pt}
		\multirow{17}{*}{\tabincell{c}{DMA\\input-\\output\\response\\based\\work}} & \cite{Shlezinger2019Dynamic} & Uplink & Yes & \tabincell{c}{NLoS} & \tabincell{c}{Closed-form solution,\\ AO algorithm} &  \tabincell{l}{Formulate a mathematical model for DMA-based uplink mMIMO systems, characterize the achievable sum-rate,\\ and design HMIMO beamforming for different DMA setups, considering frequency-flat/-selective channels.}\\
		
		\cline{2-7}
		& \cite{Wang2019Dynamic} & Downlink & Yes & \tabincell{c}{NLoS} & \tabincell{c}{Closed-form solution,\\ AO algorithm} &  \tabincell{l}{Formulate a mathematical model for DMA-based downlink mMIMO systems, and design various HMIMO\\ beamforming schemes for different DMA setups, considering frequency-flat/-selective channels.}\\
		
		\cline{2-7}
		& \tabincell{c}{\cite{wang_dynamic_2020}\\ \cite{wang_dynamic_2021}} & Uplink & Yes & \tabincell{c}{NLoS} & \tabincell{c}{Greedy algorithm,\\AO algorithm,\\ fractional programming} &  \tabincell{l}{Extend the use of DMAs in OFDM systems using low resolution ADCs, as well as jointly optimize the DMA\\ weights and the dynamic range of ADCs, under a given bit constraint.}\\
        
        \cline{2-7}
		& \cite{xu_dynamic_2021} & Uplink & Yes & \tabincell{c}{NLoS\\(Far-field)} & \tabincell{c}{Dinkelbach’s transform,\\AO algorithm} &  \tabincell{l}{Study the energy efficiency optimization of a single-cell multiuser mMIMO system by a joint optimization of\\ the UEs’ transmit precoding and the BS DMAs’ weights.}\\
		
		\cline{2-7}
		& \cite{You2022Energy} & Uplink & Yes & \tabincell{c}{NLoS\\(Far-field)} & \tabincell{c}{Dinkelbach’s transform,\\AO algorithm} &  \tabincell{l}{Examine the joint transmit precoding and DMA weights in energy efficiency maximization of a multi-user\\ uplink system, exploiting either instantaneous or statistical CSI}\\

		\cline{2-7}
		& \cite{zhang_beam_2021} & Downlink & Yes & \tabincell{c}{LoS\\(Near-field)} & \tabincell{c}{AO algorithm,\\ Riemannian gradient} &  \tabincell{l}{Formulate a mathematical model for DMA-based near-field multi-user MIMO systems, depicting both feasible\\ processing of DMAs and near-field EM wave propagation, and joint design digital precoding and DMA weights.}\\
		
		\cline{2-7}
		& \cite{Zhang2022Near} & Downlink & Yes & \tabincell{c}{LoS\\(Near-field)} & \tabincell{c}{AO algorithm,\\Riemannian gradient} &  \tabincell{l}{Present a mathematical model for DMA-based radiating near-field WPT systems, characterize the weighted\\ sum-harvested energy maximization problem, and joint design digital precoding and DMA weights.}\\
		
		\cline{2-7} 
		& \cite{Xu2022Near-Field} & Uplink & Yes & \tabincell{c}{NLoS\\(Near-field)} & \tabincell{c}{Weighted MMSE transform,\\AO algorithm,\\ minorization-maximization} &  \tabincell{l}{Formulate the DMA-based beam combining problem for XL-MIMO communications, and design the beam\\ combining for the near-field wideband XL-MIMO uplink transmissions assisted by DMAs}\\

		\Xhline{0.6pt}
		\multirow{5}{*}{\tabincell{c}{Mutual\\coupling\\based\\work}} & \cite{Williams2020Communication} & Downlink & No & \tabincell{c}{LoS\\ (Near-field)} & \tabincell{c}{Closed-form solution \\(MF)} &  \tabincell{l}{Formulate a mathematical model of mutual coupling for LIS, and provide a coupling-aware MF scheme for\\ isotropic and planar antenna elements.}\\
		
		\cline{2-7}
		& \cite{Williams2021Multiuser} & Downlink & Yes & \tabincell{c}{LoS\\ (Near-field)} & \tabincell{c}{Closed-form solution \\(MF, weighted MMSE)} &  \tabincell{l}{Present coupling-aware MF and weighted MMSE designs for LIS-based multi-user system, accounting for\\super-directivity and mutual coupling effects along with near field propagation}\\
		
		\cline{2-7}
		& \cite{Han2022Coupling} & / & / & \tabincell{c}{LoS\\ (Far-field)} & \tabincell{c}{Closed-form solution} &  \tabincell{l}{Propose a coupling-aware beamforming design for achieving super-directivity, and present a feasible method\\ to obtain the coupling matrix.}\\

		\Xhline{0.6pt}
		\multirow{12}{*}{\tabincell{c}{Holographic\\principle\\based\\work}} & \cite{di_reconfigurable_2021} & Downlink & No & \tabincell{c}{NLoS\\(Far-field)} & \tabincell{c}{AO algorithm} &  \tabincell{l}{Maximize the achievable data rate via a joint design of hybrid digital and holographic beamforming, while\\ mitigating the beam squint loss.}\\
		
		\cline{2-7}
		& \cite{Deng2021ReconfigurableBeamforming} & Downlink & Yes & \tabincell{c}{NLoS\\(Far-field)} & \tabincell{c}{Closed-form solution,\\ AO algorithm} &  \tabincell{l}{Propose to jointly optimize a hybrid digital-holographic beamforming scheme to maximize the sum-rate.}\\
		
		\cline{2-7}
		& \cite{Deng2022Reconfigurable} & Downlink & Yes & \tabincell{c}{NLoS\\(Far-field)} & \tabincell{c}{Closed-form solution,\\fractional programming,\\AO algorithm} &  \tabincell{l}{Propose a joint digital-holographic beamforming and receive combining optimization scheme to maximize\\ the sum-rate in an HMIMO surface-assisted downlink multi-user communication system.}\\
		
		\cline{2-7}
		& \cite{deng_hdma_2022} & Downlink & Yes & \tabincell{c}{LoS\\(Near-field)} & \tabincell{c}{Closed-form solution,\\(Lagrangian multiplier)} &  \tabincell{l}{Propose a new type of space-division multiple access, HDMA, design a holographic beamforming scheme to\\ maximize the capacity in an {HMIMO surface}-assisted downlink multi-user communication system.}\\
		
		\cline{2-7}
		& \cite{Deng2022Holographic} & Downlink & Yes & \tabincell{c}{LoS\\(Far-field)} & \tabincell{c}{Closed-form solution (ZF),\\dynamic programming} &  \tabincell{l}{Present a joint digital and holographic beamforming scheme to maximize the sum-rate for an uplink HMIMO\\ surface-aided communication system comprising one UE and multiple LEO satellites.}\\
		
		\cline{2-7}
		& \cite{Zhang2022Holographic} & Downlink & Yes & \tabincell{c}{NLoS\\(Far-field)} & \tabincell{c}{Semidefinite relaxation,\\ AO algorithm} &  \tabincell{l}{Propose a hybrid digital-holographic beamforming scheme that jointly performs sensing and communication,\\ and theoretically provide a lower bound on the maximum beampattern gain.}\\

		\Xhline{0.6pt} 
		\multirow{5}{*}{\tabincell{c}{EM level\\model\\based\\work}} & \tabincell{c}{\cite{Sanguinetti2021Wavenumber}} & / & No & \tabincell{c}{LoS\\(Far-field)} & \tabincell{c}{Closed-form solution\\(Fourier basis)} &  \tabincell{l}{Formulate a mathematical model in EM level for continuous aperture based point-to-point HMIMO system, and\\ design a suboptimal WDM scheme realizing a practical HMIMO communication.}\\
		
		\cline{2-7}
		& \tabincell{c}{\cite{Zhang2022Pattern}\\ \cite{Zhang2022Pattern_1}} & Downlink & Yes & \tabincell{c}{LoS\\(Far-field)} & \tabincell{c}{Weighted MMSE transform,\\ AO algorithm} &  \tabincell{l}{Formulate a mathematical model in EM level for continuous aperture based multi-user HMIMO system, and\\ design the corresponding patterns for UEs that maximize the sum-rate.}\\
		
		\cline{2-7}
		& \tabincell{c}{\cite{Wei2022Tri}} & Downlink & Yes & \tabincell{c}{LoS\\(Near-field)} & \tabincell{c}{Gaussian elimination,\\ block diagonalization} &  \tabincell{l}{Propose a near-field LoS channel model using dyadic Green's function, and present two precoding schemes\\ for cross-polarization and inter-user interference eliminations.}\\

		\Xhline{0.6pt}
		\multirow{5}{*}{\tabincell{c}{Others}} & \cite{bjornson_utility-based_2019} & Uplink & Yes & \tabincell{c}{LoS\\(Far-field)} & \tabincell{c}{Closed-form solution\\ (MRT, ZF)} &  \tabincell{l}{Explore if MRT is sufficient when transmitting to multiple UEs or if more advanced methods, such as ZF, is\\ needed to deal with interference when transmitting from an LIS.}\\
		
		\cline{2-7}
		& \cite{Jesus2020Near} & Uplink & Yes & \tabincell{c}{LoS\\(Near-field)} & \tabincell{c}{Closed-form solution\\(MRC, MMSE)} &  \tabincell{l}{Signal and interference terms are numerically analyzed as a function of the position of the transmitting devices\\ with both MRC and MMSE schemes.}\\
		
		\cline{2-7}
		& \cite{Wu2022Distance} & Downlink & No & \tabincell{c}{LoS\\(Near-field)} & \tabincell{c}{AO algorithm,\\greedy searching} &  \tabincell{l}{Formulate a mathematical model of XL-MIMO based near-field distance-aware precoding system, and jointly\\ design the analog and digital beamformers.}\\
		
		\Xhline{0.6pt} 
	\end{tabular}}
\end{table*}

\subsubsection{Others}
Some other studies were developed as well. For example, \cite{bjornson_utility-based_2019} studied the beamwidth and sidelobes of a transmitting LIS, which can be considered as a continuous surface. A comparison has been made between employing MRT and ZF schemes, in order to mitigate the interference deriving from the LIS's closed spaced antennas in a two-user scenario, under the far-field, LoS wireless channel assumption. It was shown that ZF and MRT perform equally well when the number of antennas reaches it's asymptotic limit, while MRT is sub-optimal for practical LIS sizes. Following \cite{bjornson_utility-based_2019}, reference \cite{Jesus2020Near} showed that when the distance between UE and LIS is comparable to the size of LIS, the near-field assumption holds. Both MRC and MMSE combining schemes were considered to study the uplink spectrum efficiency of two UEs communicating with an LIS with varying antenna spacing, effective antenna areas and loss from polarization mismatch. MMSE combining was proved to be superior when employing an LIS. 
Besides, the authors of \cite{Wu2022Distance} examined a near-field LoS XL-MIMO communication system, and presented a distance-aware precoding structure which can flexibly configure RF chains depending on the DoF of near-field channel. A corresponding distance-aware precoding algorithm was designed to adaptively fit the DoF variation along with the change in distance.

In the final of this subsection, we list recent studies of HMIMO beamforming and beam focusing in Table \ref{tab:beamforming} for ease of reference.


\section{Comparisons and Extensions}
\label{SectionVI}


\subsection{Comparisons with Existing Technologies}
\label{subsec:CCT}

In this subsection, we will compare HMIMO communications with existing technologies, such as RIS-aided communications, XL-MIMO communications and mMIMO communications. We first briefly interpret RIS as the passive type of {HMIMO surfaces} from a macroscopic perspective, and point out the deficiency in existing works of RIS. We then present XL-MIMO as a special case of HMIMO, where extra aspects induced in HMIMO are provided. We finally put an emphasis on the comparison between HMIMO and mMIMO in regard to the hardware, directivity, coverage, capacity, energy efficiency and other miscellaneous aspects. We detail each content as follows.

\subsubsection{Comparison with RIS}
Complying with \cite{Huang2020Holographic}, we consider RIS as the passive type of {HMIMO surfaces} which are capability-limited in sensing and computing, and are generally controlled by the BS over a dedicated control link. They are mainly used as passive reflectors with deployments between transceivers for realizing an intelligent and environmentally programmable communication system. In the following, we interpret RIS to comply with {HMIMO surfaces} from a holographic working principle perspective. As demonstrated in Section \ref{SectionIII}, the feeds used for exciting reference waves can be placed in different positions, namely integrated into the surface and located externally of the surface. RIS belongs to the latter scenario. Specifically, from a more macroscopic perspective, the reference wave can be EM signals emitted by one/more BSs or other communication nodes, which are refracted or reflected by the {HMIMO surface} to generate specific radiations. Under such an interpretation, the existing RIS and their expansions can be regarded as {HMIMO surfaces}, such that the introduced ``transmitter-RIS-receiver" link enables the underlying communications to be viewed as in a holographic principle guided fashion. Once the antenna elements of RIS are suitably adjusted, the resulting pattern constructed by effective radiating units can be considered as the required hologram for achieving a desired radiation. 

It is emphasized that most existing studies on RIS still apply the conventional information theory framework for both performance analysis and designs, which is insufficient as the surfaces become dense and large. The more physically-consistent EM information theory model, including EM wave theory and circuit theory, is more applicable to RIS for their future analysis and designs in reaching the fundamental limits of the reconfigurable environment.

\subsubsection{Comparison with XL-MIMO}
Notice that, while XL-MIMO has emerged as a promising technology recently, it is a natural evolution of the already established mMIMO. That being said, key changes need to be made in the analysis of XL-MIMO due to the increasing number of antenna elements. The primary refinements brought by XL-MIMO lie in the channel modeling, where the near-field effects \cite{Zhang20236G} and non-stationarities \cite{Carvalho2020Non} in the channel need to be carefully characterized and addressed. Spherical-wave propagation channel model is generally employed in existing XL-MIMO studies for its performance analysis and various designs, such as channel estimation and beam focusing \cite{Lu2022Communicating,Cui2022Channel,Zhang2022Beam} to name a few. It is emphasized that the theoretical analysis and design framework applied in XL-MIMO still follow Shannon's information theory. The existing works on XL-MIMO reveal that its evolution path is more likely the case that new models, e.g., spherical-wave propagation channel models, are developed and employed to fulfil system analyses and designs within a conventional ready-to-use framework. 
From the ``large" perspective, XL-MIMO coincides with HMIMO, because the large aperture area is one of the most distinctive features of HMIMO. The modeling and design ideas for near-field HMIMO can be beneficial from those of XL-MIMO. 

There is another new distinctive feature of HMIMO that cannot be found in XL-MIMO, which is from the ``dense" perspective. 
Specifically, HMIMO has a nearly continuous aperture with antenna elements densely packed over it, while the aperture of XL-MIMO is discrete with its antenna elements spacing complying with the half a wavelength requirement as in mMIMO. This distinction differentiates HMIMO from XL-MIMO, where the former opens up the possibility of EM-domain modeling and signal processing to approach the fundamental limit of wireless communications. HMIMO also has potential in exploiting new physical phenomena, such as mutual coupling of antenna elements. and massive OAM modes, to bring a multitude of new benefits, such as super-directivity and massive mode multiplexing, which seems obscure for XL-MIMO. In addition, compared with XL-MIMO, the theoretical analysis framework for HMIMO is greatly extended from the original Shannon's information theory to the EM information theory, which is a seamless blend of EM wave theory, circuit theory, and information theory. 

In summary, we consider XL-MIMO as a special case of HMIMO from the ``large" perspective. Compared with XL-MIMO, HMIMO not only brings new extra useful features to communication systems, but also has potential to harvest additionally appealing benefits. More importantly, the emergence of HMIMO possibly breaks the boundaries among different disciplines, e.g., communication theory, information theory, EM wave theory, circuit theory, etc., facilitating a promising interdisciplinary fusion.

\subsubsection{Comparison with mMIMO}
As one of the most important enablers for 5G wireless communications, mMIMO work through deploying a large number of antennas capable of supporting multiple parallel streams and achieving signal power enhancement. HMIMO can be considered as a brand-new revolution of conventional mMIMO, where various new characteristics are introduced. We compare HMIMO with conventional mMIMO, list their main diversities from a hardware perspective first, and discuss main comparisons in terms of communication metrics. It should be emphasized that the {HMIMO surfaces} we focus on here are mainly used as transceivers, which is in compliance with the working mode of conventional mMIMO antenna arrays.

\begin{figure*}[t!]
	\centering
	\includegraphics[height=5cm, width=18cm]{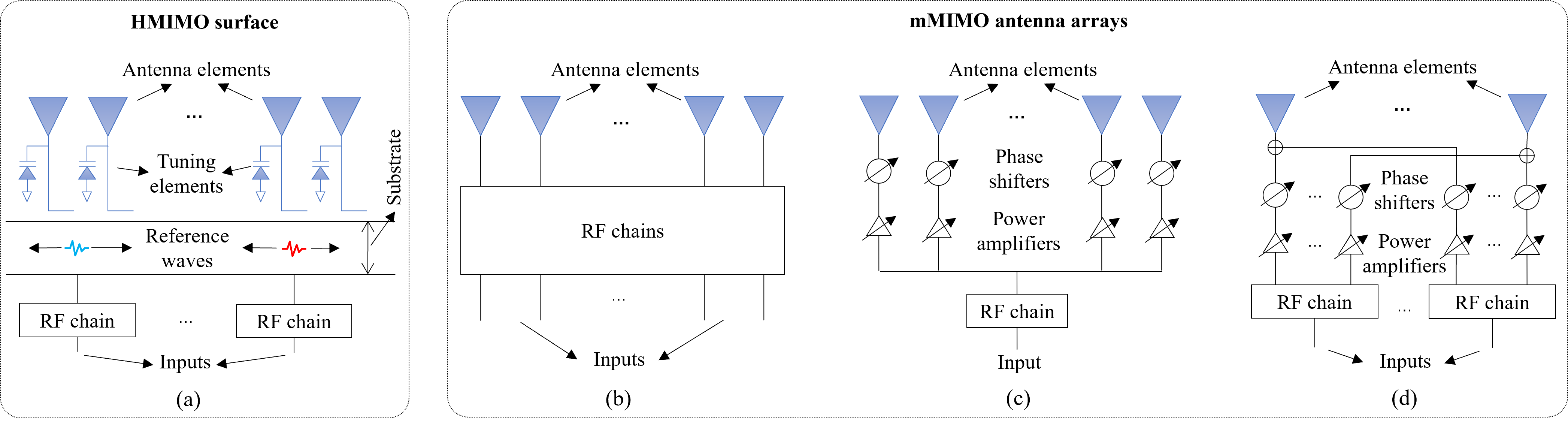}
	\caption{Block diagrams of {HMIMO surfaces} and conventional mMIMO antenna arrays: (a) {HMIMO surfaces} block diagram; (b) mMIMO antenna arrays with fully-digital processing; (c) mMIMO antenna arrays with fully-analog processing; and (d) mMIMO antenna arrays with hybrid analog-digital processing.}
	\label{fig:HMIMOS_vs_massiveMIMO}
\end{figure*}

\emph{\underline{Hardware Perspective}}: 
It should be first noted that the size of antenna elements and the spacing between two adjacent elements of {HMIMO surfaces} are in sub-wavelength size that is generally much smaller than free space wavelength \cite{Liaskos2018New}. While for conventional mMIMO antenna arrays, the spacing between adjacent antenna elements should be in half a wavelength to reduce mutual coupling, which leads to a much larger aperture size when integrating more antenna elements. From a sampling perspective, the densely packing of antenna elements results in a signal over-sampling, enabling great potential of direct manipulation of EM waves. Packing a large amount of antenna elements, {HMIMO surfaces} commonly exhibit an almost spatially continuous aperture in 2D planar shapes, and they can be potentially fabricated in almost any required shape, as shown in \ref{SectionIII_Aperture_Shape}. However, the conventional mMIMO antenna arrays with limited shape extensibility are mainly popular in spatially discrete aperture, and are mostly deployed in ULAs and moderately extended to UPAs.

Next, we show the differences in details via plain block diagrams of {HMIMO surfaces} and conventional mMIMO antenna arrays. Without loss of generality, we present block diagrams for transmitters, and omit block diagrams for receivers that can be depicted similarly. We also omit the TCA based HIMIMOS for brevity, since the processsing in these architectures takes place in the optical domain. Therefore, we mainly take the LWA based {HMIMO surfaces} for demonstration, as these schemes can straightforwardly be compared in their working principle. The block diagram of {HMIMO surfaces} is depicted at Fig. \ref{fig:HMIMOS_vs_massiveMIMO}(a), where a waveguide is included to sustain propagation of reference waves that are excited by the transmitting RF chains. Moreover, there are numerous antenna elements, each one controlled by a tuning element. The tuning elements enable a controllable capacitance that is responsible for modulating antenna elements coupled by reference waves, and manipulate both amplitude and phase of each radiating signal for generating desired beams. The mapping from inputs to radiated signals is realized by a large number of low-cost and energy efficient tuning elements (e.g., PIN diodes), and a small amount of RF chains. On the other hand, such a mapping of mMIMO antenna arrays can be implemented in three typical schematics, namely fully-digital, fully-analog and hybrid analog-digital processing, as shown in Fig. \ref{fig:HMIMOS_vs_massiveMIMO}(b)(c)(d). The fully-digital schematic relays on a large amount of RF chains to enable a fully-digital processing, with each RF chain dedicated to one antenna element. The enormous number of RF chains incurs high cost and large power consumption for mMIMO systems. Distinctively, the fully-analog schematic deals with such problems by connecting each antenna element with a phase shifter and then a power amplifier. Only one RF chain is used to supply signals to all power amplifiers. Compared to the fully-digital schematic, the fully-analog schematic mitigates the cost and power consumption problems at the expense of communication performance, i.e., from supporting multi-stream transmission to only one stream transmission. To find a balance between fully-digital and fully-analog schematics, the hybrid analog-digital framework applies phase shifters and power amplifiers at each antenna element, but one RF chain is dedicated to multiple antenna elements, thus reducing the amount of RF chains deployed\cite{Vlachos_HBF,Gong2020RF, Wang2022Two, Wang2022Hybrid, Wang2022HybridBeam}. 

Comparing conventional mMIMO antenna arrays with {HMIMO surfaces}, it is emphasized that even the fully-analog schematic, which has the least cost and power consumption, is not comparable to {HMIMO surfaces} because the phase shifters and power amplifiers are still costly and power hungry compared to the low cost and energy efficient tuning elements. Moreover, a thorough comparison between {HMIMO surfaces} and phased array antennas was conducted by the Pivotalcommware company in \cite{Pivotalcommware2019Holographic}, where the results showed that the cost and power consumption are $1/10$ and $1/3$ of that of the phased array antennas, respectively.

\emph{\underline{Directivity and Coverage}}: 
Directivity is a key performance metric for antenna arrays. It describes the concentration of antenna radiation in a particular direction. It can be used to compute the antenna array gain by further multiplying the antenna efficiency. For conventional mMIMO, the array gain is proportional to the number of packed antenna elements. However, this array gain is enhanced and has potential to reach the square of the number of antenna elements for HMIMO, due to strong mutual coupling created by almost infinite small antenna elements densely packed in sub-wavelength spacing that is much smaller than half a wavelength. The mutual coupling, considered to be harmful in conventional mMIMO design, can be potentially exploited to achieve super-directivity, a phenomenon that describes the significantly higher array gains obtained in HMIMO compared to that of mMIMO equipped with identical number of antenna elements \cite{Marzetta2019Super}. 
This phenomenon was later investigated in \cite{Williams2020Communication,Williams2021Multiuser,Han2022Coupling}, where newly designed communication models were presented, accounting for mutual coupling effects and super-directivity, based on which coupling-aware transmitter designs were developed. The results revealed that properly exploiting mutual coupling in transmitter designs indeed facilitates the realization of super-directivity, and it is important to study and design such processing frameworks, under realistic factors, e.g., antenna efficiency and ohmic losses.

The coverage of an HMIMO system is expected to be enlarged compared with conventional mMIMO systems. As stated previously, super-directivity in HMIMO can lead to higher antenna array gain over mMIMO antenna arrays. Therefore, the coverage of an HMIMO system can be further expanded over mMIMO systems with the same transmit power. To the best of our knowledge, the increase of coverage promoted by super-directivity has not been unveiled, and a future study should be performed. While prior research for reliability analysis of an LIS-based HMIMO system was carried out in \cite{Jung2019Reliability}. The authors analyzed the outage probability of LIS that can characterize the coverage, and provided a comparison of outage probabilities between LIS and mMIMO. The results showed that LIS outperforms mMIMO in terms of outage probabilities in all scenarios.

\emph{\underline{Capacity and Energy Efficiency}}: 
The capacity offered by an HMIMO communication system is envisioned to have a further increase over conventional mMIMO systems. 
Firstly, the low-cost and low power consumption of {HMIMO surfaces} make it possible to distribute more antenna elements than mMIMO for an identical aperture area. As revealed from recent studies, this can lead to super-directivity in HMIMO leading to a higher receive SNR compared with mMIMO. Following Shannon's capacity formula, the increase in receive SNR will offer an extra gain in capacity. This improvement needs to be further explored. We see that some recent studies provided discussions on array gain \cite{Hu2018LIS} and power gain \cite{Dardari2020Communicating2} for LIS-based communication systems. In another up-to-date work \cite{Wang2022HIHO} a direct comparison between LoS HMIMO and LoS mMIMO was conducted in terms of power and spectral efficiency gain in a point-to-point communication setting, which demonstrated that an up to $\pi^2$ times higher power gain and a corresponding $3.30$ bits/s/Hz spectral efficiency gain can be achieved. 
Secondly, the affordability and minimal power usage of {HMIMO surfaces} also facilitate the fabrication of extremely large apertures, which is prohibitive for conventional mMIMO antenna arrays due to power consumption and hardware cost. The extremely large apertures have a positive effect on DoF and promote communications shifting from the far-field region to the near-field region, where both distance and direction can be exploited for assisting communications. Recent research showed that the DoF of an HMIMO communication system can be higher than $1$ (i.e., the DoF of far-field mMIMO systems) in near-field LoS propagation channel \cite{Dardari2020Communicating2}, which can be exploited to promote the capacity. The extra DoF is contributed to evanescent waves in near-field region \cite{RanJi2022Extra}. 
Finally, we emphasize that the great potential of HMIMO communications can be more significant in serving multiple UEs, which is characterized in twofold. On the one hand, the enlarged coverage of an HMIMO communication network allows more UEs to be covered in an effective communication area that satisfies a required communication reliability. This will contribute an enhanced network capacity. On the other hand, the extra distance dimension introduced in near-field communications can be exploited to distinguish signals corresponding to different UEs, which can increase network capacity with an augmented capability in communicating with more UEs simultaneously. 

The energy efficiency of an HMIMO communication system is supposed to increase in comparison with mMIMO systems. As already elaborated, {HMIMO surfaces} can offer augmented performance over mMIMO, which are implemented by power-hungry phase shifters and power amplifiers, as demonstrated in \emph{Hardware Perspective}. It is natural to conclude that HMIMO communications are potentially feasible solutions in reaching energy efficient 6G networks. However, the fundamental limit of energy efficiency for HMIMO systems is not yet unveiled. Recently, some studies for realizing a DMA-enabled energy efficiency communication system were performed in \cite{xu_dynamic_2021,You2022Energy}, in which it was highlighted that DMAs can offer higher energy efficiency over mMIMO antenna arrays.

\begin{table*}[!t]
	\footnotesize
	\renewcommand{\arraystretch}{1.2}
	\caption{\textsc{Comparison between HMIMO and mMIMO.}}
	\label{tab:HMIMO_vs_mMIMO}
	\centering
	\begin{tabular}{!{\vrule width0.6pt}c|c|c!{\vrule width0.6pt}}
	
		\Xhline{0.6pt}
		\rowcolor{yellow} 
		\textbf{Metrics} & \textbf{HMIMO} & \textbf{mMIMO} \\
		
		\Xhline{0.6pt}
		\tabincell{c}{Aperture} & Nearly continuous aperture
		& Discrete aperture
		\\

        \hline 
		\tabincell{c}{{Antenna element spacing}} & {Much smaller than half a wavelength} & {Half a wavelength} \\
		
		\hline 
		Mutual coupling & \tabincell{c}{Ultra-high} & \tabincell{c}{Low (neglectable)} \\
		
		\hline 
		Antenna array gain & \tabincell{c}{High\\ (super-directivity)} & \tabincell{c}{Low} \\

        \Xhline{0.6pt}
		\tabincell{c}{Aperture area} & Extremely large & Small or moderately large \\

        \hline 
		\tabincell{c}{Communication region} & \tabincell{c}{Near-field \& \\ hybrid near-far field} & Far-field \\

		\hline 
		\tabincell{c}{Beam modes} & \tabincell{c}{Polarization and OAM modes} & \tabincell{c}{ Mainly polarization modes} \\
		
		\hline 
		\tabincell{c}{Number of beam modes} & \tabincell{c}{Infinite modes (theoretically)\\ (OAM modes)} & \tabincell{c}{Three modes\\ (linear/circular/ellipse polarization)} \\

		\hline 
		\tabincell{c}{Communication model} & \tabincell{c}{EM level model\\ (Maxwell's equations, Helmholtz\\ equation, tensor Green's function,\\ Fresnel Kirchhoff diffraction).\\ Circuit model\\ (Ohm's law, Kirchhoff's current law,\\ Kirchhoff's voltage law)} & \tabincell{c}{Mathematically abstracted model\\ (Rayleigh scattering)} \\

        \hline 
		\tabincell{c}{Signal processing domain} & \tabincell{c}{EM-domain \& \\ hybrid EM-digital domain} & Digital domain \\

		\hline 
		\tabincell{c}{Multiplexing space} & Nearly infinite and continuous
		& Limited and discontinuous
		\\
		
		\hline 
		\tabincell{c}{Multiplexing resolution} & \tabincell{c}{High\\ (follow diffraction limit)} & \tabincell{c}{Low\\ (limited by bandwidth \& beam width)} \\

		\hline 
		\tabincell{c}{Sampling domain} & Spatial sampling & Nyquist time/frequency sampling \\

		\hline 
		\tabincell{c}{Mathematical tools} & \tabincell{c}{EM information theory,\\ Kolmogorov's information theory,\\ functional analysis,\\ random process \& probability theory} & \tabincell{c}{Shannon's information theory,\\ random process \& probability theory} \\

		\Xhline{0.6pt}
	\end{tabular}
\end{table*}

\emph{\underline{Other Comparisons and Summary}}: 
Beyond the above comparisons, various differences between HMIMO and mMIMO arise and the full potential of HMIMO communications is on the way to be unveiled. To present a more panoramic view, we list a complete comparison table between HMIMO and mMIMO in Table \ref{tab:HMIMO_vs_mMIMO}. Besides the mostly mentioned differences in aperture, mutual coupling and antenna array gain, we would like to emphasize that the beam modes supported in HMIMO systems will be much more abundant than mMIMO, tending to support from conventional polarization modes to newly OAM modes owing to the powerful capabilities of {HMIMO surfaces} in EM wave manipulations. This will contribute to the revolution of wireless communications from mMIMO to massive modes, enhancing system capacity significantly in specific situations \cite{Ren2017Line, Ni2020Degrees}. In addition, the modeling of HMIMO communication systems should encompass physical phenomena, such as, EM wave propagation, mutual coupling, which are generally neglected in mMIMO modeling. For example, mMIMO channel modeling is mostly performed from a mathematical perspective by depicting wireless environment based on Rayleigh scattering. This, however, is invalid in HMIMO communications where essential physical phenomena need to be incorporated. One possible solution is modeling from an EM perspective, where EM waves follow Maxwell's and Helmholtz's equations; wireless channels can be described via Green's function; and Fresnel Kirchhoff diffraction models the light propagation. Another direction is via circuit modeling that can capture mutual coupling effects of HMIMO communication systems, providing a possible analysis tool. Future HMIMO communications are mostly considered to take place in near-field or hybrid near-far field regions, which differentiates from mMIMO, where the conventional far-field approximation holds. Compared with mMIMO with limited and discontinuous multiplexing spaces, as well as low multiplexing resolutions limited by signal bandwidth and beam width, HMIMO communications have the potential to reach nearly infinite and continuous multiplexing space with high multiplexing resolution (following the diffraction limit). This mainly contributes to the powerful capability of {HMIMO surfaces} in recording and reconstruction of EM waves in a holographic fashion. Since HMIMO opens the era of signal processing in EM-domain, conventional digital domain signal processing of mMIMO tends to face a paradigm shift. Signal processing of HMIMO communications can be performed in EM-domain or hybrid EM-digital domain. As such, sampling theory, mainly applied to time and frequency domains in mMIMO, will be possibly moved to the spatial domain. For instance, wavenumber domain channel modeling of HMIMO should be 
discretized by proper spatial sampling. This is also necessary to pattern designs of EM level system model for maximizing the communication performance. Lastly, for fully understanding, analyzing and designing HMIMO communications, new mathematical tools are essential. Shannon's information theory, random process and probability theory, popular mathematical tools in mMIMO analyses and designs, are not enough for HMIMO communications. EM information theory, Kolmogorov's information theory and functional analysis are expected to be beneficial in unveiling the full potential of HMIMO communications.

\subsection{Various Extensions}
The powerful capabilities of {HMIMO surfaces} and the proven benefits introduced in HMIMO communications bring great potentials when integrating them into various extensions. In the following of this section, we elaborate several potential HMIMO applications relevant to 6G, that are valuable for further investigation. 
The applications included but not limited to are: (i) MmWave and THz communications; (ii) WPT, wireless energy harvesting (WEH), simultaneous wireless and information power transfer (SWIPT) and wireless powered communication network (WPCN); (iii) Sensing, localization, positioning, and tracking; (iv) Satellite, UAV and vehicular communications; as well as (v) other miscellaneous applications.

\subsubsection{MmWave and THz Communications}
By deploying {HMIMO surfaces} in mmWave and THz communications, one can obtain a communication system with potential benefits, such as simplified transceiver hardware architecture, high data rates and reliable low latency communications, paving the way for seamless virtual reality (VR) experiences, as well as conquering large propagation path loss to obtain extended signal transmission distance and coverage range. 
Specifically, in \cite{Jamali2021Intelligent}, the authors studied two transmitter architectures composed of {HMIMO surfaces} that are illuminated by a few nearby active antennas each connected to a dedicated RF chain. Such an architecture facilitates an energy-efficient system capable of transmitting a phase-shifted version of incident signals from few active antennas, and exploits the array gain introduced by numerous passive antenna elements. Based upon the proposed architectures and subject to their constraints, two precoders are designed and the power consumption model considering imperfections was developed. Simulation results verified the energy efficiency and scalability of {HMIMO surfaces} for being promising candidates in building mMIMO/XL-MIMO communications.
Next, in \cite{Chaccour2020Risk}, the authors investigated a wireless VR network empowered by {HMIMO surfaces} working on THz frequencies. To meet the high requirements, a risk-based framework was suggested for data rate and reliability assurance, and a corresponding policy-based reinforcement learning (RL) algorithm enabled by a recurrent neural network (RNN) was developed for solving such a problem. The results showed high accuracy and fast convergence of the RNN. 
Enlarging signal propagation distance and coverage range in THz communications, the authors of \cite{Huang2021Multi} deployed multiple passive {HMIMO surfaces} for assisting signal transmissions between a BS and multiple single-antenna UEs, in which they propose a deep RL based hybrid digital and analog beamforming design for realizing a multi-hop communication. The deep RL based scheme is proven to be a state-of-the-art method for solving multi-hop NP-hard problems with a $50\%$ increase in coverage range compared with the zero-forcing beamforming without the assistance of {HMIMO surfaces}. 
With a particular focus on beamforming design with the consideration of the beam split problem for wideband THz communications, \cite{Yan2023Beamforming} analyzed the beam split effect with respect to different sizes, shapes, and deployments of HMIMO surfaces, and then proposed an effective beamforming scheme for alleviating the beam split effect and improving the system capacity. 
Moreover, the authors of \cite{Wan2021Terahertz} investigated {HMIMO surfaces} aided THz mMIMO communications, in which CPS and quasi-CPS apertures were considered. They theoretically derived the beam pattern with CPS apertures and revealed a decent approximation on such a beam pattern by practically feasible ultra-dense {HMIMO surfaces}. Based upon such a system, a close-loop channel estimation, including a coarse downlink together with a finer uplink channel estimation, was suggested to capture the broadband channels. The superiority of this scheme is shown in simulations.

\subsubsection{WPT, WEH, SWIPT and WPCN}
By using either active/passive HMIMO beamforming/beam focusing, {HMIMO surfaces} are capable of enhancing the strength of received signals. Such capabilities bring forth great potential in improving energy efficiency of communication systems assisted by WPT and WEH, allowing a beneficial SWIPT capability within WPCNs. 
Particularly, the authors of \cite{Luo2018Wireless} realized a highly efficient near-field WPT scheme utilizing an {HMIMO surface} whose antenna elements' layout and tuning states are configured through the holographic principle. Recently, in \cite{Zhang2022Near}, the authors presented an exploitation of spherical wavefront for near-field WPT, in which the {HMIMO surface} weights and the digital precoder are jointly designed after solving a weighted sum-harvested energy maximization problem. Simulations demonstrated the improvement in energy transfer efficiency. Furthermore, two prototypes validating near-field WPT at $5.8$ GHz are fabricated and tested in \cite{Lipworth2018Large} and \cite{Han2022Adaptively}, respectively. 
Taking advantage of WEH from information signals, an autonomous {HMIMO surface} without dedicated power supply was investigated in \cite{Ntontin2021Toward}, in which the design was implemented via dividing antenna elements into two subsets that are responsible for energy harvesting and beamforming, respectively. Efficient subset allocation policies were proposed for solving formulated problems with a suggested feasibility of autonomous {HMIMO surfaces}. 
Building up WPT and WEH capabilities, passive {HMIMO surfaces} assisted SWIPT systems are particularly investigated in \cite{Wu2020Weighted, Pan2020SWIPT}, where different optimization problems, such as the maximization of weighted sum power or the maximization of weighted sum-rate, complying with distinct constraints, e.g., signal-to-interference-plus-noise ratios of information receivers or energy harvesting requirements of energy receivers, were established and solved based upon the alternating optimization algorithm or the block coordinate descent algorithm. A further extension on multiple passive {HMIMO surfaces} aided SWIPT communications was studied in \cite{Wu2020Joint} with transmit power minimized under different quality-of-service constraints. In \cite{Xu2021WPCN}, the authors investigated passive {HMIMO surfaces} assisted WPCNs by jointly designing radio resource allocation and passive beamforming under an energy efficiency maximization problem.

\subsubsection{Sensing, Localization, Positioning, and Tracking}
With a completed exploration of their capabilities, one can expect that {HMIMO surfaces} have a great potential in assisting the realization of other attractive functionalities of 6G, such as sensing, localization and tracking. First of all, sensing is an important capability that perceives the wireless environment states. In order to mitigate spectrum congestion, the authors of \cite{Zhang2022Holographic} introduced {HMIMO surfaces} in promoting the integrated sensing and communication performance, in which an amplitude-controllable holographic beamformer is optimized, and a lower bound on the maximal beampattern gain is theoretically analyzed. They showed that more than $50\%$ increase of beamforming gain can be achieved with a reduced cost when comparing to the same size MIMO arrays. Exploring the benefit of passive {HMIMO surfaces} in spectrum sensing, the authors of \cite{Ge2022RIS} evaluated the detection probability in an asymptotic fashion. Under this setting, equivalent channel gains were provided and an insight on the required number of reflecting elements for achieving a near $100\%$ detection probability was theoretically analyzed. Integrating spectrum sensing and learning capabilities into active/semi-active multi-function and/or stacked {HMIMO surfaces} \cite{Yang2023Reconfigurable,An2023Stacked}, the framework of spectrum learning aided {HMIMO surfaces} were presented in \cite{Yang2021Intelligent, Yang2021Spectrum, Yang2022Federated}. Capitalizing on this spectrum learning ability, they are capable of reflecting useful signals while ignoring interfering signals based on an intelligent `think-and-decide' process.
Beyond wireless sensing, wireless localization and positioning are envisioned as an essential function of 6G shifting from the previous add-on feature. The approximately CPS paves the way to reach a high localization resolution. To this end, the authors of \cite{DAmico2021Cramer} studied the fundamental limits of holographic positioning in the near-field regime. The Cram{\'e}r-Rao lower bound (CRLB) was computed based on a combination of EM propagation theory and estimation theory, considering several cases, such as with/without prior knowledge of source orientation and a specific type of source location.  \cite{He20222D} utilized multiple receiving RISs comprising a new large one that acts as an HMIMO receiver, in order to perform 3D localization. Additionally, passive {HMIMO surfaces} aided localization was studied in \cite{Elzanaty2021Reconfigurable} obtaining the CRLB of ultimate localization and orientation, and providing a closed-form phase configuration for joint communication and localization. The near-field positioning was extended to near-field position and velocity tracking for a moving object in \cite{Guerra2021Near}, in which posterior CRLB was derived, and different Bayesian tracking algorithms were studied with respect to the accuracy and complexity under various cases e.g., parameter mismatches and abrupt trajectory changes. {HMIMO surfaces} are also expected to constitute an enabling technology for simultaneous localization and mapping (SLAM), where a map of the environment 
is built and the user can self-localize with respect to the map. The authors in \cite{Kim2022RIS} proposed an RIS-enabled framework for SLAM without any access points.

\subsubsection{Satellite, UAV and Vehicle Communications}
Making use of the excellent characteristics, e.g., low size, weight, cost, power consumption, and flexible aperture shapes, as well as powerful capabilities, it is prospected that {HMIMO surfaces} will assist satellite, UAV and vehicular communications by mitigating current challenges, such as power constraints, severe path loss, and hardware limitations, etc. More in detail, in \cite{Deng2022Holographic}, the authors applied {HMIMO surfaces} to LEO satellite communications to enable HMIMO communications. They designed a temporal variation law guided LEO satellite tracking scheme, eliminating the overhead of satellite positioning, and also developed a tracking error robust holographic beamforming algorithm for sum-rate maximization. In assisting the air-to-ground communication, the authors of \cite{Zhao2022Simultaneously} utilized an {HMIMO surface}, capable of simultaneously transmitting and reflecting on both sides of the aperture, to support a maximal sum-rate. A joint optimization of UAV trajectory as well as active and passive beamformings were conducted subject to the flight safety, maximum flight duration and minimum data rates of ground UEs. An RL framework and a worst-case performance guaranteed distributionally-robust RL algorithm were developed accordingly. Utilizing a similar simultaneously transmitting and reflecting {HMIMO surface} for improving double fading effect faced in vehicular communications, the authors of \cite{Chen2022Robust} jointly designed BS digital and analog beamforming based on imperfect CSI for minimizing the transmit power. A specific transmission protocol was presented for achieving the CSI acquisition with low channel training overhead. Resource allocation was optimized based on alternating optimization and constrained stochastic successive convex approximation algorithms. Furthermore, the authors of \cite{Mizmizi2022Conformal} proposed to deploy a conformal {HMIMO surface} on vehicle surface to release blockage impact. They theoretically gave the phase pattern of a cylindrical {HMIMO surface} via generalizing the conventional planar {HMIMO surfaces} and prove the benefits.

\subsubsection{Other Miscellaneous Applications}
Beyond the aforementioned extensions, the emergence of {HMIMO surfaces}, especially in their nearly passive regime, spurs a vast range of research combinations by introducing the {surfaces} to various areas, including physical layer security \cite{Yang2020Secrecy, Hong2021Robust, Khan2022Opportunities,Naeem2023Security}, index modulation \cite{Basar2020Reconfigurable, Guo2020Reflecting, Lin2021Reconfigurable,Yuan_reflection_modulation,RISindex}, non-orthogonal multiple access \cite{Hou2020Reconfigurable, Liu2020RIS, Yang2021Reconfigurable}, cognitive radio \cite{Yuan2020Intelligent, Zhang2020Intelligent, Dong2021Secure}, ambient backscattering communications \cite{Nemati2020Short, Basharat2022Reconfigurable, Liang2022Backscatter, Wang2016Ambient}, full duplex wireless communications \cite{melissa1,smida2022full, alexandropoulos2022fullduplex, alexandropoulos2021low, melissa2, Elhattab2021Reconfigurable, Peng2021Multiuser, Yang2021Optimal}, cell-free networks \cite{Gan2022Multiple, Van2022, Hao2022Robust}, mobile edge computing \cite{Chu2020Intelligent, Huang2021Reconfigurable, Cao2021Converged, Merluzzi2022MEC,Stylianopoulos_mec2022}, federated learning \cite{Liu2021RISFL, Battiloro2022Dynamic, Ni2022SemiFL}, and machine learning aided applications \cite{Huang2020Reconfigurable, Xu2022A, Liu2022Deep,Xu2022Time}. It is emphasized that these studies consider the application of {HMIMO surfaces} as passive reflectors, incorporated in conventional communication systems, e.g. MIMO and mMIMO. It is rare to see that a communication system employs {HMIMO surfaces} as active transceivers and utilizes the holographic principle, EM-domain analysis based signal processing to enable HMIMO communications. As such, theoretical analyses and practical algorithms of HMIMO communications for these cross-domain areas remain to be unveiled.

\section{Research Challenges and Future Directions}
\label{SectionVII}
We describe the recent advances of the HMIMO technology, and show their great potential in future 6G era by embracing the physical aspects of {HMIMO surfaces}, theoretical foundations and enabling technologies of HMIMO communications. It should be noted that this area is still in its initial stage, where several research challenges should be carefully addressed and mitigated for making HMIMO practical. The grand challenges will bring new opportunities as future directions. In the following of this section, we present and discuss some of the most critical research challenges of the HMIMO technology. 

\subsection{Physical Level Design and Experimentation}
To date, various {HMIMO surfaces} are designed based on different holographic design methodologies, which demonstrate the conceptual feasibility. However, numerous questions for practical implementations remain. For example, in LWA based {HMIMO surfaces}, the substrate performing as the role of a waveguide exhibits a signal loss effect in reference wave propagation, such that the radiated signals of all antenna elements leaked from the substrate experience different attenuated reference waves, which will lead to biased transmission and reception different from the lossless substrate. Furthermore, the generated fields by antenna elements have an inevitable influence back toward the reference wave propagated along the substrate. These non-ideal hardware imperfections influence the aperture size design of {HMIMO surfaces}, which will be more critical as the apertures become large. The aforementioned non-idealities should be carefully tested and compensated. Another critical question is that the dense deployment of antenna elements will unavoidably introduce strong mutual coupling that decreases the radiation efficiency and performance. The mutual coupling effect must be carefully characterized and should be suppressed or properly exploited by exploring advanced hardware structures. 
It should be further emphasized that the hardware design and optimization become challenging as the aperture size and antenna elements tend to the large regime. The complexity mainly arises from EM numerical computations that usually involve a vast amount of variables to be optimized, namely, aperture size, the number of antenna elements and their configuration and various practical imperfections. Hence, efficient methods and strategies should be designed. From the existing literature we conclude that the current {HMIMO surfaces} are mainly designed and fabricated for supporting a single or a few multiple beams that is far away from practical requirements that dozens of UEs or more should be served simultaneously through different concurrent beams. The hardware design to enable large multiplexing capability of HMIMO should be further investigated. 
On the other hand, in TCA based {HMIMO surfaces}, the key challenge in physical level design mainly emerges from the holographic RF-optical mapping, where the phases of parallel optical beams, mapped from the corresponding RF signals, are sensitive to tiny environmental variations, such as temperature, vibration, airflow and sound, which decreases the mapping performance. Stable holographic RF-optical mapping should be examined. 
To promote HMIMO for practical use, experimenting with {HMIMO surfaces} for realistic communications under various setups should be performed to validate the potential.


\subsection{HMIMO Theoretical Foundations}

\subsubsection{EM-Domain Models} 
Exploring accurate EM-domain models, especially the channel and communication models, for HMIMO is of great importance, since it allows us to unveil the fundamental limits, and create effective designs and signal processing frameworks. Existing EM-domain channel models either focus on the near-field LoS scenario or the far-field NLoS scenario, which are not enough for scenarios that the scatterers, producing NLoS paths, can possibly fall within the near-field regions of extremely large HMIMO.  Thus, the near-field NLoS channel model or even a potentially unified EM-domain channel model are expected to solve this problem. In addition, an accurate EM-domain communication model has not been fully addressed due to the incompleteness of effective characterizations in mutual coupling, and the insufficient EM-domain channel modeling. A completed depiction and integration of these EM-domain models should be further studied. 
Another important research direction is to seamlessly blend models from the conventional information/communication theory with models from the EM wave theory and circuit theory, which is not only necessary to HMIMO communications but also to EM information theory.

\subsubsection{EM-Domain Fundamental Limits} 
The classic Shannon's information theory provides a solid framework for modern wireless communications, which mathematically treats the wireless channel as a conditional probability distribution, offering a mathematical abstraction for analyzing communication systems from the time-domain. It however reaches its bottleneck in analysis of HMIMO fundamental limits due to the lack of physically-consistent characterizations of EM wave and circuit signal transmissions as well as the lack from a more general spatial-temporal-domain perspective, thereby hampering the straightforward adoption of Shannon's information theory to the HMIMO regime. 
The physically-consistent EM information theory should be established by considering the general spatial-temporal-domain signals. In an attempt to do so, the slight asymmetry between information in the temporal-domain and in the spatial-domain is retained in \cite{Migliore2021The}. Specifically, the capacity of communications could be improved with an increase of the observation time, which is similar in the spatial-domain by enlarging the size of HMIMO surfaces. 
However, different from the infinity of the observation time, the aperture size is limited, thereby, posing new challenges in deriving the performance bound of HMIMO communications from the spatial-domain perspective. Establishing a spatial-domain analysis framework and/or building a general spatial-temporal-domain framework are of great significance.

\subsubsection{Near-Field EM Wave Sampling}  
Reconstructing a continuous EM field from the spatially discrete samples by EM wave sampling is necessary for effective signal processing. Generally, a minimum number of spatial samples exists for accurately reconstructing the spatially continuous EM wave with acceptable information loss. However, the EM wave sampling is still in its infancy and a unified and accepted mathematical conclusion is not yet reached, especially for the near-field case. To that end, \cite{Pizzo2022Nyquist} investigated the Nyquist sampling in reconstructing a far-field EM wave by solving the circle/ellipse packing problems, where the authors proposed that conventional signal processing techniques for band-limited signals can be applied to EM waves due to a fundamental spatial-temporal duality. However, the wavenumber domain spectral support of near-field regions is typically infinite, thus, the mentioned sampling method in \cite{Pizzo2022Nyquist} is not applicable. Therefore, EM wave sampling in near-field with an arbitrary low reconstruction error is rather challenging. In addition, the tractable relationship between the truncation error and sampling points in near-filed is also valuable to be explored.

\subsection{HMIMO Signal Processing}

\subsubsection{HMIMO Channel Estimation} 
The extremely large number of antenna elements incorporated in {HMIMO surfaces} entails prohibitive pilot overheads, therefore, an effective channel estimation method has become a crucial issue. There are many works investigating channel estimation of massive parameters, which can be categorized mainly into two groups, i.e., estimating the general channel model at high pilot expense \cite{Demir2022Channel}, or estimating the sparse channel model at low pilot overheads \cite{Cui2022Channel}. However, both methods suffer from either expensive pilot cost or performance loss. Specifically, the former one is applicable to a general channel model since the pilots are long enough to estimate all unknown variables, while the latter one takes advantages of sparsity to adopt fewer pilots in channel estimation as the number of unknown parameters reduces. In fact, the sparsity is not available in general frameworks, resulting in estimation performance degradation. Consequently, a cost-efficient channel estimation method using lower overheads in a generic system model is tricky. Besides, developing new channel estimation approaches consistent with newly built channel models is necessary and has paramount potential in realizing high accurate channel estimations.

\subsubsection{HMIMO Beamforming and Beam Focusing}
HMIMO can possibly achieve high beamforming gains and large spatial multiplexing since it incorporates nearly spatially continuous and extremely large apertures covered with dense sub-wavelength antenna elements, which can potentially be exploited using HMIMO beamforming and beam focusing. Various angle-aware beamforming schemes were proposed in wireless communications \cite{Wu2020Joint,Mishra2019Channel,Zhang2022Holographic}, which are, however, unable to deal with the mutual coupling effects and distinguish UEs with close elevation/azimuth angles, thereby failing to offer additional beamforming and spatial multiplexing gains. Therefore, designing effective beamforming and beam focusing schemes for HMIMO systems to adapt the dense and large characteristics is demanding. 
Specifically, such frameworks need to incorporate both mutual coupling effects and distance-angle-aware functionalities, which will prove critical especially for near-field HMIMO communications. 
In addition, designing valid beamforming and beam focusing approaches, especially based on newly established theoretical foundations, such as the EM information theory, is of great significance. 
Last but not least, since {HMIMO surfaces} consist of excessively large number of antenna elements, conventional beamforming design schemes cannot be applied directly, due to the high computational complexity. Accordingly, low-complexity HMIMO beamforming and beam focusing designs are vital to enable practical systems. 

\section{Conclusions}
\label{SectionVIII}
In this paper, we have presented a comprehensive overview of the key features and recent advances of HMIMO wireless communications. We first presented a multitude of holographic applications and listed representative holographic technology roadmaps for future HMIMO communications. We then emphasized on three main components of HMIMO communications, namely, physical aspects of {HMIMO surfaces} and their theoretical foundations, as well as enabling technologies of HMIMO communications. In the first component, we embraced the physical aspects of {HMIMO surfaces} in terms of their hardware structures, holographic design methodologies, tuning mechanisms, aperture shapes, functionalities, as well as representative state-of-the-art prototypes. In the second component, we presented theoretical foundations of HMIMO communications with respect to channel modeling, DoF, and capacity limits, and overviewed the {HMIMO surfaces} capability for EM-field sampling, as well as the resulting emerging research area of EM information theory. In the last component, we presented recent advances on physical-layer HMIMO enabling technologies, and in particular, on HMIMO channel estimation and HMIMO beamforming/beam focusing. We also compared HMIMO communications with existing technologies, especially mMIMO communications, and discussed a variety of extensions of HMIMO. We finally presented a comprehensive list of technical challenges and open research directions that we believe they will drive unprecedented research promotions in the future.


%

\bibliographystyle{IEEEtran}
\bibliography{IEEEabrv,references} 
\balance

%




\end{document}